\documentclass[%
 aip,
 amsmath,amssymb,
preprint,%
]{revtex4-1}

\usepackage{graphicx}
\usepackage{dcolumn}
\usepackage{bm}
\usepackage{color}

\usepackage[utf8]{inputenc}
\usepackage[T1]{fontenc}
\usepackage{mathptmx}

\usepackage[left]{lineno}
\newcommand*\patchAmsMathEnvironmentForLineno[1]{%
  \expandafter\let\csname old#1\expandafter\endcsname\csname #1\endcsname
  \expandafter\let\csname oldend#1\expandafter\endcsname\csname end#1\endcsname
  \renewenvironment{#1}%
     {\linenomath\csname old#1\endcsname}%
     {\csname oldend#1\endcsname\endlinenomath}}%
\newcommand*\patchBothAmsMathEnvironmentsForLineno[1]{%
  \patchAmsMathEnvironmentForLineno{#1}%
  \patchAmsMathEnvironmentForLineno{#1*}}%
\AtBeginDocument{%
\patchBothAmsMathEnvironmentsForLineno{equation}%
\patchBothAmsMathEnvironmentsForLineno{align}%
\patchBothAmsMathEnvironmentsForLineno{flalign}%
\patchBothAmsMathEnvironmentsForLineno{alignat}%
\patchBothAmsMathEnvironmentsForLineno{gather}%
\patchBothAmsMathEnvironmentsForLineno{multline}%
}

\allowdisplaybreaks[1]

\begin{document}


\title[
]{Role of various scale-similarity models in stabilized mixed subgrid-scale model}

\author{Kazuhiro Inagaki}%
 \email{kinagaki@iis.u-tokyo.ac.jp}%
 \affiliation{Institute of Industrial Science, The University of Tokyo, 4-6-1 Komaba, Meguro-ku, Tokyo 153-8505, Japan}%

\author{Hiromichi Kobayashi}%
\affiliation{Department of Physics \& Research and Education Center for Natural Sciences, Hiyoshi Campus, Keio University, 4-1-1 Hiyoshi, Kohoku-ku, Yokohama 223-8521, Japan}%

\date{\today}

\begin{abstract}
We investigate the physical role of various scale-similarity models in the stabilized mixed model [K. Abe, Int. J. Heat Fluid Flow, \textbf{39}, 42 (2013); M. Inagaki and K. Abe, Int. J. Heat Fluid Flow, \textbf{64}, 137 (2017)] and evaluate their performance in turbulent channel flows. Among various models in the present study, the original model combined with the scale-similarity model for the subgrid-scale (SGS)-Reynolds term yields the best prediction for the anisotropy of the grid-scale (GS) velocity fluctuations and the SGS stress, even in coarse grid resolutions. Moreover, it successfully predicts large intensities of the spectra close to the cut-off scale in accordance with the filtered direct numerical simulation, whereas other models predict a rapid decay of the spectra in the low-wavelength region. To investigate the behavior of the models close to the cut-off scale, we analyze the budget equation for the GS Reynolds stress spectrum. The result shows that the scale-similarity model for the SGS-Reynolds term plays a role in the enhancement of the wall-normal velocity fluctuation close to the cut-off scale. Thereby, it activates turbulence close to the cut-off scale, leading to a reproduction of the proper streak structures observed in wall-bounded turbulent flows. The reproduction of velocity fluctuations close to the cut-off scale and turbulent structures is a key element for further development of SGS models.
\end{abstract}

\maketitle

\section{\label{sec:level1}Introduction}

Large-eddy simulation (LES) is an essential tool employed to predict high-Reynolds-number turbulent flows. LES solves large-scale or grid-scale (GS, resolved scale, or super-filter scale) eddies in turbulent flows. Meanwhile, effects of subgrid-scale (SGS, unresolved, or sub-filter scale) turbulent eddies are modeled. This procedure is referred to as SGS modeling. SGS modeling was studied for half a century since the pioneering work by Smagorinsky\cite{smagorinsky1963}. Numerous practical SGS models, including the Smagorinsky model\cite{smagorinsky1963}, are based on the linear eddy-viscosity assumption, which models the effect of SGS eddies as an effective viscosity. To date, several eddy-viscosity type models have been proposed, e.g., dynamic models\cite{germanoetal1991,lilly1992,meneveauetal1996}, one-equation models\cite{yh1985,germanoetal1991,ghosaletal1995}, and modified local eddy-viscosity models\cite{nd1999,vreman2004,kobayashi2005}. Recently, LES approach is also applied to the lattice-Boltzmann method, employing the eddy-viscosity concept\cite{ac2010}.

Although the eddy-viscosity models are simple and handy, their physical reliability are not sufficient. Several studies showed that the principal axis of the exact SGS stress tensor does not generally align with that of the strain rate tensor\cite{clarketal1979,bardinaetal1983,liuetal1994,taoetal2002,horiuti2003}. Hence, the eddy-viscosity models cannot reproduce the exact property of the SGS stress tensor; however, they are reasonable for estimation of the energy transfer rate from the GS to SGS fields\cite{clarketal1979,kobayashi2005}. A traditional approach to improve the SGS model is to employ scale-similarity models\cite{bardinaetal1983}. Scale-similarity models were shown to yield a better correlation with the exact SGS stress than eddy-viscosity models\cite{clarketal1979,bardinaetal1983,taoetal2002,horiuti2003,liuetal1994,kobayashi2018}. 
However, such scale-similarity models are not sufficiently dissipative to be employed by themselves for a stable performance of the LES. Moreover, scale-similarity models cause backscatter or energy transfer from the SGS to GS fields\cite{horiuti1989,liuetal1994,horiuti1997jpsj,taoetal2002}, which can induce numerical instability. A remedy for these difficulties is to combine scale-similarity models with the eddy-viscosity model. The resulting model is referred to as the mixed model, which was first proposed by Bardina \textit{et al}.\cite{bardinaetal1983}. Several types of mixed models have been suggested to date (see, Refs.~ \onlinecite{piomellietal1988,zangetal1993,liuetal1994,vremanetal1994,sb1995,horiuti1997,am1999,sarghinietal1999,mv2001}). However, the backscatter caused by scale-similarity models still makes the mixed model difficult to apply it to engineering problems with complex geometries. Furthermore, Anderson and Domaradzki\cite{ad2012} showed that the conventional scale-similarity model yields an excessive dissipation directly from the largest resolved scales, which is unphysical in the sense of localness in scale of energy transfer. In this sense, the physics of scale-similarity models itself also needs to be discussed in detail. As another branch of recent developments of SGS modeling, algebraic stress modeling approach, which was first developed in the Reynolds-averaged Navier--Stokes (RANS) modeling (see, e.g., Refs.~\onlinecite{pope1975,tsdia,gs1993,wj2000}), was discussed and its performance was evaluated\cite{marstorpetal2009,montecchiaetal2017,montecchiaetal2019}.



For practical use of the SGS model, both applicability and physical consistency of the model are required. As a unique approach to utilize the physics of scale-similarity models in a numerically stable manner, Abe\cite{abe2013} proposed a new formalism of the mixed model, which is referred to as the stabilized mixed model (SMM). 
Further discussion on the modification or other modeling approaches of the SMM are provided in Refs.~\onlinecite{ia2017,kobayashi2018,kleinetal2020}. Surprisingly, the SMM is significantly less sensitive to the grid resolution than conventional eddy-viscosity models\cite{abe2013,oa2013,abe2014}. Moreover, overestimation of the streamwise velocity fluctuation, which is often observed in the LES at coarse grid resolutions, decreases significantly, yielding a better prediction of the anisotropy of the GS velocity fluctuations in comparison with the direct numerical simulation (DNS). However, the physical mechanism that the SMM yields better results is still under discussion. Otsuka and Abe\cite{oa2013} showed that the SMM maintains streamwise vortices even at coarse grid resolutions in turbulent channel flows, hence reproducing the mean velocity profile. 
Abe\cite{abe2019} showed that the non-eddy-viscosity term contributes significantly to the generation of the GS Reynolds shear stress in the channel flow by performing an \textit{a priori} test. 
These results suggest that a non-eddy-viscosity term is a key element to improve physical properties of SGS models.

In the SMM, Abe\cite{abe2013} adopted the scale-similarity model for the SGS-Reynolds term, following the suggestion of Horiuti\cite{horiuti1993}, on the proper velocity scale for the SGS energy. However, previous studies show that other scale-similarity models yield better correlations with the exact SGS stress\cite{clarketal1979,bardinaetal1983,taoetal2002,horiuti2003,liuetal1994,kobayashi2018}. 
In this sense, physical consistency and reliability of the SMM are still unclear, albeit its attractive performance. If the SMM is used without understanding the physical properties, the results of the SMM could not be evaluated exactly in its application to turbulent flows. Therefore, it is required to reveal physical properties of the SMM. To understand the physics of the SMM is helpful for further development of SGS modeling. Now, we pose a question regarding which scale-similarity model exhibits the best performance in predicting turbulent flows. To investigate the physics of SGS models, we should analyze their properties in their applications to numerical simulations; namely, \textit{a posteriori} tests. In the present study, we construct the SMM using various scale-similarity models and evaluate their performance in turbulent channel flows. 

The rest of this paper is organized as follows. In Sec.~\ref{sec:level2}, we describe the modeling procedure of the SMM\cite{abe2013}. In Sec.~\ref{sec:level3}, we construct various types of SMMs and evaluate their performance in turbulent channel flows. In this section, we find that the reproduction of the GS and SGS anisotropies significantly depends on SGS models, through the Lumley's invariant map. Moreover, we address that the critical difference among the models lies in the low-wavelength region of the GS Reynolds spectrum.
To investigate the physical origin of the difference in the low-wavelength region, we discuss the budget for the GS Reynolds stress spectrum\cite{mizuno2016,ka2018,lm2019} in Sec.~\ref{sec:level4}. In this section, we also discuss the relation between the spectrum and the streak structures observed in wall-bounded turbulent flows. Conclusions are given in Sec~\ref{sec:level5}.

\section{\label{sec:level2}Stabilized mixed model with various scale-similarity terms}

\subsection{\label{sec:level2.1}Governing equations and scale-similarity models}

The governing equations in LES for an incompressible fluid are the filtered continuity and Navier--Stokes equations:
\begin{gather}
\frac{\partial \overline{u}_i}{\partial x_i} = 0,
\label{eq:2.1}\\
\frac{\partial \overline{u}_i}{\partial t} = - \frac{\partial}{\partial x_j} ( \overline{u}_i \overline{u}_j + \tau^\mathrm{sgs}_{ij} )
- \frac{\partial \overline{p}}{\partial x_i} + \frac{\partial}{\partial x_j} (2 \nu \overline{s}_{ij}),
\label{eq:2.2}
\end{gather}
where $\overline{\cdot}$ denotes the filtering operation, $\overline{u}_i$ is the GS velocity, $\tau^\mathrm{sgs}_{ij} (= \overline{u_i u_j} - \overline{u}_i \overline{u}_j)$ is the SGS stress, $\overline{p}$ is the pressure divided by the density, $\nu$ is the kinematic viscosity, and $\overline{s}_{ij} [ = (\partial \overline{u}_i/\partial x_j + \partial \overline{u}_j/\partial x_i)/2 ]$ is the GS strain rate. To close the equations, we must model the SGS stress $\tau^\mathrm{sgs}_{ij}$. 
A conventional model for $\tau^\mathrm{sgs}_{ij}$ is an eddy-viscosity type model:
\begin{align}
\tau^\mathrm{sgs}_{ij} = \frac{1}{3} \tau^\mathrm{sgs}_{\ell \ell} \delta_{ij} - 2 \nu^\mathrm{sgs} \overline{s}_{ij},
\label{eq:2.3}
\end{align}
where $\nu^\mathrm{sgs}$ denotes the SGS eddy viscosity. A widely accepted expression for $\nu^\mathrm{sgs}$ was proposed by Smagorinsky\cite{smagorinsky1963}, which expresses $\nu^\mathrm{sgs}$ through the filter width $\overline{\Delta}$ and the GS strain rate $\overline{s}_{ij} \overline{s}_{ij}$. The dynamic approach\cite{germanoetal1991,lilly1992,meneveauetal1996} extends applicability of the Smagorinsky model\cite{smagorinsky1963} to more complex turbulent flows. However, these models involve an intrinsic shortfall of negative viscosity, which causes exponential divergence\cite{germanoetal1991,horiuti1997}. As a remedy, other local expressions for $\nu^\mathrm{sgs}$ have been proposed\cite{nd1999,vreman2004,kobayashi2005}.

Although the eddy-viscosity models are simple and handy, several studies showed that the strain rate tensor does not necessarily align with the exact SGS stress\cite{clarketal1979,bardinaetal1983,liuetal1994,taoetal2002,horiuti2003}. An approach to improve this problem is to employ scale-similarity models\cite{bardinaetal1983}. Conventionally, the SGS stress is decomposed into the following three terms\cite{popebook}:
\begin{subequations}
\begin{gather}
\tau^\mathrm{sgs}_{ij} = \mathcal{L}_{ij} + \mathcal{C}_{ij} + \mathcal{R}_{ij}, 
\label{eq:2.4a} \\
\mathcal{L}_{ij} = \overline{\overline{u}_i \overline{u}_j} - \overline{u}_i \overline{u}_j, 
\label{eq:2.4b} \\
\mathcal{C}_{ij} = \overline{\overline{u}_i u_j''} + \overline{u_i'' \overline{u}_j}, 
\label{eq:2.4c} \\
\mathcal{R}_{ij} = \overline{u_i'' u_j''},
\label{eq:2.4d}
\end{gather}
\end{subequations}
where $u_i'' = u_i - \overline{u}_i$. We refer to $\mathcal{L}_{ij}$, $\mathcal{C}_{ij}$, and $\mathcal{R}_{ij}$ as the Leonard, cross, and SGS-Reynolds terms, respectively. Note that the sum of the Leonard and cross terms satisfy the Galilean invariance, although the individual terms do not\cite{speziale1985}. The scale-similarity assumption\cite{bardinaetal1983} gives the approximation $\overline{u_i'' u_j} \simeq \overline{u_i''} \ \overline{u}_j$. Under this scale-similarity assumption, the cross and SGS-Reynolds terms, Eqs.~(\ref{eq:2.4c}) and (\ref{eq:2.4d}), yield

\begin{subequations}
\begin{gather}
\mathcal{C}_{ij} \simeq \overline{\overline{u}}_i \overline{u_j''} + \overline{u_i''} \overline{\overline{u}}_j
= \overline{\overline{u}}_i (\overline{u}_j - \overline{\overline{u}}_j) + (\overline{u}_i - \overline{\overline{u}}_i ) \overline{\overline{u}}_j, 
\label{eq:2.5a} \\
\mathcal{R}_{ij} \simeq \overline{u_i''} \ \overline{u_j''} = (\overline{u}_i - \overline{\overline{u}}_i) (\overline{u}_j - \overline{\overline{u}}_j). 
\label{eq:2.5b}
\end{gather}
\end{subequations}
The sum of $\mathcal{L}_{ij}$, $\mathcal{C}_{ij}$, and $\mathcal{R}_{ij}$ under the scale-similarity assumption reads
\begin{align}
\mathcal{L}_{ij} + \mathcal{C}_{ij} + \mathcal{R}_{ij} \simeq \mathcal{L}^\mathrm{m}_{ij}
= \overline{\overline{u}_i \overline{u}_j} - \overline{\overline{u}}_i \overline{\overline{u}}_j,
\label{eq:2.6}
\end{align}
where $\mathcal{L}^\mathrm{m}_{ij}$ is referred to as the modified Leonard term, which is a redefined Leonard term employed to satisfy the Galilean invariance\cite{germano1986}. Employing the Gaussian or top-hat filter as the test filter, the Taylor expansion of the filtered velocity yields
\begin{align}
\widehat{\overline{u}}_i = \overline{u}_i + \frac{\widehat{\Delta}_\ell^2}{24} \frac{\partial^2 \overline{u}_i}{\partial x_\ell \partial x_\ell} + O (\widehat{\Delta}^4),
\label{eq:2.7}
\end{align}
where the test filter operation $\widehat{\cdot}$ is used to explicitly show which filter is expanded, and $\widehat{\Delta}_i$ is the filter width in the $x_i$ direction accompanied with $\widehat{\cdot}$. Hence, the modified Leonard term is expanded as
\begin{align}
\mathcal{L}^\mathrm{m}_{ij}
& = C_{ij} - \frac{\overline{\Delta}_\ell^2}{24} \frac{\partial^2 \overline{u}_i}{\partial x_\ell \partial x_\ell} \frac{\overline{\Delta}_m^2}{24} \frac{\partial^2 \overline{u}_j}{\partial x_m \partial x_m} + O (\overline{\Delta}^4) \nonumber \\
& = C_{ij} + O (\overline{\Delta}^4),
\label{eq:2.8} \\
C_{ij} & = \frac{\overline{\Delta}_\ell^2}{12} \frac{\partial \overline{u}_i}{\partial x_\ell} \frac{\partial \overline{u}_j}{\partial x_\ell},
\label{eq:2.9}
\end{align}
where $C_{ij}$ is referred to as the Clark term. The second term on the first line of Eq.~(\ref{eq:2.8}) corresponds to the Taylor expansion of the scale-similarity model for the SGS-Reynolds term $\mathcal{R}_{ij}$;
\begin{align}
\mathcal{R}_{ij} \simeq \frac{\overline{\Delta}_\ell^2}{24} \frac{\partial^2 \overline{u}_i}{\partial x_\ell \partial x_\ell} \frac{\overline{\Delta}_m^2}{24} \frac{\partial^2 \overline{u}_j}{\partial x_m \partial x_m}. 
\label{eq:2.10}
\end{align}
In this paper, we refer to terms expressed by Eqs.~(\ref{eq:2.5a}), (\ref{eq:2.5b}), (\ref{eq:2.6}), and (\ref{eq:2.9}) as scale-similarity models.

\subsection{\label{sec:level2.2}Stabilized mixed model}

Abe\cite{abe2013} proposed the following mixed model, referred to as the SMM:
\begin{gather}
\tau^\mathrm{sgs}_{ij} = \frac{2}{3} k^\mathrm{sgs} \delta_{ij} - 2 \nu^\mathrm{sgs} \overline{s}_{ij} + \tau^\mathrm{eat}_{ij}, \nonumber \\
\tau^\mathrm{eat}_{ij} = 2k^\mathrm{sgs} \frac{\tau^\mathrm{a}_{ij}|_\mathrm{tl} + 2 \nu^\mathrm{a} \overline{s}_{ij}}{\tau^\mathrm{a}_{\ell \ell}}, \ \ 
\nu^\mathrm{a} = - \frac{ \tau^\mathrm{a}_{ij}|_\mathrm{tl} \overline{s}_{ij}}{2 \overline{s}_{\ell m} \overline{s}_{\ell m}},
\nonumber \\
\nu^\mathrm{sgs} = C_\mathrm{sgs} f_\nu \overline{\Delta} \sqrt{k^\mathrm{sgs}},
\label{eq:2.11}
\end{gather}
Here, $k^\mathrm{sgs} (= \tau^\mathrm{sgs}_{\ell \ell}/2)$ is the SGS kinetic energy, $\nu^\mathrm{sgs}$ is the SGS eddy viscosity, $A_{ij}|_\mathrm{tl} = A_{ij} - A_{\ell \ell} \delta_{ij}/3$, $\tau^\mathrm{a}_{ij}$ denotes an additional term, and $C_\mathrm{sgs}$ is a constant. $f_\nu$ is the wall damping function based on the Kolmogorov scale\cite{inagaki2011,ia2017}:
\begin{gather}
f_\nu = 1 - \exp [- (d_\varepsilon/A_0)^{2/(1+C_0)} ], 
\nonumber \\
d_\varepsilon = \frac{u_\varepsilon n}{\nu} \left(\frac{n}{\overline{\Delta}} \right)^{C_0}, \ \ 
u_\varepsilon = (\nu \varepsilon^\mathrm{sgs})^{1/4}, 
\nonumber \\
\varepsilon^\mathrm{sgs} = C_\varepsilon \frac{(k^\mathrm{sgs})^{3/2}}{\overline{\Delta}} + \varepsilon^\mathrm{wall}, \ \ 
\varepsilon^\mathrm{wall} = \frac{2\nu k^\mathrm{sgs}}{n^2},
\label{eq:2.12}
\end{gather}
where $n$ denotes the distance from the nearest solid wall, $\varepsilon^\mathrm{sgs}$ denotes the SGS energy dissipation rate, and $C_\varepsilon$, $A_0$, and $C_0$ are constants. We refer to $\tau^\mathrm{eat}_{ij}$ in Eq.~(\ref{eq:2.11}) as the extra anisotropic term (EAT). In the original model\cite{abe2013}, an unusual filter scale, $\overline{\Delta} = \sqrt{\text{max}(\Delta x \Delta y, \Delta y \Delta z, \Delta z \Delta x)}$ in Cartesian coordinates, is adopted. Inagaki and Abe\cite{ia2017} modified this unusual filter scale to the conventional one, $\overline{\Delta} = (\Delta x \Delta y \Delta z)^{1/3}$ in Cartesian coordinates, and set the model constants in $C_\mathrm{sgs} = 0.075$, $C_\varepsilon = 0.835$, $A_0 = 13$, and $C_0 = 1/3$. The SGS kinetic energy $k^\mathrm{sgs}$ is obtained by numerically solving the following model equation:
\begin{align}
\frac{\partial k^\mathrm{sgs}}{\partial t} & = - \frac{\partial }{\partial x_j} ( \overline{u}_j k^\mathrm{sgs})
- \tau^\mathrm{sgs}_{ij} \overline{s}_{ij} - \varepsilon^\mathrm{sgs}
\nonumber \\
& \hspace{1em}
+ \frac{\partial}{\partial x_j} \left[ \left( \nu + C_k f_\nu \overline{\Delta} \sqrt{k^\mathrm{sgs}} \right) \frac{\partial k^\mathrm{sgs}}{\partial x_j} \right], 
\label{eq:2.13}
\end{align}
where $f_\nu$ and $\varepsilon^\mathrm{sgs}$ are defined in Eq.~(\ref{eq:2.12}) and $C_k$ is set as $C_k = 0.1$.

There are two notable features on the SMM. First, the $\nu^\mathrm{a}$ related term is introduced to remove the backscatter and stabilize the model; namely, we have
\begin{align}
\overline{s}_{ij} \tau^\mathrm{eat}_{ij} = 
2k^\mathrm{sgs} \frac{\overline{s}_{ij} \tau^\mathrm{a}_{ij}|_\mathrm{tl} + 2 \nu^\mathrm{a} \overline{s}_{ij} \overline{s}_{ij}}{\tau^\mathrm{a}_{\ell \ell}} = 0,
\label{eq:2.14}
\end{align}
which indicates that the last term on the first line of Eq.~(\ref{eq:2.11}) does not contribute to the energy transfer between the GS and SGS fields. Thereby, the second term on the right-hand side of Eq.~(\ref{eq:2.13}) yields
\begin{align}
- \tau^\mathrm{sgs}_{ij} \overline{s}_{ij} = 2 \nu^\mathrm{sgs} \overline{s}_{ij} \overline{s}_{ij} \ge 0,
\label{eq:2.15}
\end{align}
which indicates the absence of backscatter because $\nu^\mathrm{sgs}\ge0$. Hence, we calculate Eq.~(\ref{eq:2.13}) in a numerically stable manner when utilizing scale-similarity models. Because Eq.~(\ref{eq:2.13}) does not necessarily guarantee the positiveness of $k^\mathrm{sgs}$, we must clip the negative value of $k^\mathrm{sgs}$ in the numerical simulation. Otherwise, $\nu^\mathrm{sgs}$, which is proportional to $\sqrt{k^\mathrm{sgs}}$, cannot be calculated. 

Second, the EAT is expressed by a form of the normalized anisotropy tensor:
\begin{align}
\tau^\mathrm{eat}_{ij}
= 2k^\mathrm{sgs} \left( \frac{\tau^\mathrm{a}_{ij} + 2 \nu^\mathrm{a} \overline{s}_{ij}}{\tau^\mathrm{a}_{\ell \ell} +  2 \nu^\mathrm{a} \overline{s}_{\ell \ell}} - \frac{1}{3} \delta_{ij} \right) = 2k^\mathrm{sgs} b_{ij}^\mathrm{a}.
\label{eq:2.16}
\end{align}
The advantage of this normalized form is that the model can predict the anisotropy of the SGS field even when a component of $\tau^\mathrm{a}_{ij}$ becomes small. This is because the denominator $\tau^\mathrm{a}_{\ell \ell}$ decreases at the same rate as $\tau^\mathrm{a}_{ij}$. Owing to these two features, we can make full use of scale-similarity models in a numerically stable manner. 


In this study, we note that Abe\cite{abe2013} or Inagaki and Abe\cite{ia2017} adopted
\begin{align}
\tau^\mathrm{a}_{ij} = (\overline{u}_i - \widehat{\overline{u}}_i) (\overline{u}_j - \widehat{\overline{u}}_j),
\label{eq:2.17}
\end{align}
for the EAT. Equation~(\ref{eq:2.17}) corresponds to the scale-similarity model for the SGS-Reynolds term (\ref{eq:2.5b}), although the repeated filter operation of $\overline{\cdot}$ is replaced with the test filter operation $\widehat{\cdot}$. Because the stabilization procedure of Abe\cite{abe2013} is independent of the choice of $\tau^\mathrm{a}_{ij}$, we can construct various types of SMMs, by for example using the modified Leonard term $\mathcal{L}^\mathrm{m}_{ij}$ (\ref{eq:2.6}) or the Clark term $C_{ij}$ (\ref{eq:2.9}). Considering the results of numerous previous studies addressing the correlation between scale-similarity models and the exact SGS stress\cite{clarketal1979,bardinaetal1983,liuetal1994,taoetal2002,horiuti2003,kobayashi2018}, we expect that the modified Leonard or Clark terms yield a better performance in the reproduction of the physics of turbulent flows in the LES than the scale-similarity model for the SGS-Reynolds term. In the following section, we construct the SMM using various scale-similarity models and evaluate their performance in turbulent channel flows as a typical case of wall-bounded turbulent shear flows.

\section{\label{sec:level3}Statistics of various stabilized mixed models in turbulent channel flows}

\subsection{\label{sec:level3.1}Variety of stabilized mixed models}

We construct various SMMs based on the model suggested by Inagaki and Abe\cite{ia2017}(hereafter denoted as IA), which was modified from the original model\cite{abe2013} to enable the use of the conventional filter scale $\overline{\Delta} = (\Delta x \Delta y \Delta z)^{1/3}$ in Cartesian coordinates. In the construction, we fix all model constants described in Sec.~\ref{sec:level2.2} except $C_\mathrm{sgs}$ for the cases excluding the EAT and only change $\tau^\mathrm{a}_{ij}$, which is the scale-similarity model for the SGS-Reynolds term (\ref{eq:2.17}) in the original model. The applied models are listed in the followings:
\begin{enumerate}
\item Original IA model\cite{ia2017} (IA):
\begin{align*}
\tau^\mathrm{a}_{ij} = (\overline{u}_i - \widehat{\overline{u}}_i) (\overline{u}_j - \widehat{\overline{u}}_j),
\end{align*}
\item IA model with the Clark term (IA-CL):
\begin{align*}
\tau^\mathrm{a}_{ij} = \frac{\overline{\Delta}_\ell^2}{12} \frac{\partial \overline{u}_i}{\partial x_\ell} \frac{\partial \overline{u}_j}{\partial x_\ell}
\end{align*}
\item IA model with the modified Leonard term (IA-ML):
\begin{align*}
\tau^\mathrm{a}_{ij} = \overline{\overline{u}_i \overline{u}_j} - \overline{\overline{u}}_i \overline{\overline{u}}_j
\end{align*}
\item IA model with the stress based on the Laplacian of velocity (IA-LV):
\begin{align*}
\tau^\mathrm{a}_{ij} = \overline{\Delta}^4 (\nabla^2 \overline{u}_i) (\nabla^2 \overline{u}_j)
\end{align*}
\item IA model without the EAT (IA-LN):
\begin{align*}
\tau^\mathrm{a}_{ij} = 0
\end{align*}
\end{enumerate}
In addition to the conventional scale-similarity models, we test a model based on the Laplacian of velocity (IA-LV). The scale-similarity model for the SGS-Reynolds term is reduced to this LV model through the Taylor expansion using a cubic grid:
\begin{align}
(\overline{u}_i - \widehat{\overline{u}}_i) (\overline{u}_j - \widehat{\overline{u}}_j) 
& = \frac{\widehat{\Delta}_\ell^2}{24} \frac{\partial^2 \overline{u}_i}{\partial x_\ell \partial x_\ell} \frac{\widehat{\Delta}_m^2}{24}
\frac{\partial^2 \overline{u}_i}{\partial x_m \partial x_m} 
\nonumber \\ 
& = \frac{\gamma^4}{24^2} \overline{\Delta}^4 (\nabla^2 \overline{u}_i) (\nabla^2 \overline{u}_j),
\label{eq:3.1}
\end{align}
where $\gamma = \widehat{\Delta}_\alpha/\overline{\Delta}_\alpha (=\text{const.})$. Note that the SMM does not depend on the value of the coefficient $\gamma^4/24^4$ because $\tau^\mathrm{a}_{ij}$ is normalized, as shown in Eqs.~(\ref{eq:2.11}) or (\ref{eq:2.16}). For the numerical simulation of turbulent channel flows, we often employ a rectangular grid, such that IA and IA-LV yield different results.

\subsection{\label{sec:level3.2}Computational methods and numerical conditions}

We employ a Cartesian coordinate grid and set the streamwise, wall-normal, and spanwise directions as $x (=x_1)$, $y (=x_2)$, and $z (=x_3)$, respectively. We use the staggered grid system and adopt the fully conservative central finite difference scheme\cite{morinishietal1998} for the $x$ and $z$ directions with fourth-order accuracy and the conservative central finite difference scheme on non-uniform grids\cite{kajishimatairabook} for the $y$ direction with second-order accuracy, for both equations for the velocity and SGS kinetic energy $k^\mathrm{sgs}$. The boundary condition is periodic in the $x$ and $z$ directions and it is no slip in the $y$ direction. The Poisson equation for pressure is solved using a fast Fourier transformation. For time marching in the velocity field, we adopt the second-order Adams--Bashforth method. For time marching in $k^\mathrm{sgs}$, we adopt the explicit Euler method, except for the dissipation term $\varepsilon^\mathrm{sgs}$, which is treated implicitly. For the test filtering operation, we approximate it through the Taylor expansion as Eq.~(\ref{eq:2.7}). The spatial derivative for each direction in Eq.~(\ref{eq:2.7}) is discretized with second-order accuracy, i.e., we discretize $\widehat{\overline{q}}^{(I,J,K)}$ as
\begin{align}
\widehat{\overline{q}}^{(I,J,K)} & = \overline{q}^{(I,J,K)}
+ \frac{\widehat{\Delta}_x^2}{24} \frac{\overline{q}^{(I-1,J,K)} - 2\overline{q}^{(I,J,K)} + \overline{q}^{(I+1,J,K)}}{\Delta x^2}
\nonumber \\
& \hspace{4em}
+ \frac{\widehat{\Delta}_y^2}{24} \frac{1}{\Delta y^{(J)}} \left[ - \frac{- \overline{q}^{(I,J-1,K)} + \overline{q}^{(I,J,K)}}{\Delta y^{(J-1/2)}} \right.
\nonumber \\
& \hspace{9em} \left.
+ \frac{- \overline{q}^{(I,J,K)} + \overline{q}^{(I,J+1,K)}}{\Delta y^{(J+1/2)}} \right]
\nonumber \\
& \hspace{4em}
+ \frac{\widehat{\Delta}_z^2}{24} \frac{\overline{q}^{(I,J,K-1)} - 2\overline{q}^{(I,J,K)} + \overline{q}^{(I,J,K+1)}}{\Delta z^2}
\nonumber \\
& \hspace{1em}
+ O(\Delta x^4)  + O(\Delta y^4) + O(\Delta z^4),
\label{eq:3.2}
\end{align}
where the superscript $(I,J,K)$ denotes the grid point. Because $\widehat{\Delta}_i \propto \Delta x_i$, the test-filtered variables retain fourth-order accuracy in the central finite difference scheme. We set $\overline{\Delta}_i = \Delta x_i$, $\widehat{\overline{\Delta}}_i = 2 \overline{\Delta}_i$, and $\widehat{\Delta}_i = \sqrt{3} \ \overline{\Delta}_i$ to satisfy $\overline{\Delta}_\alpha^2 + \widehat{\Delta}_\alpha^2 = \widehat{\overline{\Delta}}_\alpha^2$, which is satisfied when the Gaussian filter is employed as the test filter\cite{popebook}. We perform two Reynolds number cases, $\mathrm{Re}_\tau = 180$ and $\mathrm{Re}_\tau =1000$ where $\mathrm{Re}_\tau (= u_\tau h/\nu)$ is the Reynolds number based on the channel half width $h$ and the wall friction velocity $u_\tau (=\sqrt{|\partial U_x/\partial |_\text{wall}|})$, where $U_x (= \langle \overline{u}_x \rangle)$ is the streamwise mean velocity and $\langle \cdot \rangle$ denotes the statistical average. In the present simulation, we adopt the average over the $x$--$z$ plane and time to obtain the statistical average. For $\mathrm{Re}_\tau = 180$, we perform a DNS with the same discretization method as the present LES.

\begin{table*}
\caption{\label{tab:1}Flow cases and numerical parameters.}
\begin{ruledtabular}
\begin{tabular}{lccccccc}
Case & $\mathrm{Re}_\tau$ & $L_x \times L_y \times L_z$ & $N_x \times N_y \times N_z$ & $\Delta x^+$ & $\Delta y^+$ & $\Delta z^+$ & $C_\mathrm{sgs}$ \\ \hline
IA180LR & 180 & $4\pi h \times 2 h \times 4\pi h/3$ & $24 \times 64 \times 16$ & 94 & 1.1--11 & 47 & 0.075\\
IA180MR & 180 & $4\pi h \times 2 h \times 4\pi h/3$ & $48 \times 64 \times 32$ & 47 & 1.1--11 & 24 & 0.075\\
IA180HR & 180 & $4\pi h \times 2 h \times 4\pi h/3$ & $96 \times 64 \times 64$ & 24 & 1.1--11 & 12 & 0.075\\
IA180LD & 180 & $16\pi h \times 2 h \times 16\pi h/3$ & $96 \times 64 \times 64$ & 94 & 1.1--11 & 47 & 0.075\\
IA-CL180LR & 180 & $4\pi h \times 2 h \times 4\pi h/3$ & $24 \times 64 \times 16$ & 94 & 1.1--11 & 47 & 0.075\\
IA-CL180LD & 180 & $16\pi h \times 2 h \times 16\pi h/3$ & $96 \times 64 \times 64$ & 94 & 1.1--11 & 47 & 0.075\\
IA-ML180LR & 180 & $4\pi h \times 2 h \times 4\pi h/3$ & $24 \times 64 \times 16$ & 94 & 1.1--11 & 47 & 0.075\\
IA-LV180LR & 180 & $4\pi h \times 2 h \times 4\pi h/3$ & $24 \times 64 \times 16$ & 94 & 1.1--11 & 47 & 0.075\\
IA-LN180LR & 180 & $4\pi h \times 2 h \times 4\pi h/3$ & $24 \times 64 \times 16$ & 94 & 1.1--11 & 47 & 0.075\\
IA-LNcs42-180LR & 180 & $4\pi h \times 2 h \times 4\pi h/3$ & $24 \times 64 \times 16$ & 94 & 1.1--11 & 47 & 0.042\\
IA-LNcs42-180MR & 180 & $4\pi h \times 2 h \times 4\pi h/3$ & $48 \times 64 \times 32$ & 47 & 1.1--11 & 24 & 0.042\\
IA-LNcs42-180HR & 180 & $4\pi h \times 2 h \times 4\pi h/3$ & $96 \times 64 \times 64$ & 24 & 1.1--11 & 12 & 0.042\\
IA-LNcs42-180LD & 180 & $16\pi h \times 2 h \times 16\pi h/3$ & $96 \times 64 \times 64$ & 94 & 1.1--11 & 47 & 0.042\\
DSM180LR & 180 & $4\pi h \times 2 h \times 4\pi h/3$ & $24 \times 64 \times 16$ & 94 & 1.1--11 & 47 & -\\
DSM180MR & 180 & $4\pi h \times 2 h \times 4\pi h/3$ & $48 \times 64 \times 32$ & 47 & 1.1--11 & 24 & -\\
DSM180HR & 180 & $4\pi h \times 2 h \times 4\pi h/3$ & $96 \times 64 \times 64$ & 24 & 1.1--11 & 12 & -\\
no-model180LR & 180 & $4\pi h \times 2 h \times 4\pi h/3$ & $24 \times 64 \times 16$ & 94 & 1.1--11 & 47 & -\\
no-model180MR & 180 & $4\pi h \times 2 h \times 4\pi h/3$ & $48 \times 64 \times 32$ & 47 & 1.1--11 & 24 & -\\
no-model180HR & 180 & $4\pi h \times 2 h \times 4\pi h/3$ & $96 \times 64 \times 64$ & 24 & 1.1--11 & 12 & -\\
DNS & 180 & $4\pi h \times 2 h \times 4\pi h/3$ & $256 \times 128 \times 256$ & 8.8 & 0.23--6.8 & 2.9 & -\\
IA1000VLR & 1000 & $2\pi h \times 2 h \times \pi h$ & $48 \times 96 \times 32$ & 130 & 1.0--58 & 98 & 0.075\\
IA1000LR & 1000 & $2\pi h \times 2 h \times \pi h$ & $96 \times 96 \times 64$ & 65 & 1.0--58 & 49 & 0.075\\
IA-LN1000LR & 1000 & $2\pi h \times 2 h \times \pi h$ & $96 \times 96 \times 64$ & 65 & 1.0--58 & 49 & 0.075\\
DSM1000LR & 1000 & $2\pi h \times 2 h \times \pi h$ & $96 \times 96 \times 64$ & 65 & 1.0--58 & 49 & -\\
no-model1000LR & 1000 & $2\pi h \times 2 h \times \pi h$ & $96 \times 96 \times 64$ & 65 & 1.0--58 & 49 & -\\
\end{tabular}
\end{ruledtabular}
\end{table*}

Flow cases and numerical parameters are listed in Table~\ref{tab:1}. For $\mathrm{Re}_\tau = 180$, we evaluate the grid resolution dependency of the representative models and set three grid resolution cases; low resolution (LR), where $\Delta x^+ = 94$ and $\Delta z^+ = 47$; medium resolution (MR), where $\Delta x^+ = 47$ and $\Delta z^+ = 24$; high resolution (HR), where $\Delta x^+ = 94$ and $\Delta z^+ = 47$, while $\Delta y^+$ is fixed. Values with a superscript `$+$' denote those normalized by $u_\tau$ and $\nu$. Further, we perform the simulations with the same resolution as LR, but in a large domain (LD), where $L_x = 16\pi h$ and $L_z = 16\pi h/3$ to obtain smooth profile spectra. For IA-LN in LR at $\mathrm{Re}_\tau = 180$, the value $C_\mathrm{sgs} = 0.075$, which is optimized for IA\cite{ia2017} is too large to sustain the GS turbulent fluctuation. Hence, we establish a reference linear eddy-viscosity model case with a smaller coefficient $C_\mathrm{sgs} = 0.042$ (IA-LNcs42), which is optimized for an one-equation eddy-viscosity model \cite{inagaki2011}. To evaluate the filtered values in the \textit{a priori} test through the DNS result, we adopt the sharp cut-off filter in the Fourier space in the $x$ and $z$ directions, while no filtering operation is applied in the $y$ direction. The filter scale is set in $\overline{\Delta}_x^+ = 94$ and $\overline{\Delta}_z^+ = 47$ to the same value as the LR cases in LES. 

To investigate a higher Reynolds number case, we perform LES at $\mathrm{Re}_\tau = 1000$. At this Reynolds number, we perform a low resolution (LR) simulation, where $\Delta x^+ = 65$ and $\Delta z^+ = 49$, in which the spanwise resolution is almost the same as in LR at $\mathrm{Re}_\tau = 180$. For IA, we additionally perform a very low resolution (VLR) case, where $\Delta x^+ = 130$ and $\Delta z^+ = 98$.

\subsection{\label{sec:level3.3}Basic statistics}

\subsubsection{\label{sec:level3.3.1}Mean velocity for basic models}

\begin{figure*}
  \begin{minipage}{0.49\hsize}
   \centering
   \includegraphics[width=\textwidth]{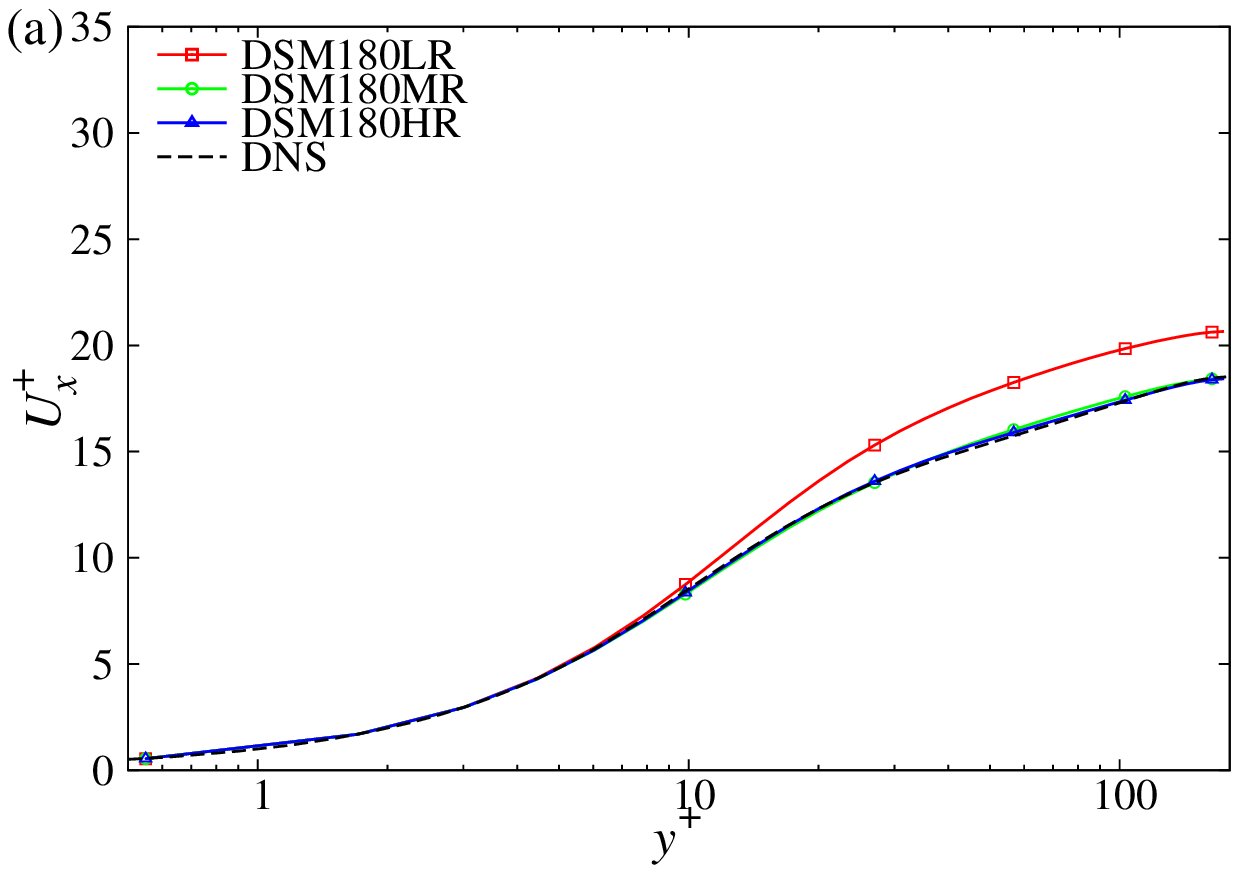}
  \end{minipage}
  \begin{minipage}{0.49\hsize}
   \centering
   \includegraphics[width=\textwidth]{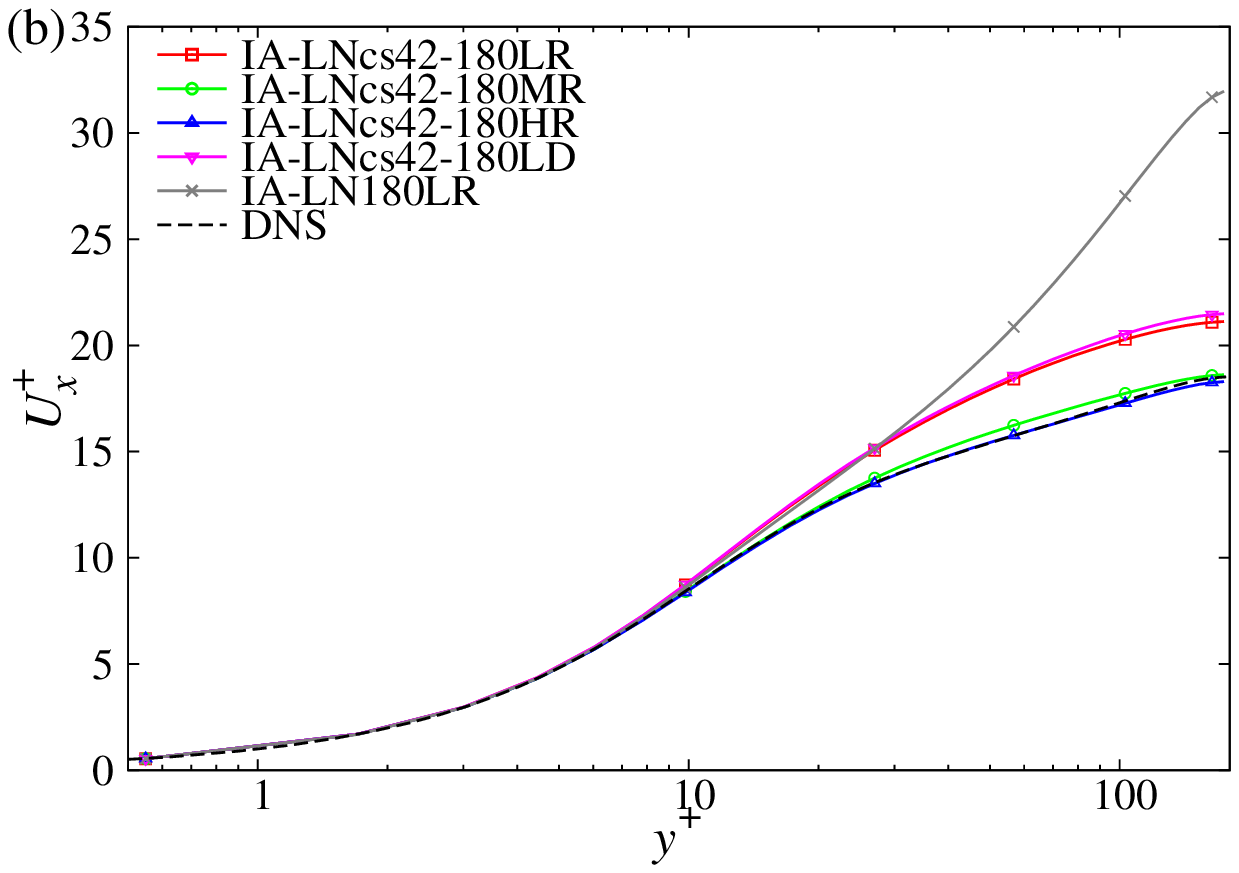}
  \end{minipage} \\

  \begin{minipage}{0.49\hsize}
   \centering
   \includegraphics[width=\textwidth]{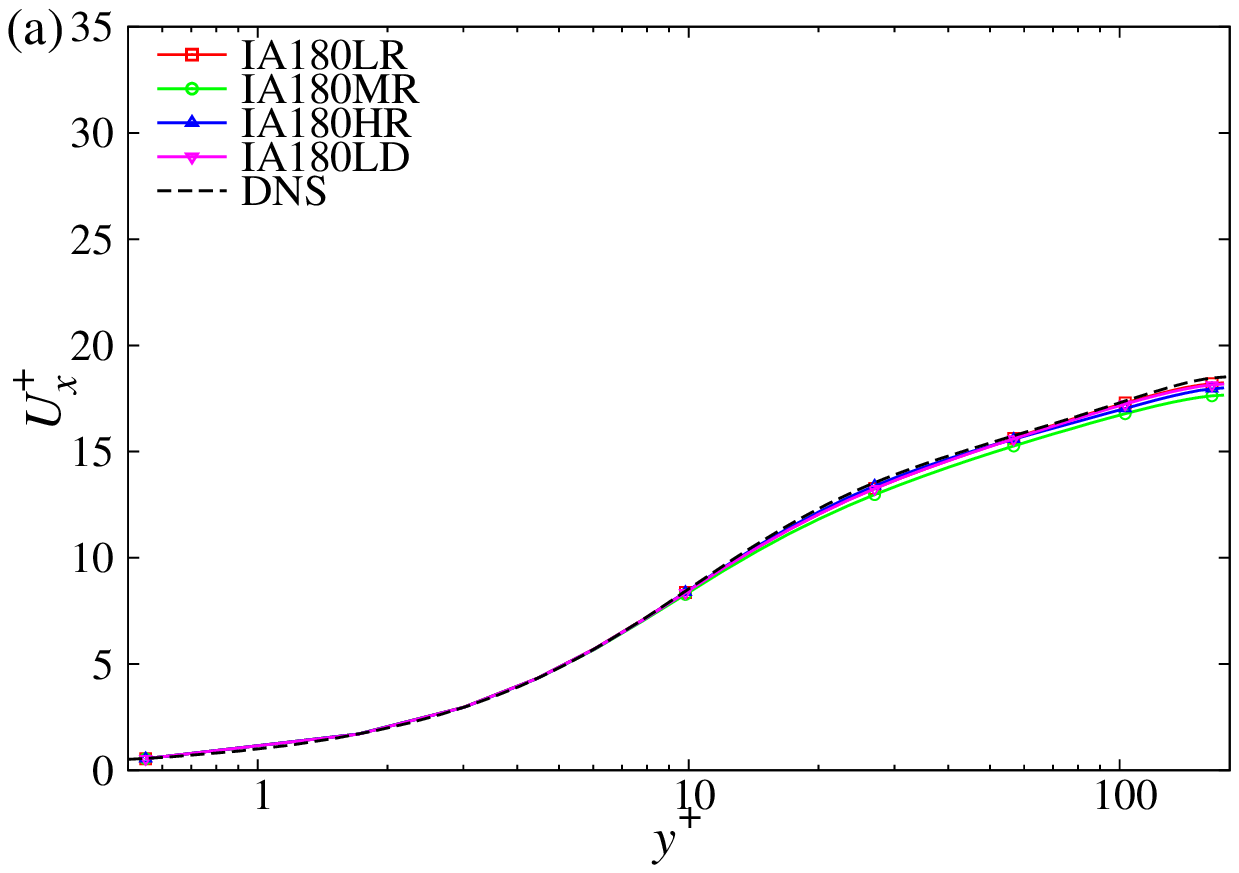}
  \end{minipage}
  \begin{minipage}{0.49\hsize}
   \centering
   \includegraphics[width=\textwidth]{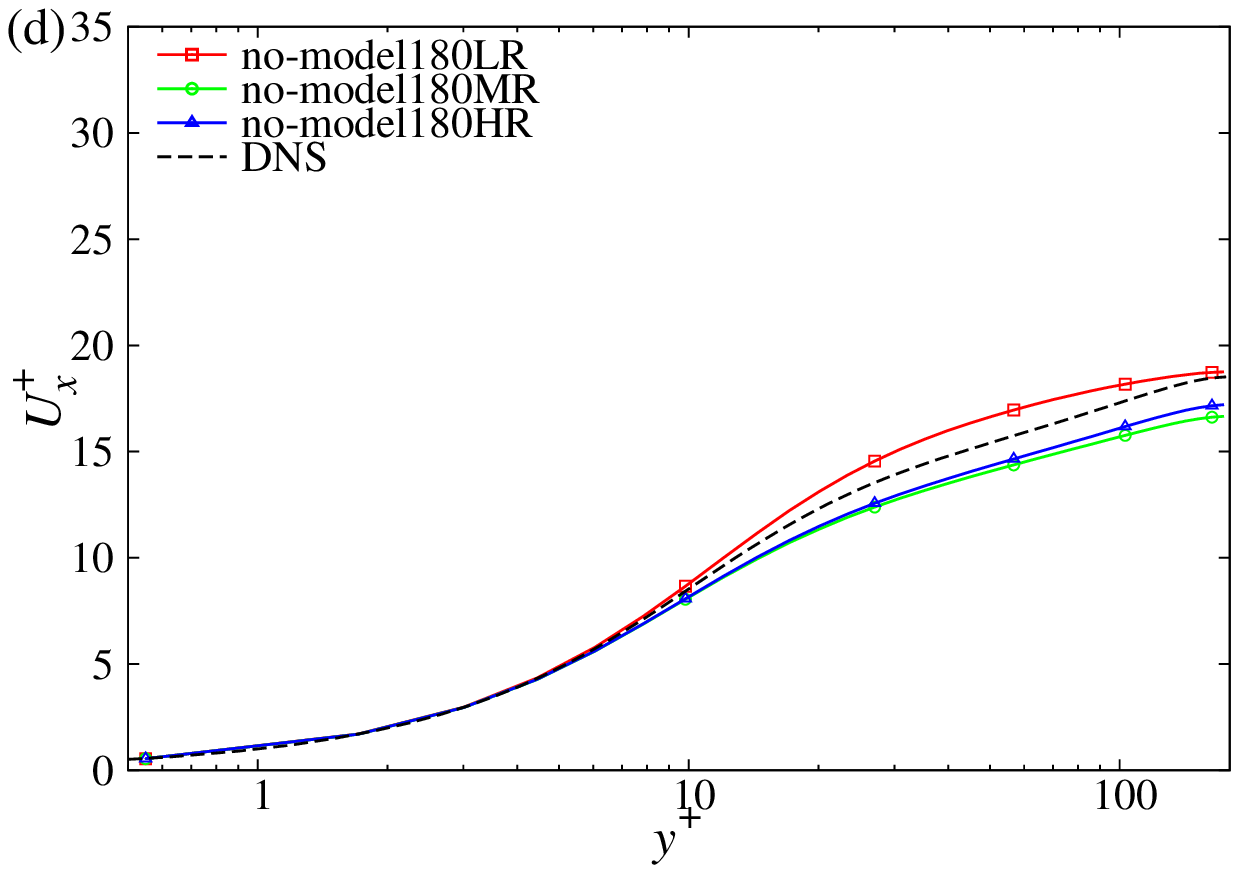}
  \end{minipage}
\caption{\label{fig:1} Mean velocity profile for (a) DSM, (b) IA-LN, (c) IA, and (d) no-model at various resolutions or domain sizes for $\mathrm{Re}_\tau = 180$.}
\end{figure*}

Figure~\ref{fig:1} shows the mean velocity profile for (a) DSM, (b) IA-LN, and (c) IA at various grid resolutions or domain sizes at $\mathrm{Re}_\tau = 180$. We plot the no-model result in Fig.~\ref{fig:1}(d) for reference. In LR, all cases except for IA overestimate the mean velocity. The IA model provides a good prediction irrespective of the grid resolution. In Fig.~\ref{fig:1}(c), IA-LN180LR excessively overestimates the mean velocity. This is because the GS velocity fluctuations disappear. However, this does not lead to a laminar or parabolic profile because the SGS eddy viscosity $\nu^\mathrm{sgs}$ does not vanish due to the mean value of $k^\mathrm{sgs}$. Overestimation of the mean velocity is also observed in other eddy-viscosity models\cite{abe2013,ia2017}. Hence, the EAT makes IA insensitive to the grid resolution. The results in LD overlap those in LR for IA and IA-LNcs42 almost perfectly, which suggests that the statistics of the present simulation depend not on the domain size, but the grid resolution.

\begin{figure}
\centering 
\includegraphics[width=0.5\textwidth]{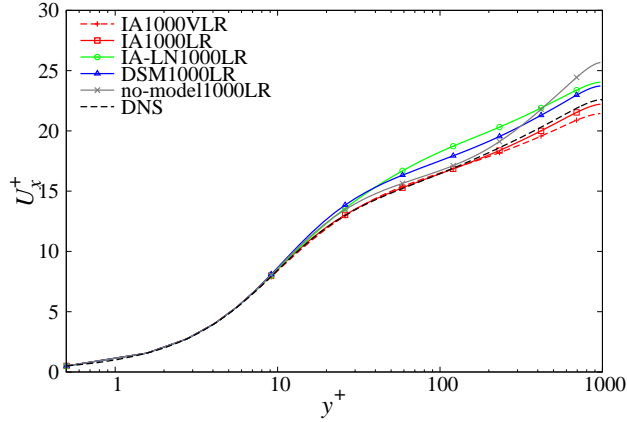}\\
\caption{\label{fig:2}Mean velocity profile for representative cases at $\mathrm{Re}_\tau = 1000$.}
\end{figure}

Figure~\ref{fig:2} shows the mean velocity profile for representative cases at $\mathrm{Re}_\tau = 1000$. The reference DNS was performed by Lee and Moser\cite{lm2015}. IA yields a good prediction, while other cases result in overestimation. Surprisingly, IA yields a reasonable prediction even in VLR, where the spanwise grid size is $\Delta z^+ = 98$, which is close to the distance between the streak structure observed in the near-wall region in wall-bounded turbulent flows\cite{klineetal1967}. Notably, IA succeeds in predicting the mean velocity profile in the near-wall region $y^+<100$, irrespective of the grid resolution. These results suggest that the physical feature of the SMM lies in the near-wall region.

\begin{figure}
\centering 
\includegraphics[width=0.5\textwidth]{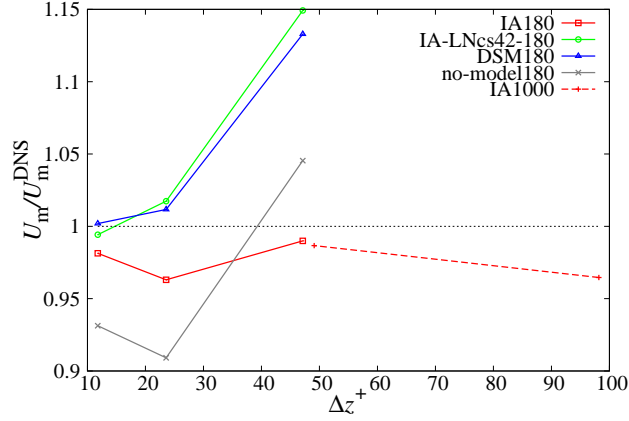}
\caption{\label{fig:3}Bulk mean velocity normalized by DNS value for representative cases with respect to spanwise grid size.}
\end{figure}

Figure~\ref{fig:3} shows the bulk mean velocity normalized by the DNS value for representative cases with respect to the spanwise grid size. The bulk mean velocity is defined by
\begin{align}
U_\mathrm{m} = \frac{1}{2} \int_0^{2h} \mathrm{d}y \ U_x (y),
\label{eq:3.3}
\end{align}
where $y=0, 2h$ corresponds to the solid wall boundary. Figure~\ref{fig:3} indicates that IA is the least sensitive to the grid resolution, and the error is within 5\% for both Reynolds numbers. Interestingly, the no-model overestimates the mean velocity in LR at $\mathrm{Re}_\tau=180$, while it results in underestimation in MR and HR, as observed in Fig.~\ref{fig:1}(d) and \ref{fig:3}. This suggests that the mechanism producing the turbulent momentum transfer or the Reynolds shear stress alters between LR and MR at $\mathrm{Re}_\tau=180$. Generally, insufficient grid resolution without the use of any SGS model makes the system less dissipative, leading to the increase in velocity fluctuations. Consequently, the mean velocity is decreased due to the excessive turbulent momentum transfer. In contrast, in LR, the lack of grid resolution may result in the loss of a generation mechanism of the turbulent or Reynolds shear stress, leading to overestimation of the mean velocity. Considering this situation, it is worth noting that IA succeeds in reproducing the effective momentum transfer due to turbulence even in LR at $\mathrm{Re}_\tau=180$. Therefore, we focus on the properties of models in LR at $\mathrm{Re}_\tau=180$.

\subsubsection{\label{sec:level3.3.2}Mean velocity for various SMMs}

\begin{figure}
\centering 
\includegraphics[width=0.5\textwidth]{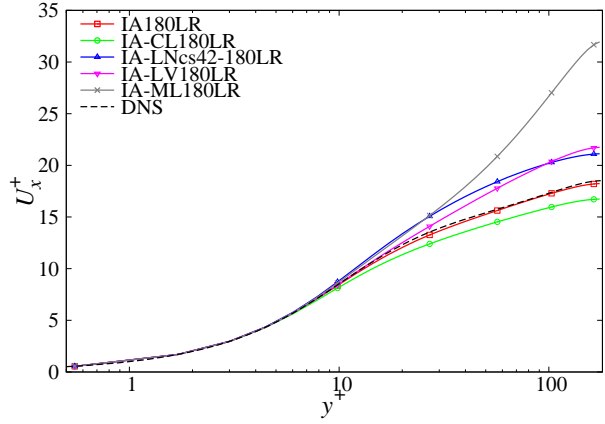}\\
\caption{\label{fig:4} Mean velocity profile for various SMMs in LR at $\mathrm{Re}_\tau = 180$.}
\end{figure}

Figure~\ref{fig:4} shows the mean velocity profile for variable stabilized mixed models including the no EAT model (IA-LNcs42) in LR. The excessive overestimation of IA-ML results from the vanishing GS velocity fluctuations, similar to that shown with IA-LN180LR in Fig.~\ref{fig:1}(c). The reason behind the failure of IA-ML may be attributed to the scale-similarity model for the SGS-Reynolds term entering the modified Leonard term involves with a negative coefficient, as seen in Eqs.~(\ref{eq:2.8}) and (\ref{eq:2.10}). 
As both IA and IA-CL yield a reduction of the flow rate compared with other models, the Clark and the SGS-Reynolds terms have similar effect. Moreover, $\overline{\Delta}_i$ is large in LR. Thereby, the first and second terms on the first line of Eq.~(\ref{eq:2.8}) are canceled each other in the modified Leonard term, yielding the weak contribution of the EAT to the SGS stress for IA-ML. Then, IA-ML leads to a laminarization due to the strong eddy viscosity, as shown in Fig.~\ref{fig:1}(b).
IA-LNcs42 and IA-LV overestimate the mean velocity, while IA-CL underestimates it. The prediction of the mean velocity can be refined by tuning model parameters, e.g., $C_\mathrm{sgs}$. However, this is beyond of scope of the present study. A notable point is that IA-CL and IA-LV succeed in sustaining the turbulence even in LR,  where it leads to laminarization without the EAT. IA-LV cannot provide a good prediction of the mean velocity, even though the rank of the spatial derivative is the same as the scale-similarity model for the SGS-Reynolds term. Under the scale-similarity assumption, $\overline{u}_i - \widehat{\overline{u}}_i$ is evaluated as 
\begin{align}
& \overline{u}_i^{(I,J,K)} - \widehat{\overline{u}}_i^{(I,J,K)}
\nonumber \\
& = \frac{1}{8} \left[
\left( \overline{u}_i^{(I-1,J,K)} - 2\overline{u}_i^{(I,J,K)} + \overline{u}_i^{(I+1,J,K)} \right) \right.
\nonumber \\
& \hspace{2em}
+ \Delta y^{(J)} \left( - \frac{- \overline{u}_i^{(I,J-1,K)} + \overline{u}_i^{(I,J,K)}}{\Delta y^{(J-1/2)}} \right.
\nonumber \\
& \hspace{6em} \left.
+ \frac{- \overline{u}_i^{(I,J,K)} + \overline{u}_i^{(I,J+1,K)}}{\Delta y^{(J+1/2)}} \right)
\nonumber \\
& \hspace{2em} + \left.
\left( \overline{u}_i^{(I,J,K-1)} - 2\overline{u}_i^{(I,J,K)} + \overline{u}_i^{(I,J,K+1)} \right) \right],
\label{eq:3.4}
\end{align}
while the Laplacian of the velocity for the finite difference with second-order accuracy yields
\begin{align}
& \nabla^2 \overline{u}_i^{(I,J,K)} 
\nonumber \\
& = \frac{1}{(\Delta y^{(J)})^2} \left[
\left(\frac{\Delta y}{\Delta x}\right)^2 \left( \overline{u}_i^{(I-1,J,K)} - 2\overline{u}_i^{(I,J,K)} + \overline{u}_i^{(I+1,J,K)} \right) \right.
\nonumber \\
& \hspace{5em}
+ \Delta y^{(J)} \left( - \frac{- \overline{u}_i^{(I,J-1,K)} + \overline{u}_i^{(I,J,K)}}{\Delta y^{(J-1/2)}} \right.
\nonumber \\
& \hspace{9em} \left.
+ \frac{- \overline{u}_i^{(I,J,K)} + \overline{u}_i^{(I,J+1,K)}}{\Delta y^{(J+1/2)}} \right)
\nonumber \\
& \hspace{5em} + \left.
\left(\frac{\Delta y}{\Delta z}\right)^2 \left( \overline{u}_i^{(I,J,K-1)} - 2\overline{u}_i^{(I,J,K)} + \overline{u}_i^{(I,J,K+1)} \right) \right].
\label{eq:3.5}
\end{align}
Note that the SMM does not depend on the coefficient $1/(\Delta y^{(J)})^2$, because the EAT is expressed by the form of the normalized tensor. In the present simulation, $\Delta x$ and $\Delta z$ are at least about 5 to 10 times larger than $\Delta y$. Hence, the $x$- and $z$-derivative parts in Eq.~(\ref{eq:3.5}) contributes little to the EAT. This represents a critical difference between IA and IA-LV. We confirm that the result does not change for IA when the test filter is adopted only to the $x$ and $z$ directions (not shown). Namely, the test filter in the $x$ and $z$ directions is essential in the scale-similarity model for the SGS-Reynolds term in turbulent channel flows.

\subsubsection{\label{sec:level3.3.3}Reynolds stress}

\begin{figure*}
  \begin{minipage}{0.49\hsize}
   \centering
   \includegraphics[width=\textwidth]{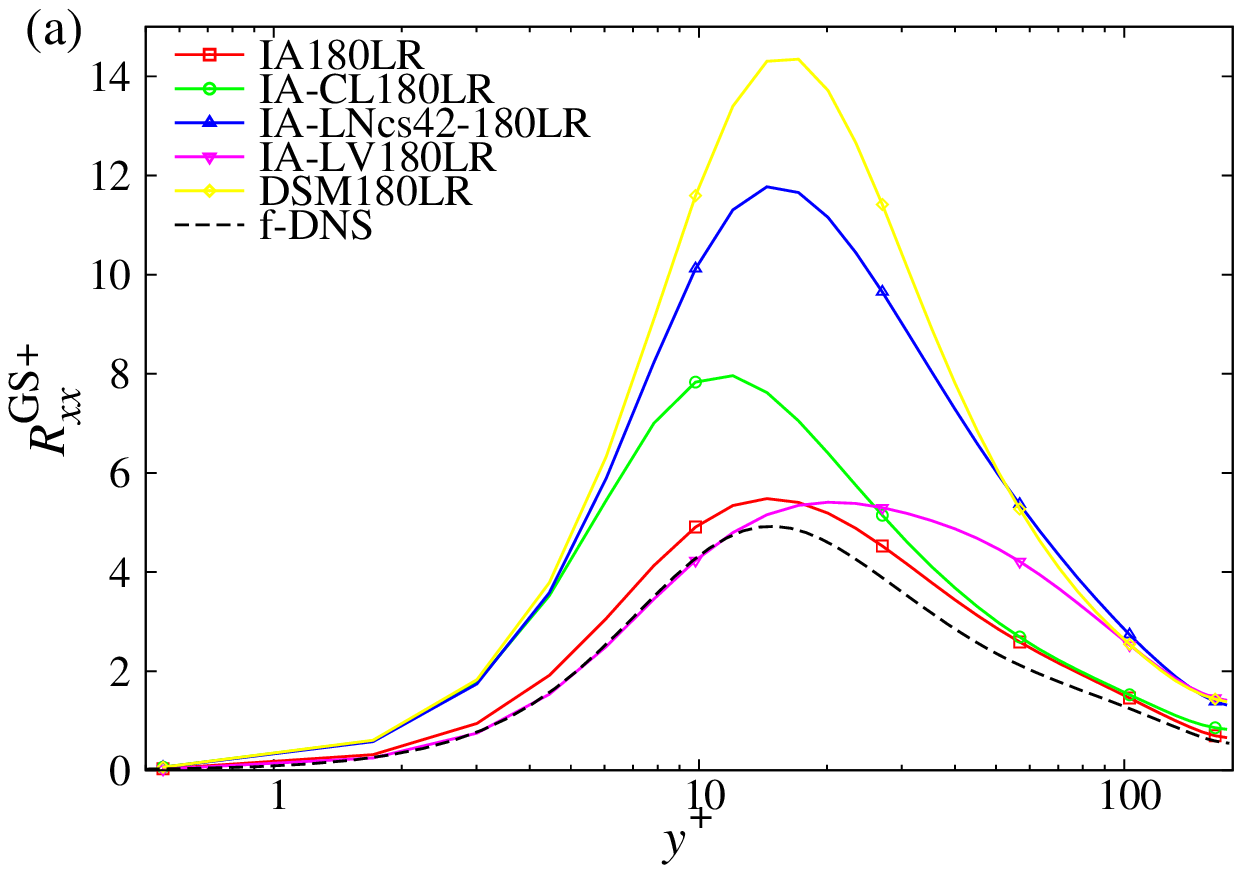}
  \end{minipage}
  \begin{minipage}{0.49\hsize}
   \centering
   \includegraphics[width=\textwidth]{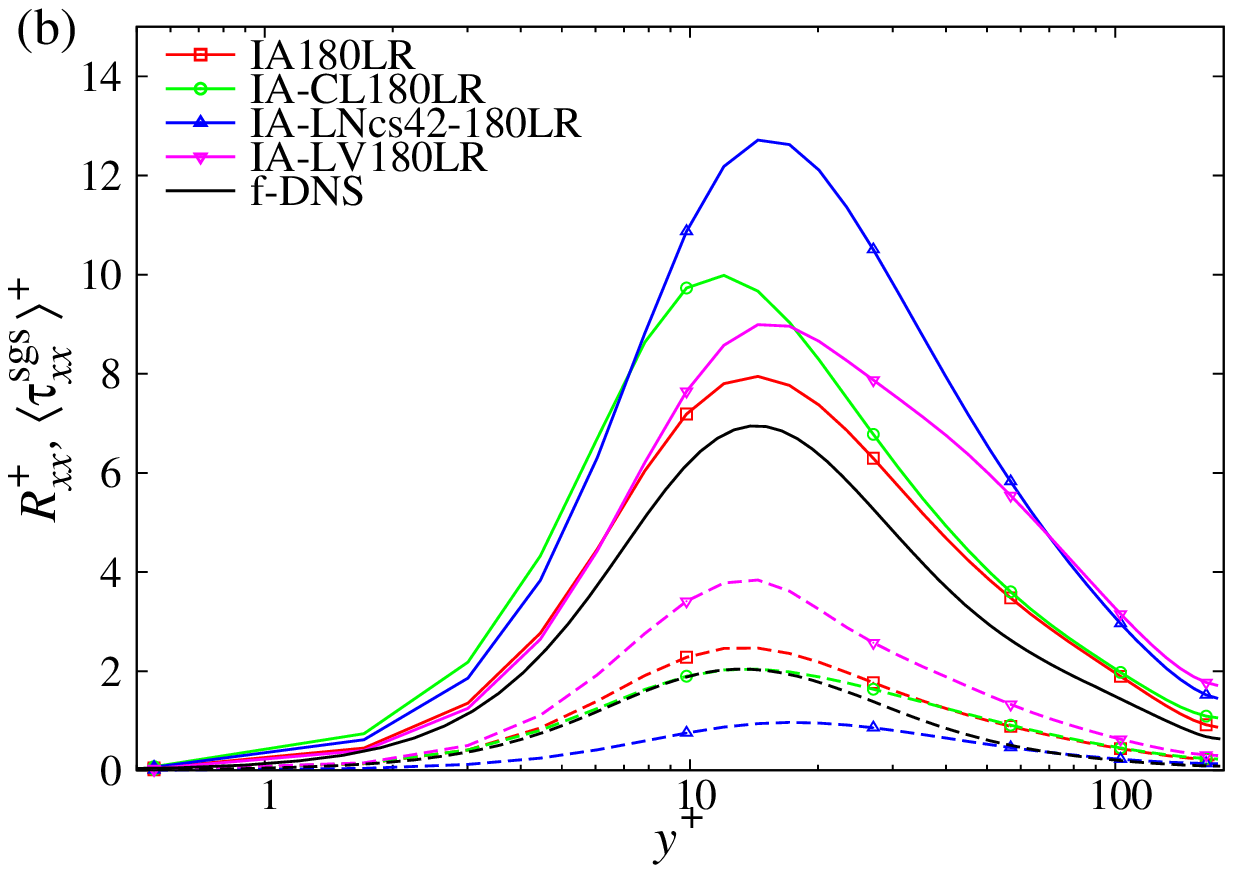}
  \end{minipage} \\

  \begin{minipage}{0.49\hsize}
   \centering
   \includegraphics[width=\textwidth]{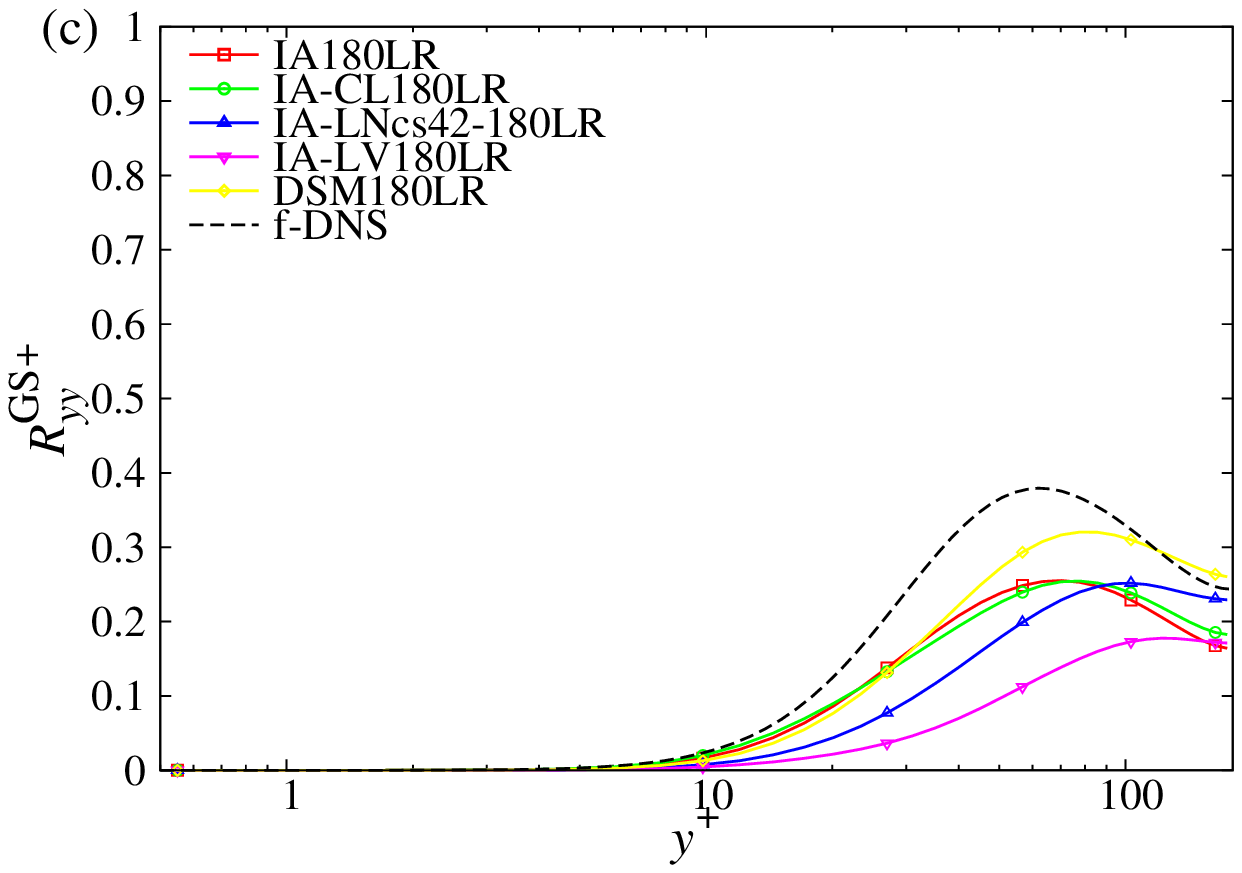}
  \end{minipage}
  \begin{minipage}{0.49\hsize}
   \centering
   \includegraphics[width=\textwidth]{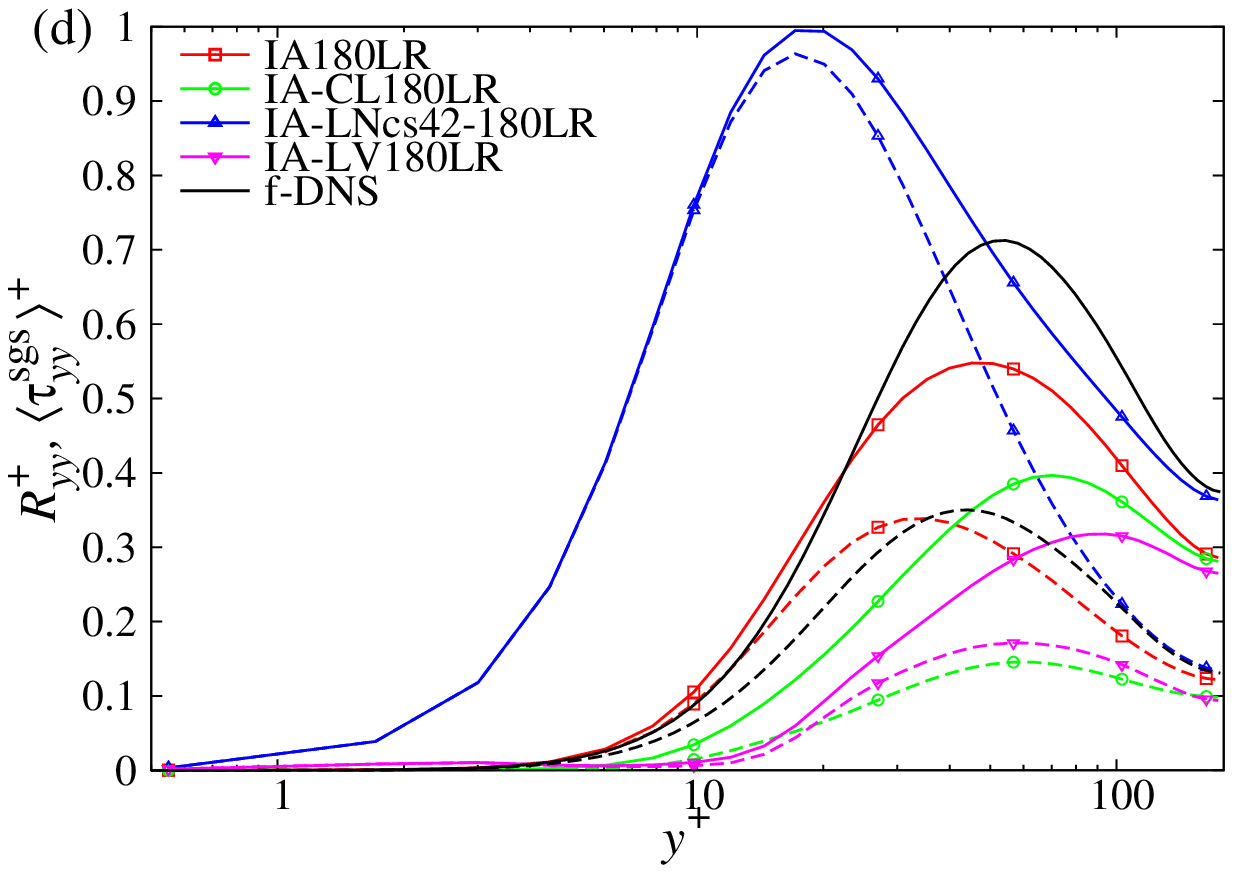}
  \end{minipage}
\caption{\label{fig:5} Profiles of the Reynolds stress of (a) $R^\mathrm{GS}_{xx}$, (b) $R_{xx}$ and $\langle \tau^\mathrm{sgs}_{xx} \rangle$, (c) $R^\mathrm{GS}_{yy}$, and (d) $R_{yy}$ and $\langle \tau^\mathrm{sgs}_{yy} \rangle$ at $\mathrm{Re}_\tau = 180$. In (b) and (d), solid lines depict the total value, while dashed lines depict the SGS value.}
\end{figure*}

In LES, the Reynolds stress $R_{ij}$ can be defined by
\begin{align}
R_{ij} = R^\mathrm{GS}_{ij} + \langle \tau^\mathrm{sgs}_{ij} \rangle, \ \ 
R^\mathrm{GS}_{ij} = \langle \overline{u}_i' \overline{u}_j' \rangle,
\label{eq:3.6}
\end{align}
where $q' (=q - \langle q \rangle)$ is the fluctuation of $q$ around the mean value. For the present SMM, we explicitly solve the SGS kinetic energy $k^\mathrm{sgs}$. Thereby, we can calculate the SGS part of the Reynolds stress, which involves $k^\mathrm{sgs}$ as the isotropic part. Figure~\ref{fig:5} shows the profile of the streamwise and wall-normal components of the Reynolds stress for various SMMs in LR at $\mathrm{Re}_\tau = 180$. The result for the filtered DNS is likewise plotted, and it is denoted by f-DNS. For DSM, we obtain only the GS value, which is plotted in Fig.~\ref{fig:5}(a) and (c).

In Fig.~\ref{fig:5}(a), DSM and IA-LNcs42 overestimate the streamwise component of the GS Reynolds stress $R_{xx}^\mathrm{GS}$, which is twice as large as that of f-DNS. $R_{xx}^\mathrm{GS}$ contributes the most to the GS turbulent kinetic energy $K^\mathrm{GS} (=R^\mathrm{GS}_{ii}/2)$ in turbulent channel flows. Such overestimation of the GS turbulent energy is often observed in the LES with coarse grid resolutions (see, e.g., Morinishi and Vasilyev\cite{mv2001}). IA predicts $R^\mathrm{GS}_{xx}$ better than other models. Although IA-CL succeeds in reducing $R^\mathrm{GS}_{xx}$, it still results in overestimation. IA-LV overestimates $R^\mathrm{GS}_{xx}$ for $y^+ > 20$. Interestingly, IA-CL overestimates $R^\mathrm{GS}_{xx}$, while it underestimates the mean velocity, as shown in Fig.~\ref{fig:4}. This suggests that the overestimation of $R^\mathrm{GS}_{xx}$ is not necessarily caused by the overestimation of the velocity gradient. The overestimation of $R^\mathrm{GS}_{xx}$ leads to overestimation of the total Reynolds stress $R_{xx}$. In Fig.~\ref{fig:5}(b), IA gives a reasonable prediction of the total value $R_{xx}$, while all other models overestimate this value. IA and IA-CL provide a reasonable prediction of the SGS component $\langle \tau^\mathrm{sgs}_{xx} \rangle$. IA-LNcs42 underestimates $\langle \tau^\mathrm{sgs}_{xx} \rangle$, whereas IA-LV overestimates it. Note that the eddy-viscosity part in the SGS stress contributions little to the normal stress; namely, $\langle \tau^\mathrm{eat}_{\alpha \alpha} \rangle \gg -2 \langle \nu^\mathrm{sgs} \overline{s}_{\alpha \alpha} \rangle \simeq 0$ (see, Appendix~\ref{sec:a}). This is in agreement with the statement that the eddy-viscosity model is an `isotropic' model. Thus, we can recognize that the non-eddy-viscosity term is required to express the anisotropy of the SGS stress in turbulent flows.

For $R^\mathrm{GS}_{yy}$ in Fig.~\ref{fig:5}(c), DSM seems to provide the best prediction; however, it overestimates $R^\mathrm{GS}_{xx}$. Namely, DSM is more anisotropic than f-DNS. IA-LNcs42 is likewise excessively anisotropic. In the present cases, IA yields a good prediction of the wall-normal component of the GS Reynolds stress $R^\mathrm{GS}_{yy}$ while also efficiently predicting the streamwise component $R^\mathrm{GS}_{xx}$. In Fig.~\ref{fig:5}(d), IA-LNcs42 provides the profile of $R_{yy}$ or $\langle \tau^\mathrm{sgs}_{yy} \rangle$ far from f-DNS. As presented in Appendix~\ref{sec:a}, the eddy-viscosity part in the SGS stress contributes little to the normal stress, such that $\langle \tau^\mathrm{sgs}_{yy} \rangle \simeq 2 \langle k^\mathrm{sgs} \rangle/3$ in IA-LNcs42. For IA, IA-CL, and IA-LV, the wall-normal component of the EAT is negative, i.e., $\langle \tau^\mathrm{eat}_{yy} \rangle < 0$. Hence, these SMMs predict a smaller value of the wall-normal Reynolds stress compared with the eddy-viscosity models. In Fig.~\ref{fig:5}(d), IA seems to yield the best prediction, although it underestimates the GS and the total value of the wall-normal stress.

\begin{figure}[h]
 \centering
 \includegraphics[width=0.5\textwidth]{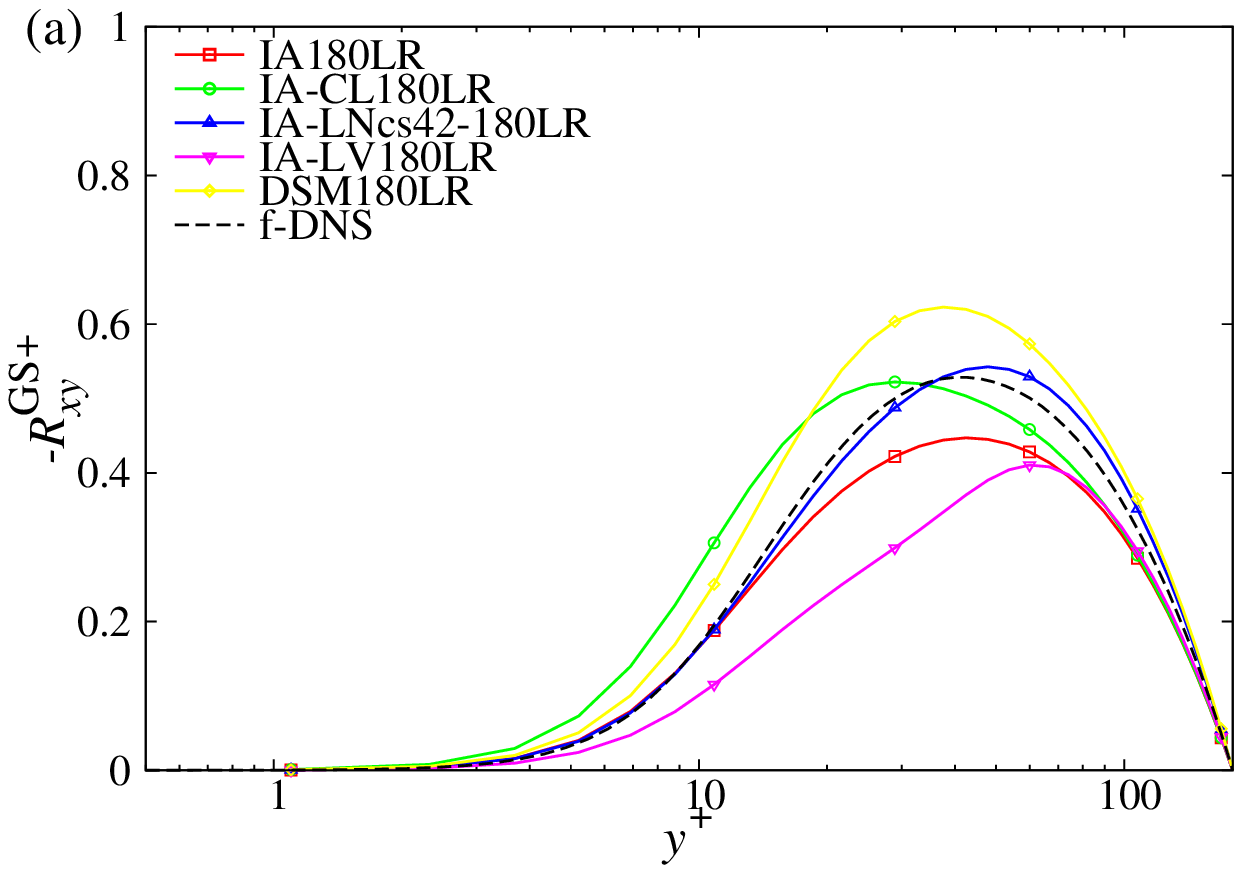}
 \includegraphics[width=0.5\textwidth]{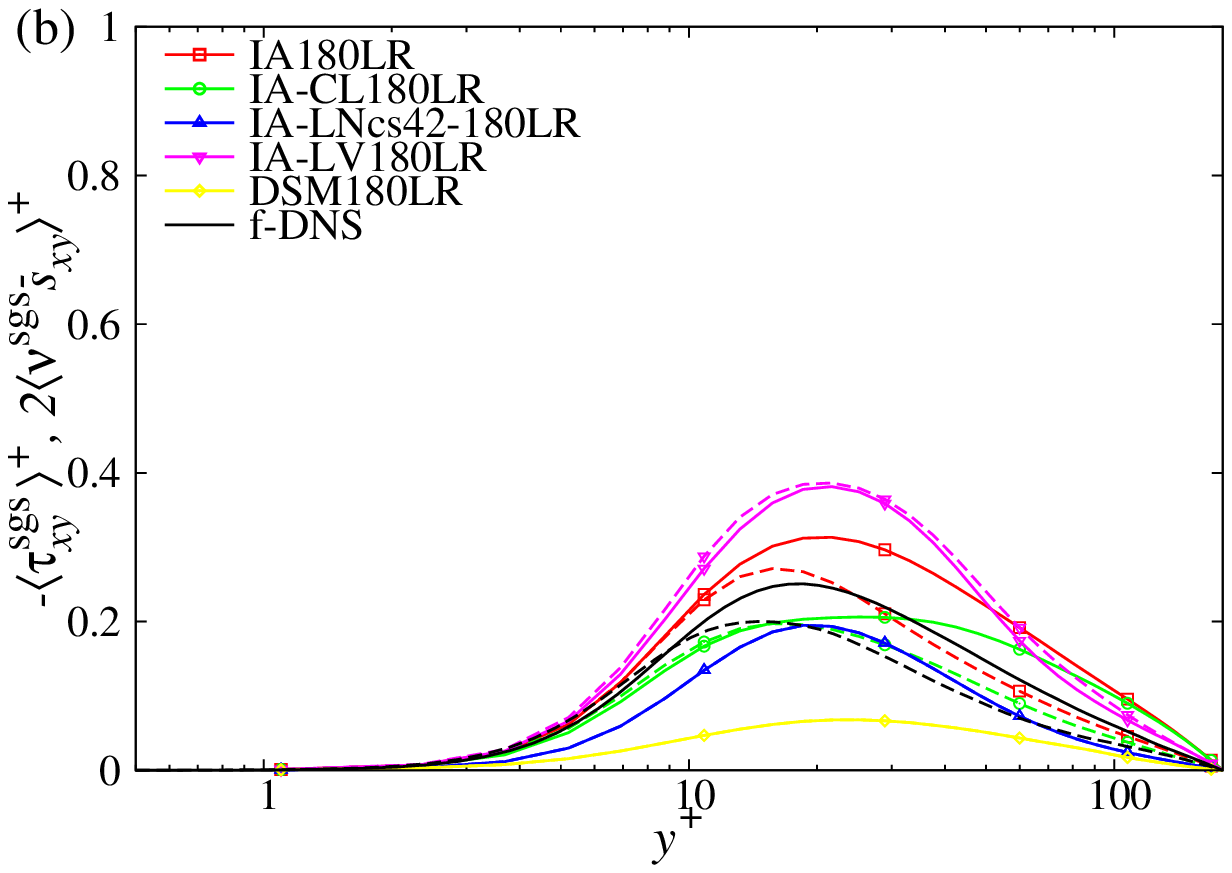}
 \includegraphics[width=0.5\textwidth]{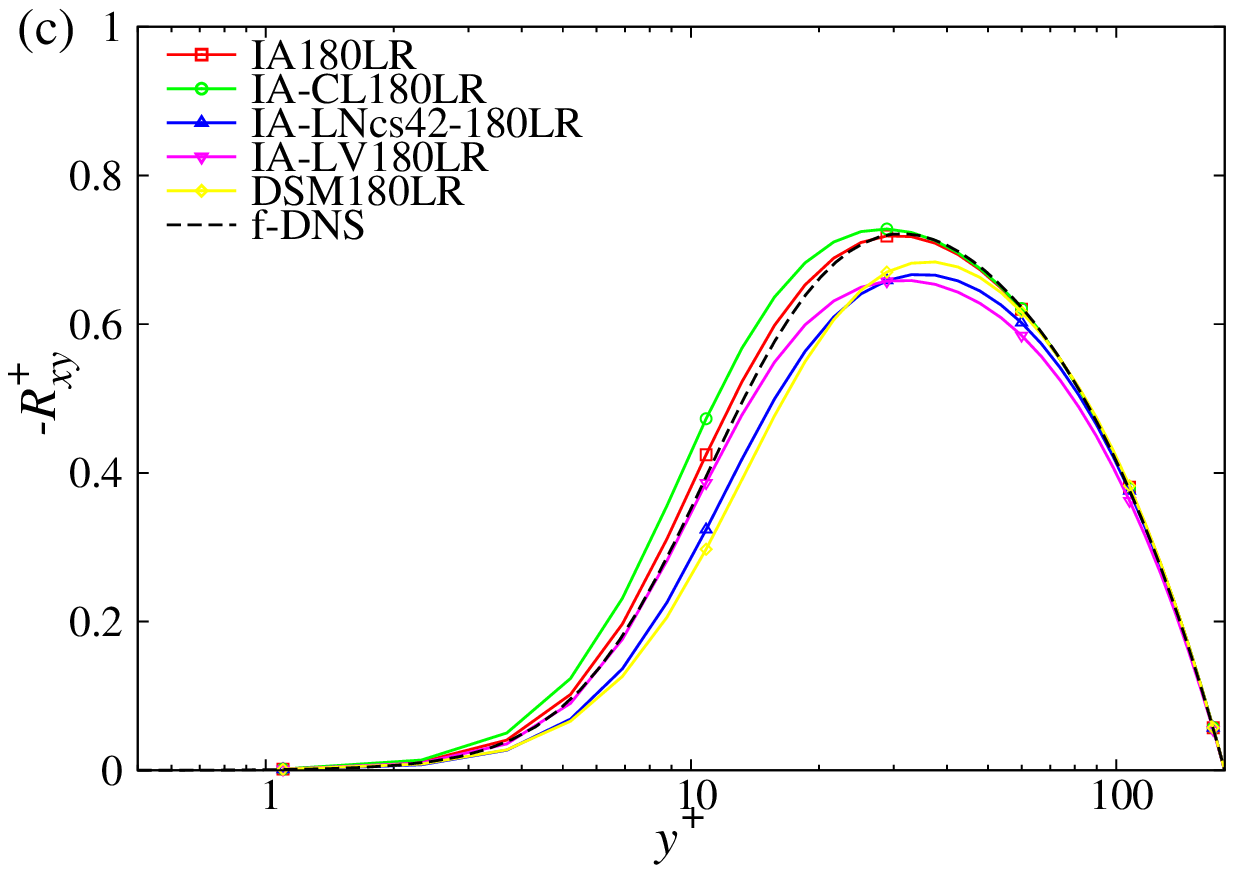}
\caption{\label{fig:6} Profiles of the Reynolds shear stress for (a) GS $-R^\mathrm{GS}_{xy}$, (b) SGS $-\langle \tau^\mathrm{sgs}_{xy} \rangle$, and (c) total components $-R_{xy}$ at $\mathrm{Re}_\tau = 180$. In (b), solid lines depict the value of $-\langle \tau^\mathrm{sgs}_{xy} \rangle$, while dashed lines depict the contribution from the eddy-viscosity term $2 \langle \nu^\mathrm{sgs} \overline{s}_{xy} \rangle$.}
\end{figure}

Figure~\ref{fig:6} shows the profiles of the Reynolds shear stress for various cases in LR at $\mathrm{Re}_\tau = 180$. The contribution from the eddy-viscosity term is plotted in Fig.~\ref{fig:6}(b). For f-DNS, the SGS eddy viscosity $\nu^\mathrm{sgs}$ is evaluated through
\begin{align}
\nu^\mathrm{sgs} = -\frac{\tau^\mathrm{sgs}_{ij} \overline{s}_{ij}}{2 \overline{s}_{\ell m} \overline{s}_{\ell m}},
\label{eq:3.7}
\end{align}
which is the same approach used in Abe\cite{abe2019}. Note that $\nu^\mathrm{sgs}$ given by Eq.~(\ref{eq:3.7}) is not necessarily positive. Hence, it allows the backscatter through the eddy-viscosity term, while $\nu^\mathrm{sgs}$ in the present LES is established to be positive. Although this is not a unique approach to determine $\nu^\mathrm{sgs}$, we can decompose the SGS stress into two parts through this procedure, where one plays the role of energy transfer between GS and SGS components through the eddy viscosity $\nu^\mathrm{sgs}$, the other plays the role of SGS forcing apart from the energy transfer. This decomposition is consistent with the concept of the SMM described in Sec.~\ref{sec:level2.2}.

In Fig.~\ref{fig:6}(a), DSM overestimates the GS component $R^\mathrm{GS}_{xy}$. IA-CL and IA-LNcs42 efficiently predict $R^\mathrm{GS}_{xy}$, whereas they overestimate the streamwise velocity fluctuation $R^\mathrm{GS}_{xx}$ in the same manner as DSM. This suggests that the statistical profiles of the mean velocity and the GS Reynolds shear stress alone cannot account for the overestimation of $R^\mathrm{GS}_{xx}$ observed in Fig.~\ref{fig:5}(a), indicating the presence of some effect related to turbulent structures. We discuss this issue further in Sec.~\ref{sec:level4}. IA and IA-LV underestimate $R^\mathrm{GS}_{xy}$; however, IA complements the underestimation of the GS value with the contribution from the SGS part as shown in Fig.~\ref{fig:6}(c). In Fig.~\ref{fig:6}(b), we find that the eddy-viscosity term contributes significantly to the SGS shear stress in f-DNS. IA and IA-CL reproduce this trend. In this sense, the value of the eddy-viscosity coefficient $C_\mathrm{sgs} = 0.075$ may not be considered an artificially large value, although it is too large to sustain the turbulent fluctuation without the EAT as observed in Fig.~\ref{fig:1}(b). In Fig.~\ref{fig:6}(c), IA yields the best prediction of the total Reynolds shear stress $R_{xy}$, leading to the best prediction of the mean velocity, as shown in Figs.~\ref{fig:1} and \ref{fig:4}. For IA-CL, the overestimation of $-R_{xy}$ in the near-wall region $y^+<30$ results in the underestimation of the mean velocity, as shown in Fig.~\ref{fig:4}.

\subsubsection{\label{sec:level3.3.4}Lumley's invariant map}

\begin{figure}[t]
 \centering
 \includegraphics[width=0.5\textwidth]{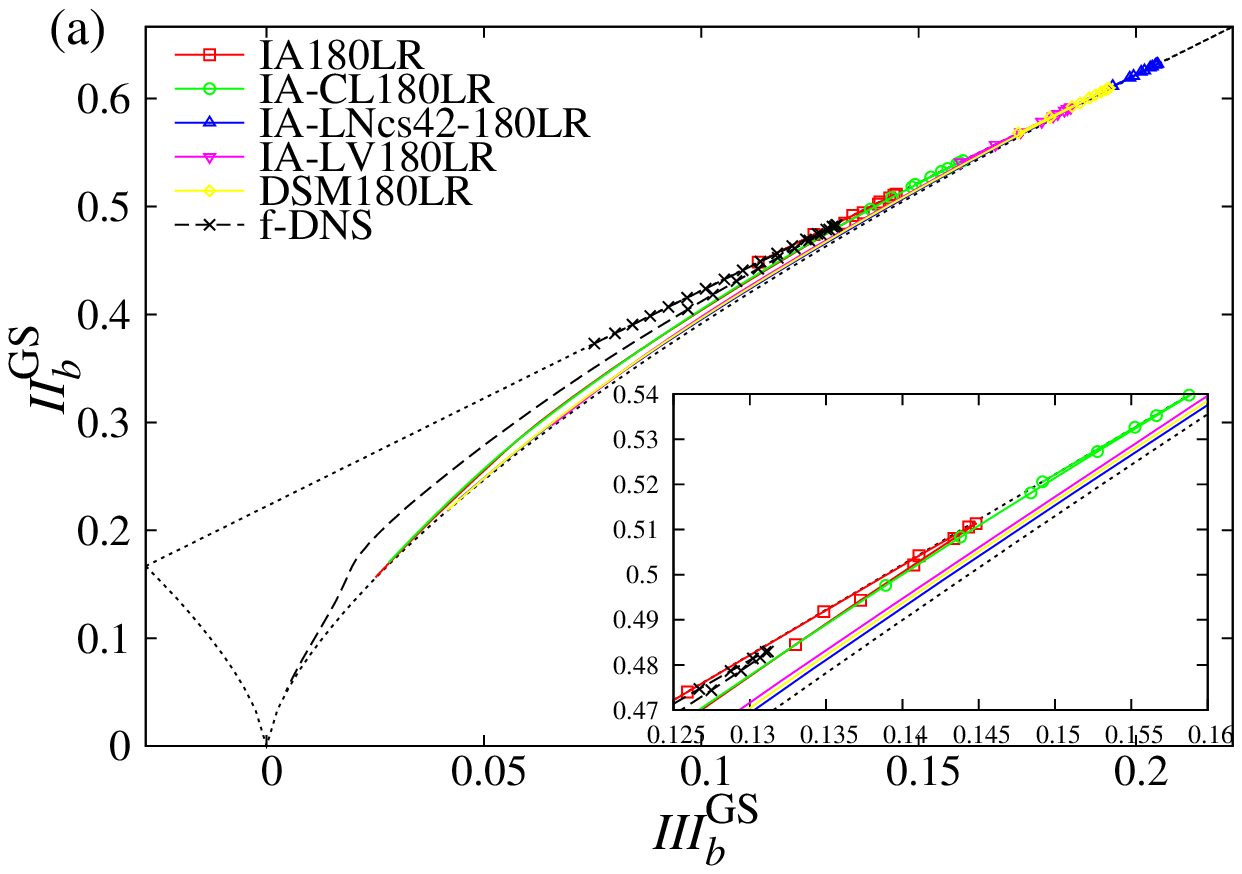}
 \includegraphics[width=0.5\textwidth]{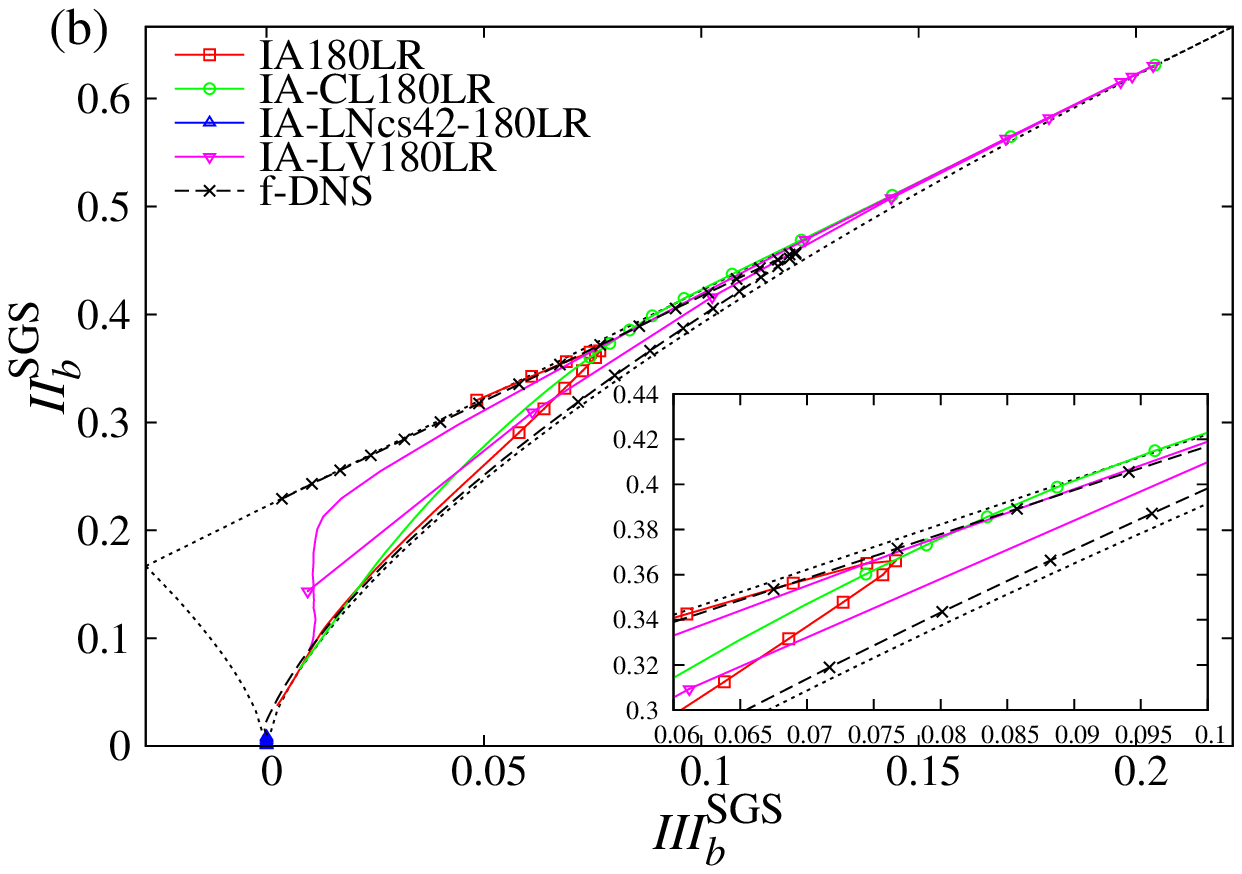}
\caption{\label{fig:7} Lumley's invariant map for (a) GS and (b) SGS normalized anisotropy tensor at $\mathrm{Re}_\tau=180$. Values for $y^+ < 20$ are plotted with symbols. The inset shows the enlarged view in which IA turns around. The channel center corresponds to the end of lines closest to the origin. As it approaches the solid wall, f-DNS proceeds toward the right top, then turns around and asymptotically approaches the top line of the two-component turbulence, in both (a) and (b).}
\end{figure}

To quantitatively evaluate the anisotropy of the turbulent stress for various models, we investigate Lumley's invariant map (see, e.g., Hanjali\'c and Launder\cite{hanjaliclaunderbook}). We define the GS and SGS normalized anisotropy tensor $b^\mathrm{GS}_{ij}$ and $b^\mathrm{SGS}_{ij}$, respectively, by
\begin{align}
b^\mathrm{GS}_{ij} = \frac{R_{ij}^\mathrm{GS}}{R^\mathrm{GS}_{\ell \ell}} - \frac{1}{3} \delta_{ij}, \ \ 
b^\mathrm{SGS}_{ij} = \frac{\langle \tau^\mathrm{sgs}_{ij} \rangle}{\langle \tau^\mathrm{sgs}_{\ell \ell} \rangle} - \frac{1}{3} \delta_{ij}.
\label{eq:3.8}
\end{align}
Their second and third invariants read
\begin{gather}
II^\mathrm{A}_b = b^\mathrm{A}_{ij} b^\mathrm{A}_{ij}/2, \ \ 
III^\mathrm{A}_b = b^\mathrm{A}_{ij} b^\mathrm{A}_{j\ell} b^\mathrm{A}_{\ell i}/3, 
\label{eq:3.9}
\end{gather}
where $\mathrm{A} = \mathrm{GS}, \mathrm{SGS}$. The realizability conditions\cite{schumann1977,vremanetal1994realizability} for the Reynolds or SGS stress indicate that the invariants (\ref{eq:3.9}) lie in the following region\cite{hanjaliclaunderbook}:
\begin{align}
II^\mathrm{A}_b \le 2 III^\mathrm{A}_b + \frac{2}{9}, \ \ 
(II^\mathrm{A}_b)^3 \ge 6 ( III^\mathrm{A}_b)^2.
\label{eq:3.10}
\end{align}
Figure~\ref{fig:7} shows the invariant map for various cases at $\mathrm{Re}_\tau=180$. The values for $y^+ < 20$ are plotted with symbols. In this map, the upper line, $II^\mathrm{A}_b = 2 III^\mathrm{A}_b +2/9$, denotes the two-component turbulence. In turbulent channel flows, this corresponds to the condition that the wall-normal stress is negligible relative to the streamwise and spanwise components. The upper right tip corresponds to the one-component turbulence (only the streamwise component), whereas the left tip corresponds to the two-component isotropic turbulence. The origin is the three-component isotropic condition. 

In Fig.~\ref{fig:7}(a), IA-LNcs42, IA-LV, and DSM approach the top right tip as they proceed toward the solid wall, which indicates that the GS velocity fluctuation becomes nearly one component. The profiles for IA-CL and IA approach that for f-DNS; namely, they recover the asymptotic behavior of the GS turbulence anisotropy in the vicinity of the wall. The anisotropy of the SGS component exhibit more evident differences between the models than the GS component. In Fig.~\ref{fig:7}(b), it should be noted that IA-LNcs42 lies at the origin, which indicates that the SGS stress is isotropic. This reflects the property of the eddy-viscosity model as the `isotropic' model. IA-LV exhibits strange behavior, and it does not asymptotically approach the two-component turbulence in the vicinity of the wall. This is because IA-LV fails to reproduce the asymptotics of the wall-normal stress, $\langle \tau^\mathrm{sgs}_{yy} \rangle \sim O(y^4)$. IA-CL approaches the two-component turbulence in the vicinity of the wall; however it approaches the one-component turbulence tip and does not turn around. Only IA turns around in the near-wall region, such that it reproduces the asymptotic trend of the SGS stress in the vicinity of the wall, although it does not perfectly correspond to f-DNS. In summary, IA yields the best prediction among the presented models on the anisotropy of both the GS and SGS stress.

\subsubsection{\label{sec:level3.3.5}Reynolds stress spectrum}

According to Abe\cite{abe2019}, the SMM employing the scale-similarity model for the SGS-Reynolds term recovers the energy spectrum close to the cut-off scale. The energy spectrum reflects information concerning turbulent structures such as streak structures observed in wall-bounded turbulence. We define the spectrum of the GS Reynolds stress by
\begin{gather}
E^\mathrm{GS}_{ij} (k_x,y,k_z) = \Re \langle \tilde{\overline{u}}_i \tilde{\overline{u}}_j^* \rangle, 
\label{eq:3.11} \\
R^\mathrm{GS}_{ij} (y) = \sum_{k_x=0}^{k_x^\mathrm{max}} \sum_{k_z=0}^{k_z^\mathrm{max}} E^\mathrm{GS}_{ij} (k_x,y,k_z) \Delta k_x \Delta k_z,
\label{eq:3.12}
\end{gather}
where $k_\alpha^\mathrm{max} = \pi N_\alpha/L_\alpha$, $\Delta k_\alpha = 2\pi/L_\alpha$, $\alpha=x,z$, the superscript `$*$' denotes the complex conjugate, and $\tilde{q}$ is the Fourier coefficient of an instantaneous variable $q$, defined by
\begin{align}
\tilde{q}^{(J)} (k_x,k_z) = \frac{1}{N_x N_z} \sum_{I=1}^{N_x} \sum_{K=1}^{N_z} q^{(I,J,K)} \mathrm{e}^{-\mathrm{i} (k_x I L_x/N_x+ k_z K L_z/N_z)}.
\label{eq:3.13}
\end{align}
Hereafter, we focus on the streamwise spectrum $E^\mathrm{GS}_{ij}(k_x,y)$, which is defined by 
\begin{align}
E^\mathrm{GS}_{ij}(k_x,y) = \sum_{k_z=0}^{k_z^\mathrm{max}} E^\mathrm{GS}_{ij} (k_x,y,\allowbreak k_z) \Delta k_z
\label{eq:3.14}
\end{align}
Because the difference between the models is evident at $y^+ = 20$ for both the mean velocity in Fig.~\ref{fig:4} and the Reynolds stress in Figs.~\ref{fig:5} and \ref{fig:6}, we focus on the spectrum at that plane.

\begin{figure}[t]
 \centering
 \includegraphics[width=0.5\textwidth]{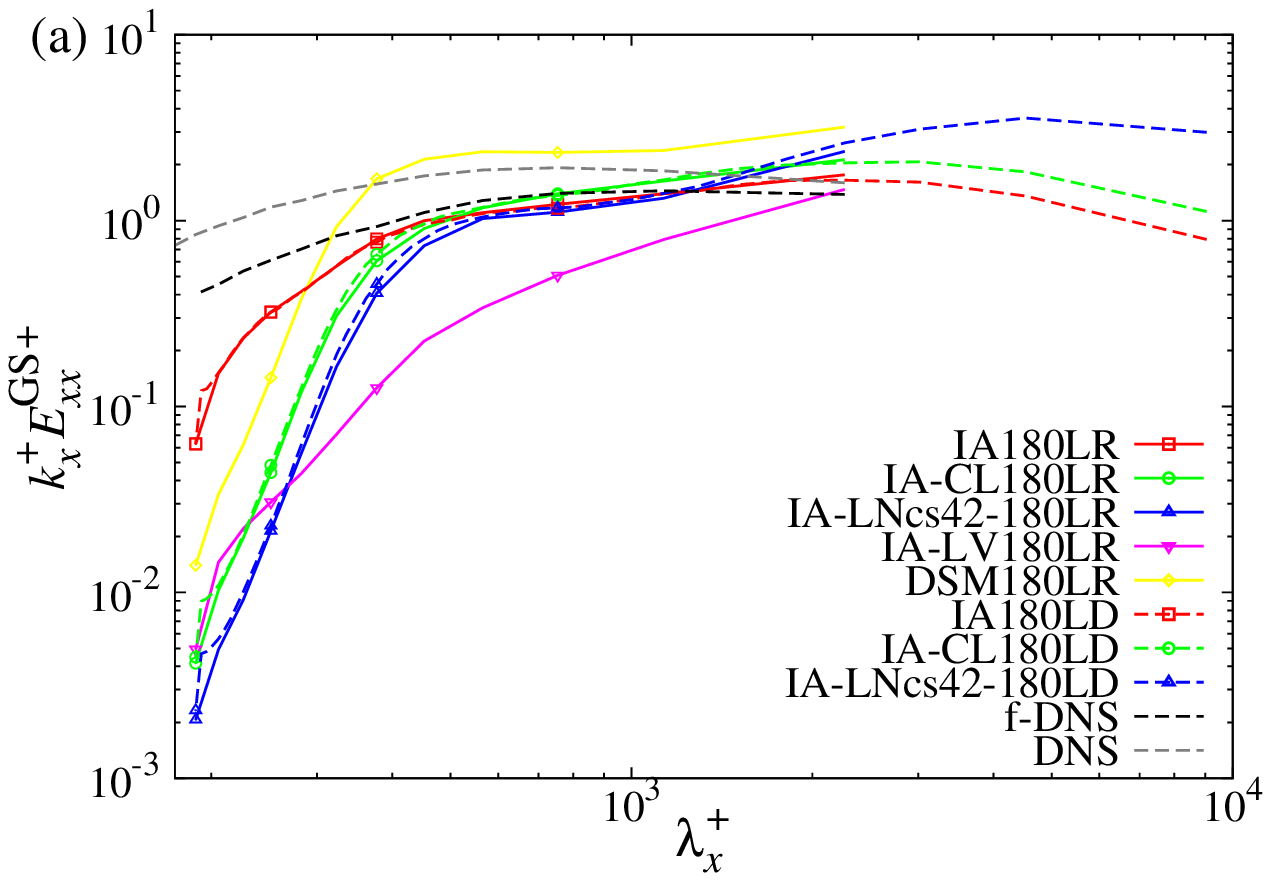}
 \includegraphics[width=0.5\textwidth]{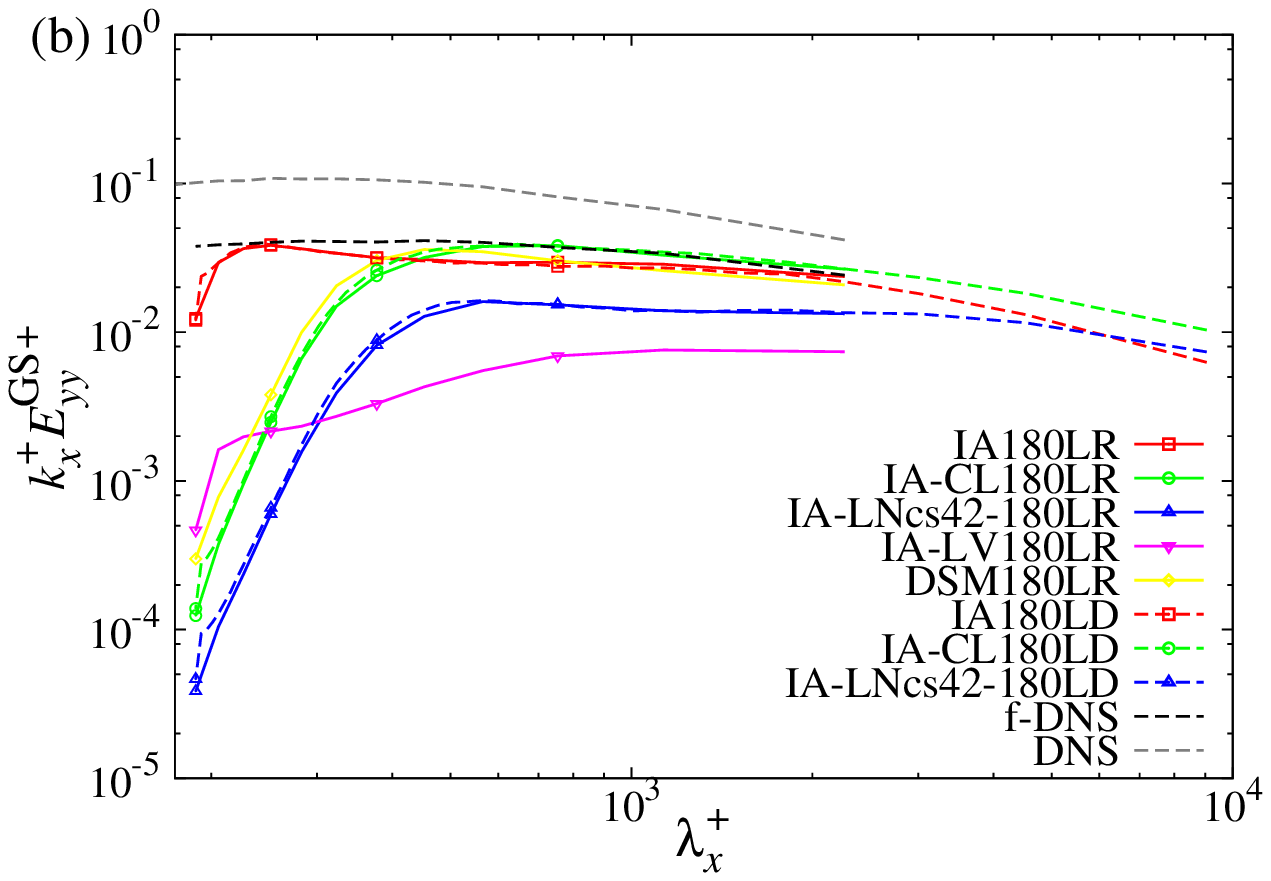}
 \includegraphics[width=0.5\textwidth]{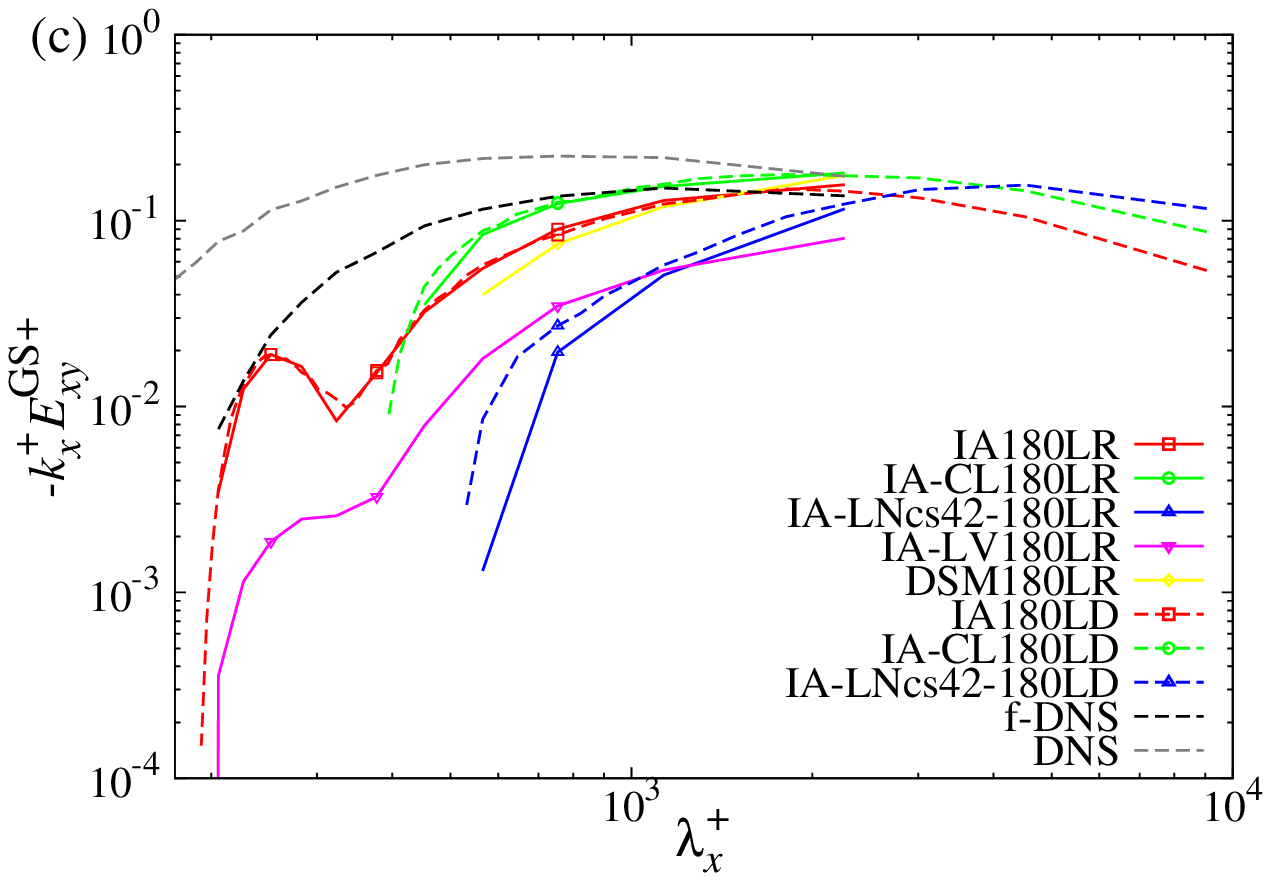}
\caption{\label{fig:8} Profiles of the GS Reynolds stress spectrum for (a) streamwise $E^\mathrm{GS}_{xx}$, (b) wall-normal $E^\mathrm{GS}_{yy}$, and (c) shear components $-E^\mathrm{GS}_{xy}$ at $y^+=20$ for $\mathrm{Re}_\tau = 180$.}
\end{figure}

Figure~\ref{fig:8} shows the streamwise spectrum of the GS Reynolds stress for various cases in LR or LD at $y^+=20$ for $\mathrm{Re}_\tau = 180$. We plot the result for the non-filtered DNS for reference. To emphasize the high-wavenumber or low-wavelength region, we set the horizontal axis in the wavelength $\lambda_x (=2\pi/k_x)$ instead of the wavenumber $k_x$. First, the results in LD almost overlap those in LR, in the same manner as the mean velocity in Fig.~\ref{fig:1}. Hence, the statistics in the present LES do not depend on the domain size, but on the grid resolution. In Fig.~\ref{fig:8}(a), as shown by Abe\cite{abe2019}, IA recovers the intensity of the GS streamwise velocity spectrum close to the cut-off wavelength scale. The spectrum of the GS streamwise velocity is related to the near-wall streak structure. The spectrum accumulated in the high-wavelength region such as IA-LNcs42 or DSM, indicates that the structure elongated in the streamwise direction is considerably more dominant than the short scale structures. For IA-CL, the slope of the spectrum at the high-wavelength region is gentle. This indicates that IA-CL succeeds in reproducing the length of the streak structure to some extent. This is further discussed in Sec.~\ref{sec:level4.2}. Several studies discussed the relation between the breakdown of the streaks and the sustaining process of turbulent shear flows\cite{hamiltonetal1995,waleffe1997}. In this sense, IA may succeed in reproducing the sustaining process of turbulent shear flow with the coarse grid by the enhancement of the turbulence close to the cut-off scale. As shown in Fig.~\ref{fig:8}(b), IA also predicts the GS wall-normal velocity spectrum in the entire wavelength range. Other models, including IA-CL, cannot reproduce the spectrum close to the cut-off wavelength scale. The difference between IA and IA-CL suggests that the EAT, based not on the first-order but on the second-order spatial derivative of the velocity field, is useful in restoring the behavior of the spectra close to the cut-off scale.

For $E^\mathrm{GS}_{xy}$ in Fig.~\ref{fig:8}(c), the lines disappear at the low-wavelength region for some models. This is because $-E^\mathrm{GS}_{xy}$ is negative at wavelength regions below that scale. In contrast, IA predicts the positive $-E^\mathrm{GS}_{xy}$ up to the cut-off wavelength as efficiently as f-DNS. Although IA-LV likewise predicts a positive $-E^\mathrm{GS}_{xy}$ value in the entire wavelength region, the value is smaller than that obtained by IA or f-DNS. Furthermore, IA-LV cannot predict $E^\mathrm{GS}_{xx}$ and $E^\mathrm{GS}_{yy}$ at the low-wavelength region. For IA in Fig.~\ref{fig:8}(c), the spectrum bends around $\lambda_x^+ = 300$. This corresponds to the contribution of the EAT on the budget of $E^\mathrm{GS}_{yy}$, as described later in Sec.~\ref{sec:level4.1.2}. However, the success of IA suggests that the reproduction of the GS Reynolds shear stress spectrum $E^\mathrm{GS}_{xy}$ close to the cut-off scale is an essence of further development of SGS models.

\section{\label{sec:level4}Discussion}

\subsection{\label{sec:level4.1}Budget equation for the GS Reynolds stress spectrum}

As shown in Sec.~\ref{sec:level3}, the SGS models showed evident differences on anisotropy and structures represented by the Reynolds stress spectrum, in the near-wall region. To investigate the effect of the SGS stress on both anisotropy and structures of turbulence in shear flows, it is useful to analyze the budget equation for the Reynolds stress spectrum\cite{mizuno2016,ka2018,lm2019}. In the use of SGS models, the budget equation for the GS Reynolds stress spectrum yields
\begin{align}
& \frac{\partial E^\mathrm{GS}_{ij}}{\partial t} + \frac{\partial}{\partial x_\ell} (U_\ell E^\mathrm{GS}_{ij})
\nonumber \\
& = \check{P}^\mathrm{GS}_{ij} - \check{\varepsilon}^\mathrm{GS}_{ij}
+ \check{D}^\mathrm{t,GS}_{ij} + \check{\Phi}^\mathrm{GS}_{ij} + \check{D}^\mathrm{p,GS}_{ij}
+ \check{D}^\mathrm{v,GS}_{ij}
+ \check{T}^\mathrm{GS}_{ij}
\nonumber \\
& \hspace{1em}
+ \check{\xi}^\mathrm{SGS}_{ij} + \check{D}^\mathrm{SGS}_{ij}.
\label{eq:4.1}
\end{align}
Each term on the right-hand side is referred to as the production $\check{P}^\mathrm{GS}_{ij}$, destruction $\check{\varepsilon}^\mathrm{GS}_{ij}$, turbulent diffusion $\check{D}^\mathrm{t,GS}_{ij}$, pressure--strain correlation $\check{\Phi}^\mathrm{GS}_{ij}$, pressure diffusion $\check{D}^\mathrm{p,GS}_{ij}$, viscous diffusion $\check{D}^\mathrm{v,GS}_{ij}$, inter-scale transfer $\check{T}^\mathrm{GS}_{ij}$, SGS destruction $\check{\xi}^\mathrm{SGS}_{ij}$, and SGS diffusion $\check{D}^\mathrm{SGS}_{ij}$, respectively. When the turbulent field is inhomogeneous only in the $y$ direction, they are defined as follows:
\begin{subequations}
\begin{align}
\check{P}^\mathrm{GS}_{ij} & = - E^\mathrm{GS}_{iy} \frac{\partial U_j}{\partial y}
- E^\mathrm{GS}_{jy} \frac{\partial U_i}{\partial y}, 
\label{eq:4.2a} \\
\check{\varepsilon}^\mathrm{GS}_{ij} & = 2 \nu \Re \left< \tilde{\overline{s}}_{i\ell}' (\tilde{\partial}_\ell \tilde{\overline{u}}_j')^*
+ \tilde{\overline{s}}_{j\ell}' (\tilde{\partial}_\ell \tilde{\overline{u}}_i')^* \right>, 
\label{eq:4.2b} \\
\check{D}^\mathrm{t,GS}_{ij} & = -\frac{\partial}{\partial y} \Re \left< \tilde{\overline{u}}_y' (\widetilde{\overline{u}_i' \overline{u}_j'})^* \right>, 
\label{eq:4.2c} \\
\check{\Phi}^\mathrm{GS}_{ij} & = 2 \Re \left< \tilde{\overline{p}}^\mathrm{total}{}' \tilde{\overline{s}}_{ij}'{}^* \right>, 
\label{eq:4.2d} \\
\check{D}^\mathrm{p,GS}_{ij} & = - \frac{\partial}{\partial y} \left[ \Re \left< \tilde{\overline{p}}^\mathrm{total}{}' \tilde{\overline{u}}_i'{}^* \right> \delta_{jy} + \Re \left< \tilde{\overline{p}}^\mathrm{total}{}' \tilde{\overline{u}}_j'{}^* \right> \delta_{iy} \right], 
\label{eq:4.2e} \\
\check{D}^\mathrm{v,GS}_{ij} & = 2 \nu \frac{\partial}{\partial y} \Re \left< \tilde{\overline{s}}_{jy}' \tilde{\overline{u}}_i'{}^* +  \tilde{\overline{s}}_{iy}' \tilde{\overline{u}}_j'{}^* \right>, 
\label{eq:4.2f} \\
\check{T}^\mathrm{GS}_{ij} & = \Re \left< \tilde{\overline{u}}_j' \tilde{N}_i^* + \tilde{\overline{u}}_i' \tilde{N}_j^* \right> - \check{P}^\mathrm{GS}_{ij} - \check{D}^\mathrm{t,GS}_{ij},
\label{eq:4.2g} \\
\check{\xi}^\mathrm{SGS}_{ij} & =\Re \left< \tilde{\tau}^\mathrm{sgs}_{i\ell}{}'|_\mathrm{tl} (\tilde{\partial}_\ell \tilde{\overline{u}}_j')^*
+ \tilde{\tau}^\mathrm{sgs}_{j\ell}{}'|_\mathrm{tl} (\tilde{\partial}_\ell \tilde{\overline{u}}_i')^* \right>, 
\label{eq:4.2h} \\
\check{D}^\mathrm{SGS}_{ij} & = - \frac{\partial}{\partial y} \Re \left< \tilde{\tau}^\mathrm{sgs}_{jy}{}'|_\mathrm{tl} \tilde{\overline{u}}_i'{}^* + \tilde{\tau}^\mathrm{sgs}_{iy}{}'|_\mathrm{tl} \tilde{\overline{u}}_j'{}^* \right>,
\label{eq:4.2i}
\end{align}
\end{subequations}
where $\tilde{\partial}_j = (\mathrm{i} k_x, \partial /\partial y,\mathrm{i}k_z)$, and $N_i = - \partial (\overline{u}_i \overline{u}_j)/\partial x_j$. $\overline{p}^\mathrm{total}$ is the sum of $\overline{p}$ and the SGS dynamic pressure $2k^\mathrm{sgs}/3$; namely, it is defined by
\begin{align}
\overline{p}^\mathrm{total} = \overline{p} + \frac{2}{3} k^\mathrm{sgs}.
\label{eq:4.3}
\end{align}
Furthermore, we decompose the SGS destruction term into two terms as follows:
\begin{align}
\check{\xi}^\mathrm{SGS}_{ij} = - \check{\varepsilon}^\mathrm{EV}_{ij} + \check{\xi}^\mathrm{EAT}_{ij},
\label{eq:4.4}
\end{align}
where
\begin{subequations}
\begin{align}
\check{\varepsilon}^\mathrm{EV}_{ij} & = 2 \Re \left< \widetilde{\nu^\mathrm{sgs} \overline{s}_{i\ell}} (\tilde{\partial}_\ell \tilde{\overline{u}}_j')^*
+ \widetilde{\nu^\mathrm{sgs} \overline{s}_{j\ell}} (\tilde{\partial}_\ell \tilde{\overline{u}}_i')^* \right>, 
\label{eq:4.5a} \\
\check{\xi}^\mathrm{EAT}_{ij} & = 
\Re \left< \tilde{\tau}^\mathrm{eat}_{i\ell}{}' (\tilde{\partial}_\ell \tilde{\overline{u}}_j')^*
+ \tilde{\tau}^\mathrm{eat}_{j\ell}{}' (\tilde{\partial}_\ell \tilde{\overline{u}}_i')^* \right>,
\label{eq:4.5b}
\end{align}
\end{subequations}
where $\tau^\mathrm{eat}_{ij}$ is defined in Eq.~(\ref{eq:2.11}). We refer to $\check{\varepsilon}^\mathrm{EV}_{ij}$ and $\check{\xi}^\mathrm{EAT}_{ij}$ as the eddy-viscosity destruction and the anisotropic redistribution terms, respectively. When the budget Eq.~(\ref{eq:4.1}) is summed over the wavenumbers, it leads to the conventional budget for the GS Reynolds stress (see, Appendix \ref{sec:b} or Abe\cite{abe2019}):
\begin{align}
A^\mathrm{GS}_{ij} = \sum_{k_x=0}^{k_x^\mathrm{max}} \check{A}^\mathrm{GS}_{ij} \Delta k_x,
\label{eq:4.6}
\end{align}
where $\check{A}^\mathrm{GS}_{ij}$ corresponds to each term in Eq.~(\ref{eq:4.2a})--(\ref{eq:4.2i}), (\ref{eq:4.5a}), and (\ref{eq:4.5b}), while $A^\mathrm{GS}_{ij}$ corresponds to each term on the right-hand side of Eq.~(\ref{eq:b1}).

\subsubsection{\label{sec:level4.1.1}Shear component}

\begin{figure*}[t]
  \begin{minipage}{0.49\hsize}
   \centering
   \includegraphics[width=\textwidth]{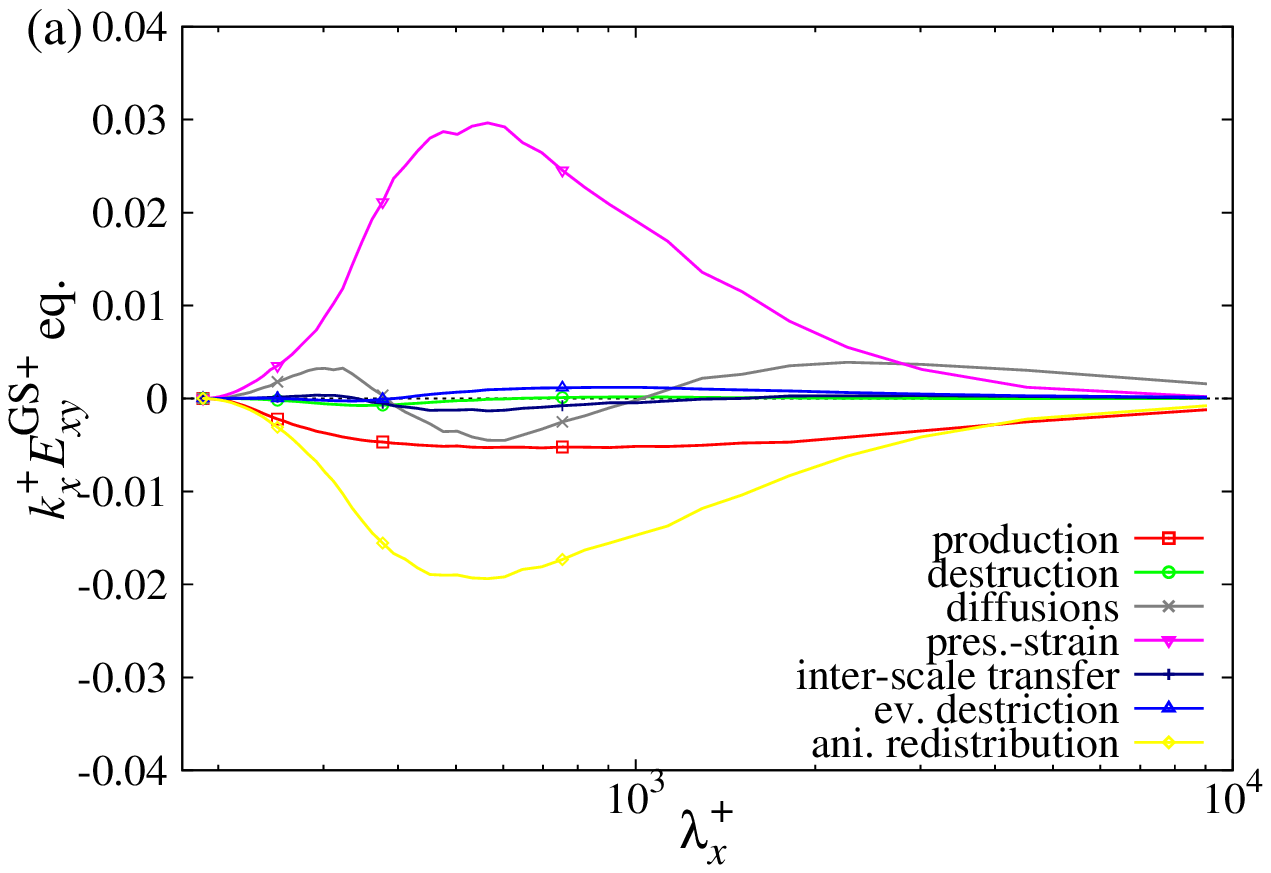}
  \end{minipage}
  \begin{minipage}{0.49\hsize}
   \centering
   \includegraphics[width=\textwidth]{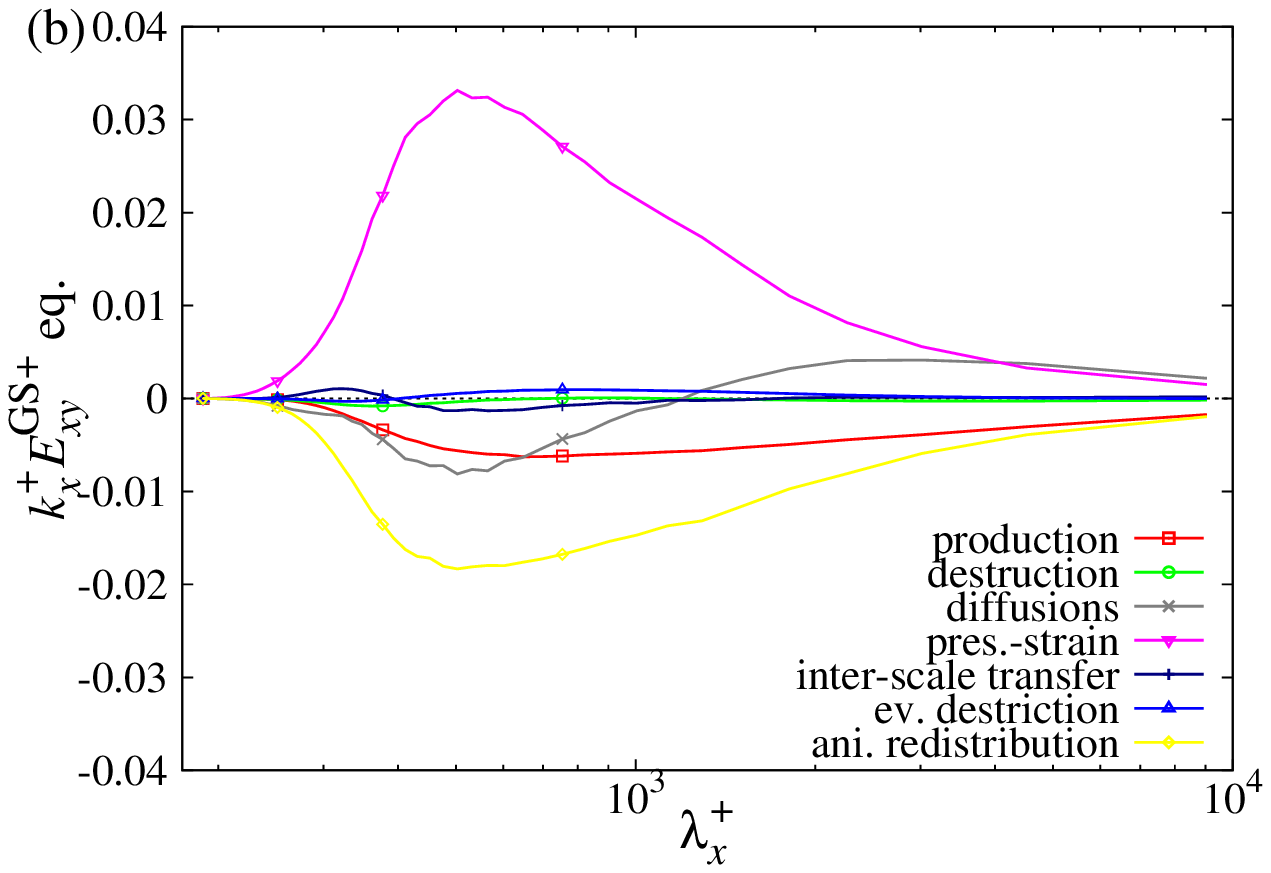}
  \end{minipage} \\

  \begin{minipage}{0.49\hsize}
   \centering
   \includegraphics[width=\textwidth]{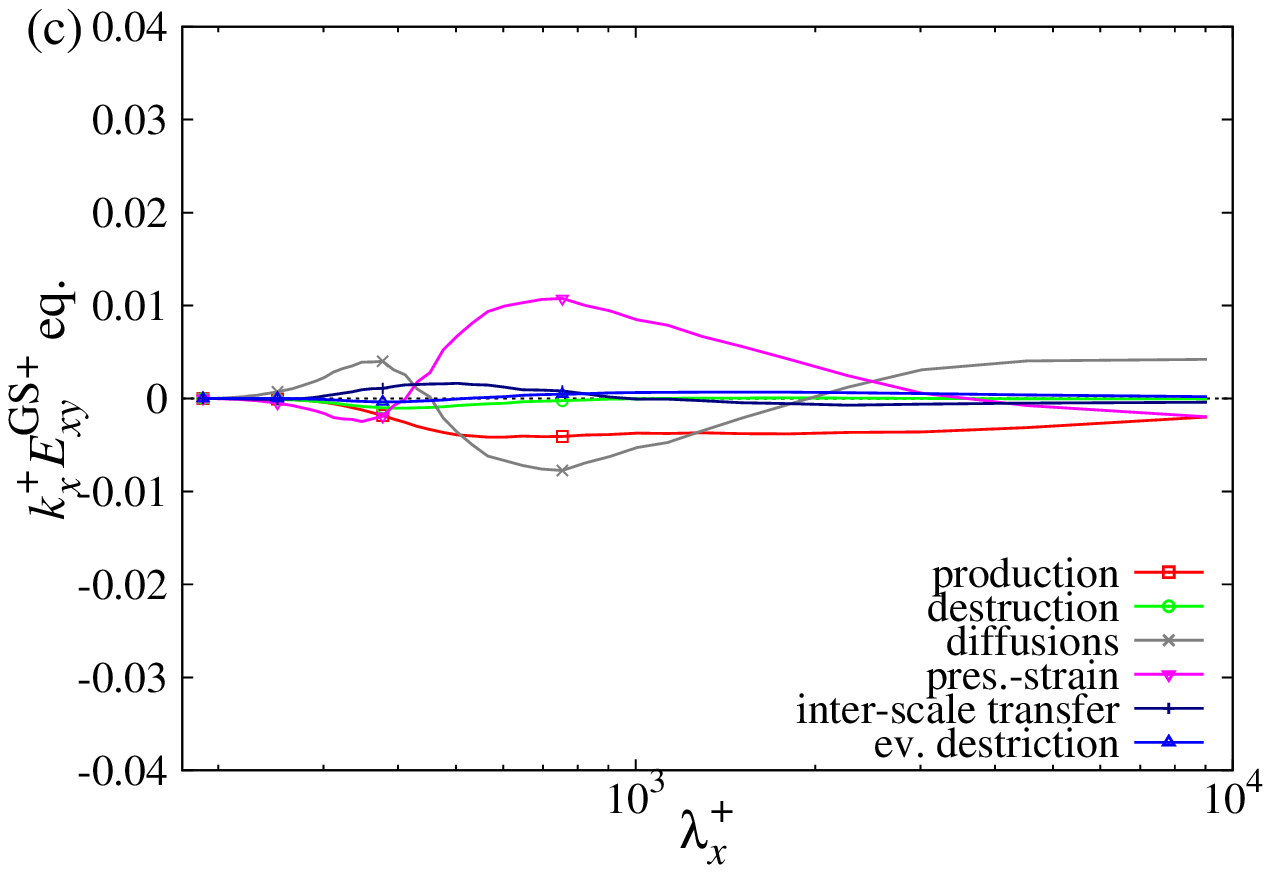}
  \end{minipage}
  \begin{minipage}{0.49\hsize}
   \centering
   \includegraphics[width=\textwidth]{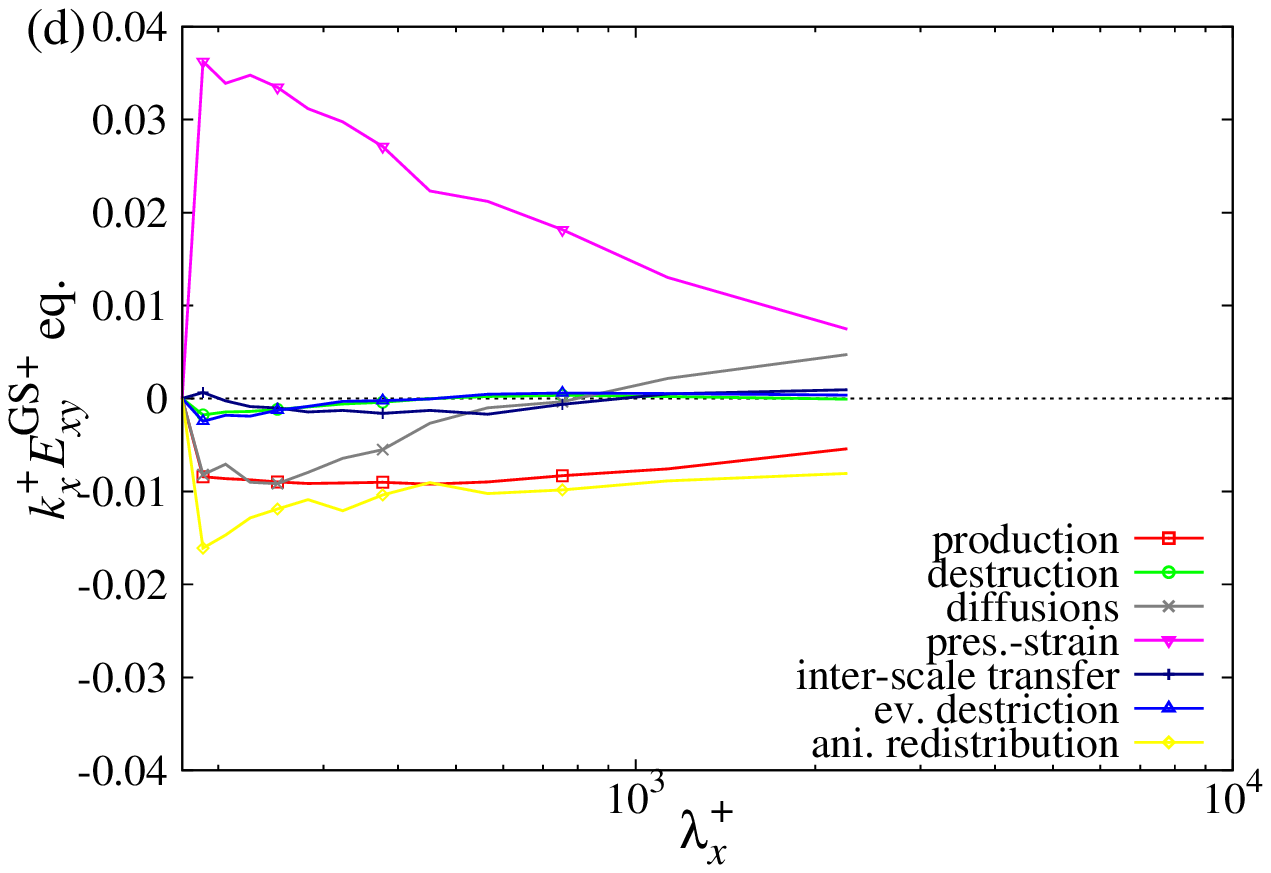}
  \end{minipage}
\caption{\label{fig:9} Budget for the GS Reynolds shear stress spectrum $E^\mathrm{GS}_{xy}$ normalized by viscous unit for (a) IA180LD, (b) IA-CL180LD, (c) IA-LNcs42-180LD, and (d) f-DNS at $y^+=20$ for $\mathrm{Re}_\tau=180$.}
\end{figure*}

A critical feature of the IA model is that it reproduces the positive $-E^\mathrm{GS}_{xy}$ in the entire wavelength region, as shown in Fig.~\ref{fig:8}(c). Abe\cite{abe2019} showed that the anisotropic redistribution term $\xi^\mathrm{EAT}_{xy}$ is indispensable to predict the profile of the SGS destruction term obtained from the filtered DNS in the budget of the GS Reynolds shear stress $R^\mathrm{GS}_{xy}$. Figure~\ref{fig:9} shows the budget for the GS Reynolds shear stress spectrum $E^\mathrm{GS}_{xy}$ for representative cases at $y^+=20$ for $\mathrm{Re}_\tau = 180$. For all cases, the eddy-viscosity destruction term contributes little to the budget in the entire wavelength region. This is a critical shortcoming of the eddy-viscosity models. IA-LNcs42 fails to predict the profile of the pressure--strain correlation $\check{\Phi}^\mathrm{GS}_{xy}$ in Fig.~\ref{fig:9}(c). IA and IA-CL qualitatively reproduce the profile of the pressure--strain correlation $\check{\Phi}^\mathrm{GS}_{xy}$ and anisotropic redistribution $\check{\xi}^\mathrm{EAT}_{xy}$ obtained from f-DNS. The difference between IA and IA-CL lies in the profile of the production term $\check{P}^\mathrm{GS}_{xy}$. $\check{P}^\mathrm{GS}_{xy}$ in IA-CL in Fig.~\ref{fig:9}(b) disappears close to the cut-off wavelength scale. This arises from the profile of $E^\mathrm{GS}_{yy}$, because $\check{P}^\mathrm{GS}_{xy}$ reads
\begin{align}
\check{P}^\mathrm{GS}_{xy} = - E^\mathrm{GS}_{yy} \frac{\partial U_x}{\partial y}.
\label{eq:4.7}
\end{align}
Hence, the reproduction of $\check{P}^\mathrm{GS}_{xy}$ close to the cut-off wavelength scale relies on the reproduction of $E^\mathrm{GS}_{yy}$.

\subsubsection{\label{sec:level4.1.2}Wall-normal component}

\begin{figure*}[t]
  \begin{minipage}{0.49\hsize}
   \centering
   \includegraphics[width=\textwidth]{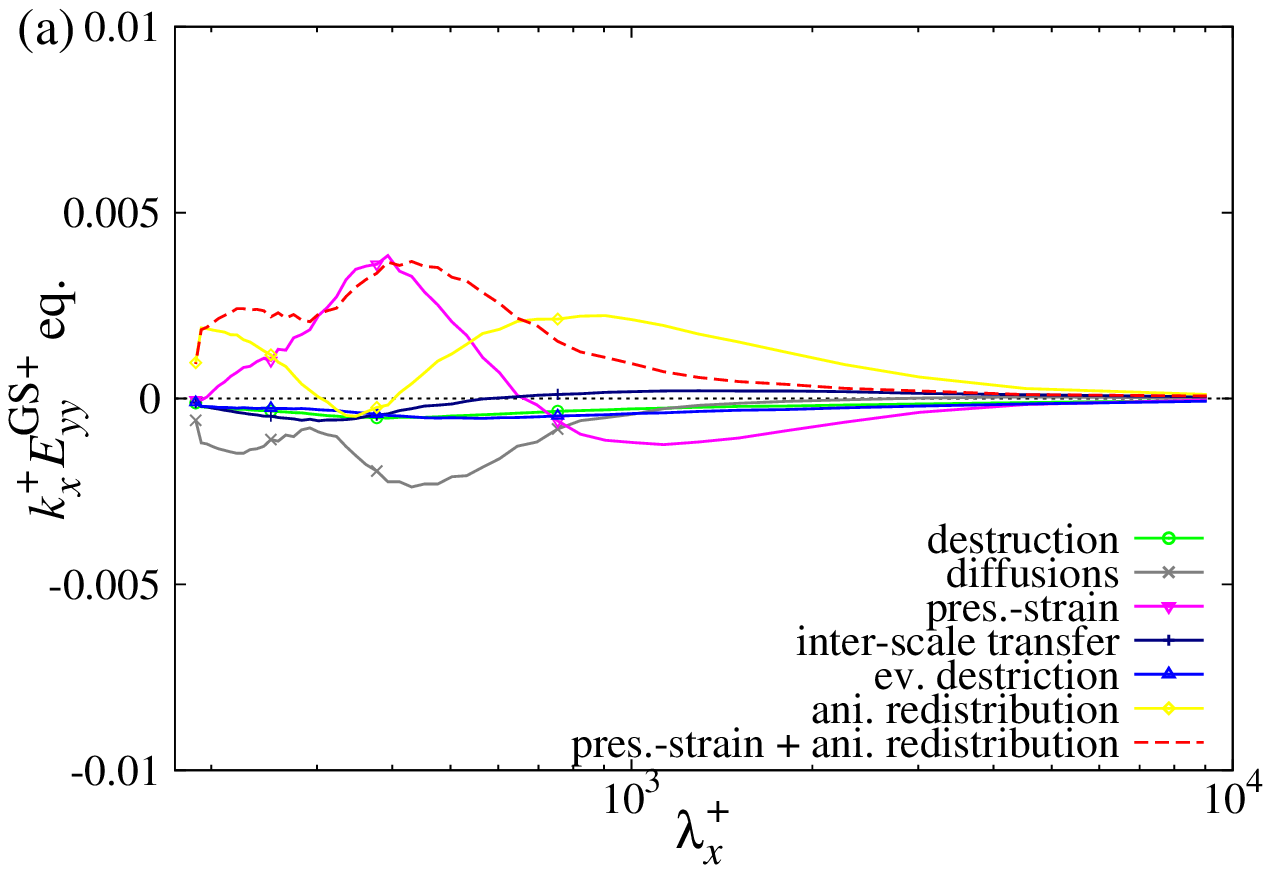}
  \end{minipage}
  \begin{minipage}{0.49\hsize}
   \centering
   \includegraphics[width=\textwidth]{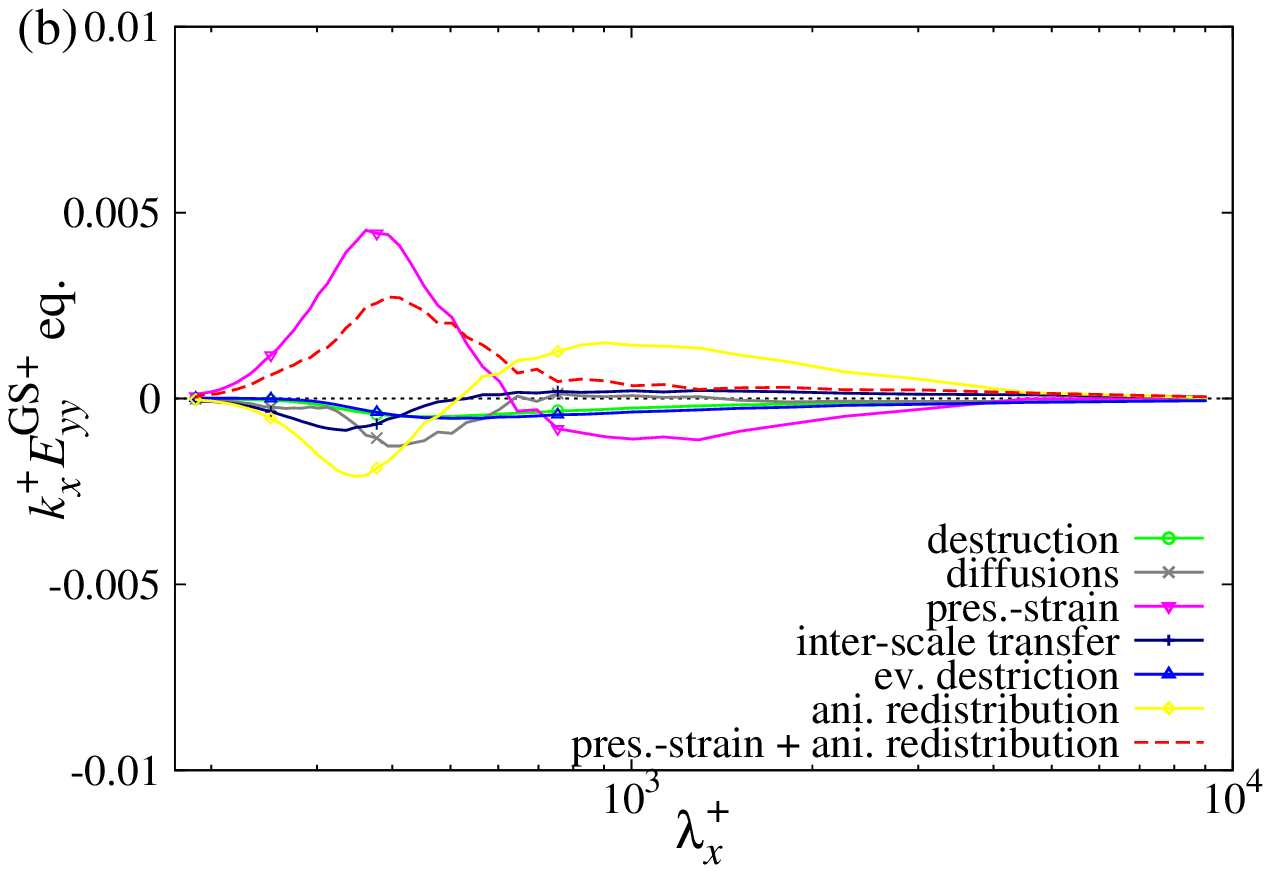}
  \end{minipage} \\

  \begin{minipage}{0.49\hsize}
   \centering
   \includegraphics[width=\textwidth]{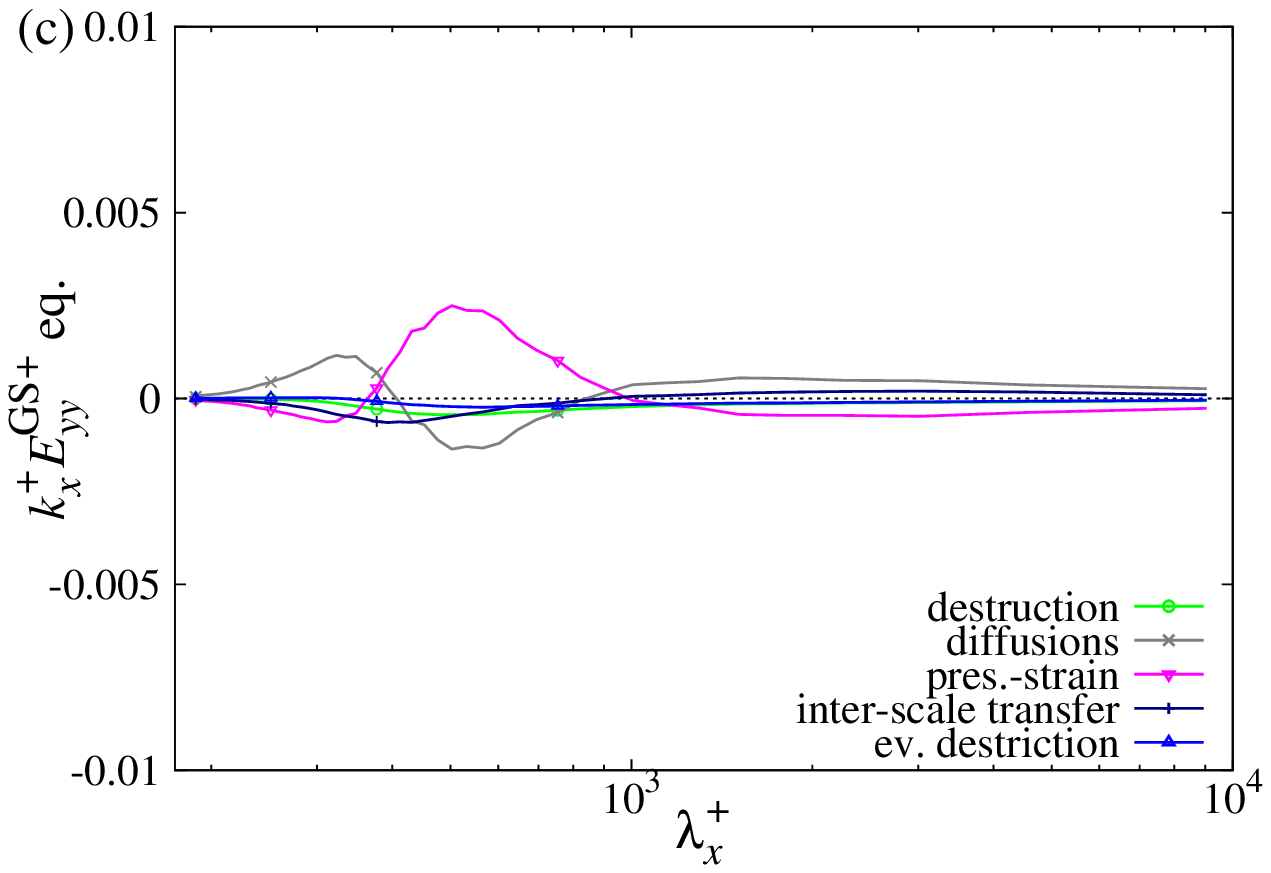}
  \end{minipage}
  \begin{minipage}{0.49\hsize}
   \centering
   \includegraphics[width=\textwidth]{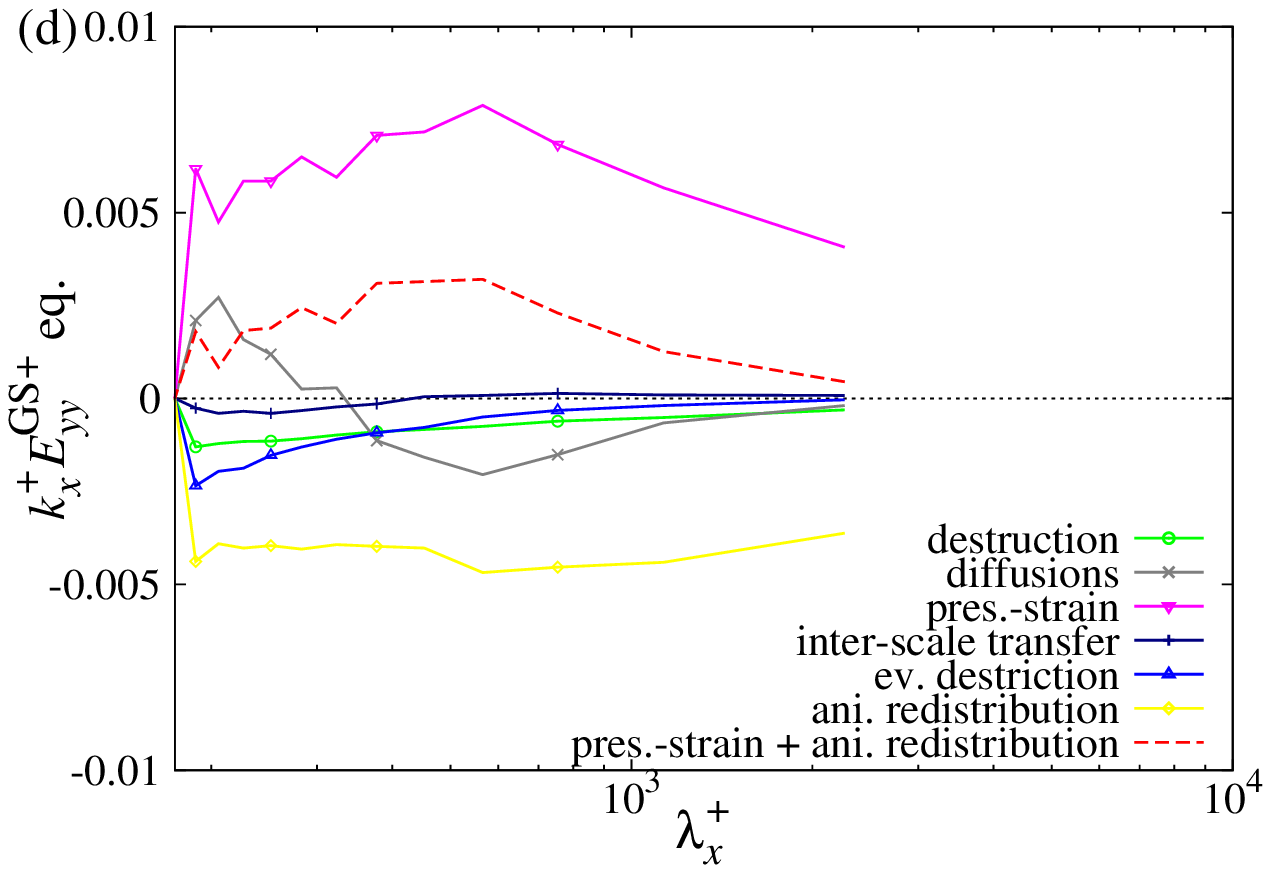}
  \end{minipage}
\caption{\label{fig:10} Budget for the wall-normal component of the GS Reynolds spectrum $E^\mathrm{GS}_{yy}$ normalized by viscous unit for (a) IA180LD, (b) IA-CL180LD, (c) IA-LNcs42-180LD, and (d) f-DNS at $y^+=20$ for $\mathrm{Re}_\tau = 180$.}
\end{figure*}

Figure~\ref{fig:10} shows the budget for the wall-normal component of the GS Reynolds stress spectrum $E^\mathrm{GS}_{yy}$ for representative cases at $y^+=20$ for $\mathrm{Re}_\tau = 180$. Both IA-CL and IA fail to reproduce the profile of the anisotropic redistribution term $\check{\xi}^\mathrm{EAT}_{yy}$ obtained from f-DNS. However, they provide a reasonable prediction of the sum of the pressure--strain correlation and the anisotropic redistribution terms $\check{\Phi}^\mathrm{GS}_{yy} + \check{\xi}^\mathrm{EAT}_{yy}$. For IA in Fig.~\ref{fig:10}{(a), $\check{\Phi}^\mathrm{GS}_{yy} + \check{\xi}^\mathrm{EAT}_{yy}$ is positive close to the cut-off wavelength scale, as is the case with f-DNS in Fig.~\ref{fig:10}(d). Hence, IA succeeds in reproducing the large intensity of the wall-normal component of the Reynolds stress spectrum $E^\mathrm{GS}_{yy}$, as shown in Fig.~\ref{fig:8}(b). For IA in Fig.~\ref{fig:10}(a), the wavelength in which the anisotropic redistribution term $\check{\xi}^\mathrm{EAT}_{yy}$ exhibits a dent corresponds to that in which $-E^\mathrm{GS}_{xy}$ is bent in Fig.~\ref{fig:8}(c). This suggests that the reproduction of the spectrum close to the cut-off scale for the shear component $-E^\mathrm{GS}_{xy}$ is realized by the increase of the wall-normal velocity fluctuation at that scale attributed to the EAT through the anisotropic redistribution term $\check{\xi}^\mathrm{EAT}_{yy}$.

\begin{figure}
 \centering
 \includegraphics[width=0.5\textwidth]{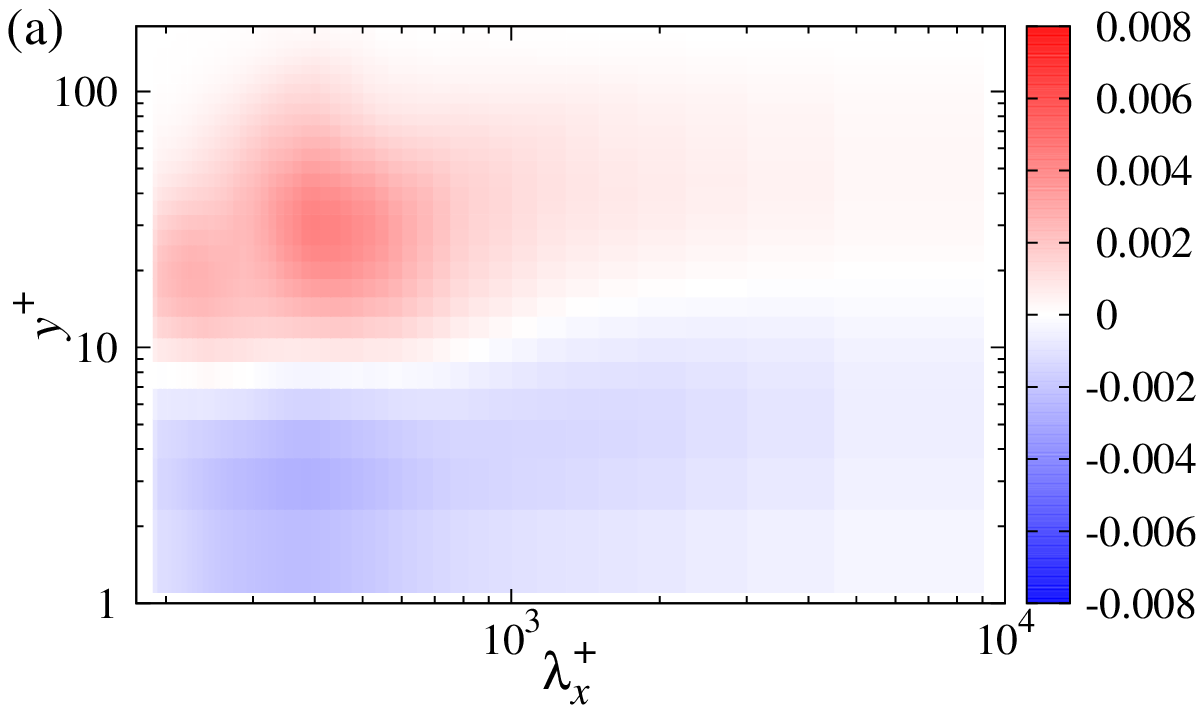}
 \includegraphics[width=0.5\textwidth]{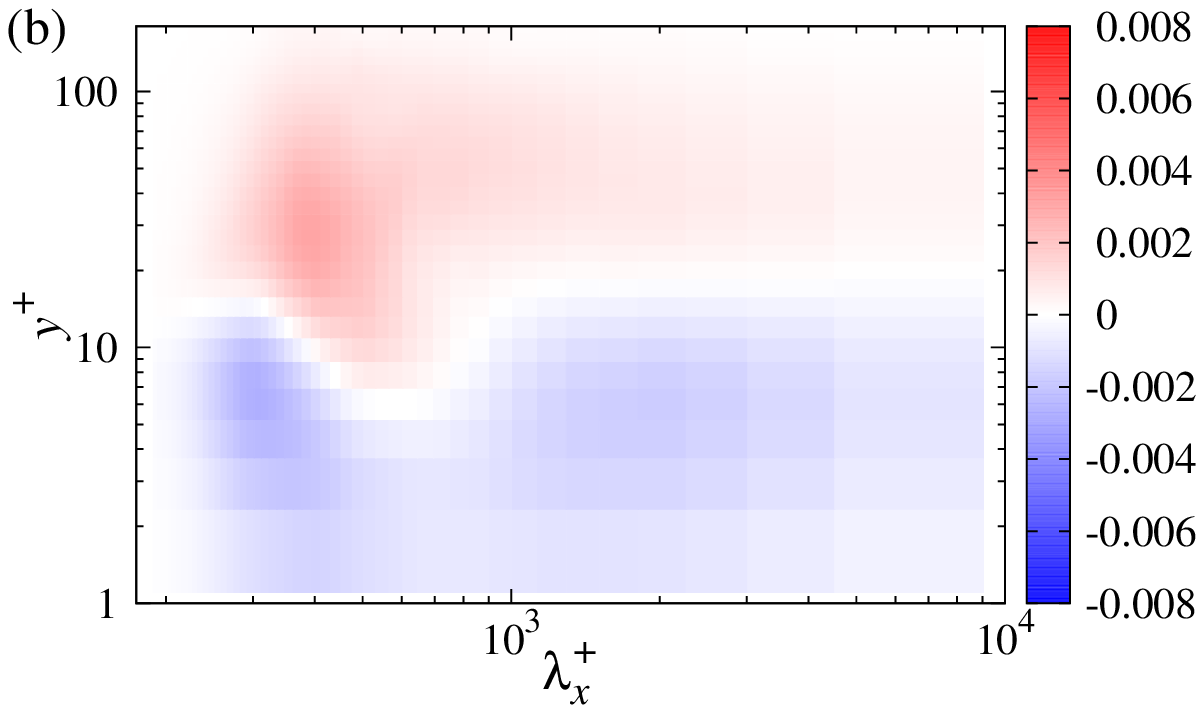}
 \includegraphics[width=0.5\textwidth]{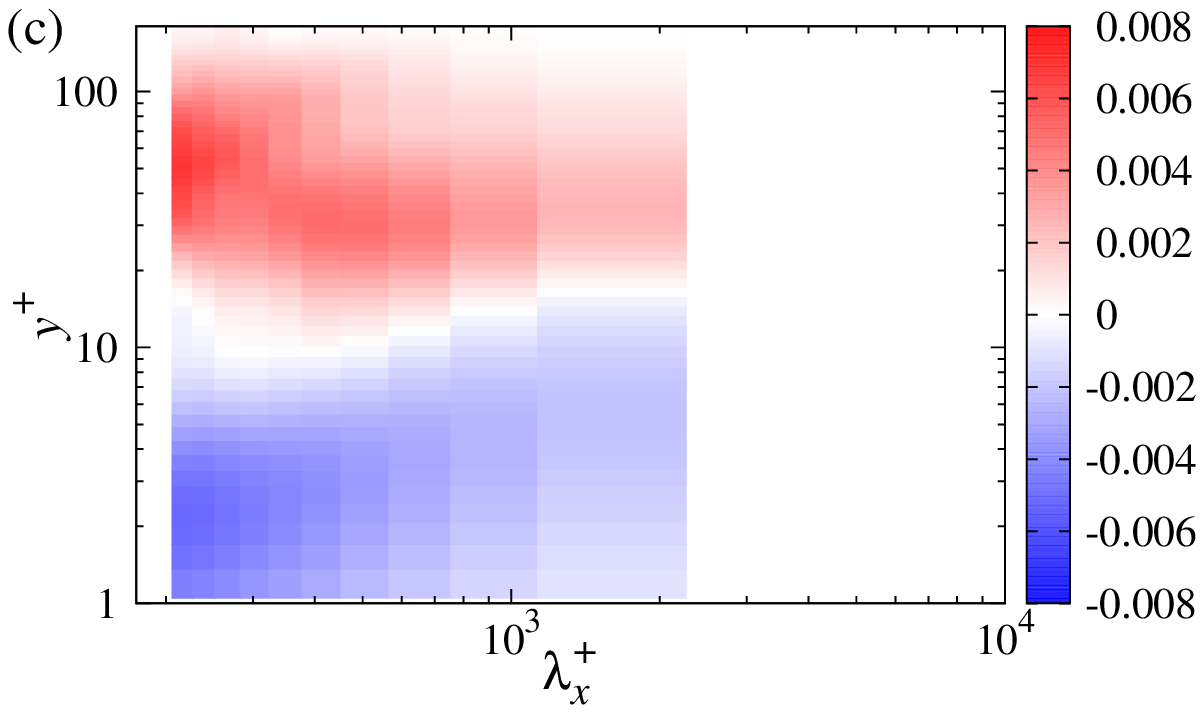}
\caption{\label{fig:11} Contour map of sum of the pressure--strain correlation and the anisotropic redistribution terms $\check{\Phi}^\mathrm{GS}_{yy} + \check{\xi}^\mathrm{EAT}_{yy}$ in budget for the wall-normal component of the GS Reynolds stress spectrum $E^\mathrm{GS}_{yy}$ normalized by viscous unit for (a) IA180LD, (b) IA-CL180LD, and (c) f-DNS at $\mathrm{Re}_\tau = 180$.}
\end{figure}

Figure~\ref{fig:11} shows the contour map of the sum of the pressure--strain correlation and the anisotropic redistribution terms $\check{\Phi}^\mathrm{GS}_{yy} + \check{\xi}^\mathrm{EAT}_{yy}$ in the budget equation for the wall-normal component of the GS Reynolds stress spectrum $E^\mathrm{GS}_{yy}$, for representative cases at $y^+=20$ for $\mathrm{Re}_\tau = 180$. All cases are similar, except for the region close to the cut-off wavelength scale, where IA exhibits a comparable value to the middle wavelength scale, as is the case with f-DNS. The negative contribution for $y^+ < 10$ corresponds to the `splatting' effect\cite{mk1982,lm2019}, leading to the two-component turbulence, as observed in Fig.~\ref{fig:7}(a). The positive contribution of $\check{\Phi}^\mathrm{GS}_{yy} + \check{\xi}^\mathrm{EAT}_{yy}$ in the region $y^+ > 10$ and $\lambda_x^+ < 300$ for IA in Fig.~\ref{fig:11}(a) leads to the reproduction of $E^\mathrm{GS}_{yy}$ close to the cut-off scale, as shown in Fig.~\ref{fig:8}(b). The reproduction of $E^\mathrm{GS}_{yy}$ close to the cut-off scale results in feedback to the budget of $E^\mathrm{GS}_{xy}$ as a source term through the production term $\check{P}^\mathrm{GS}_{xy}$.

\subsubsection{\label{sec:level4.1.3}Streamwise component}

\begin{figure}
 \centering
 \includegraphics[width=0.5\textwidth]{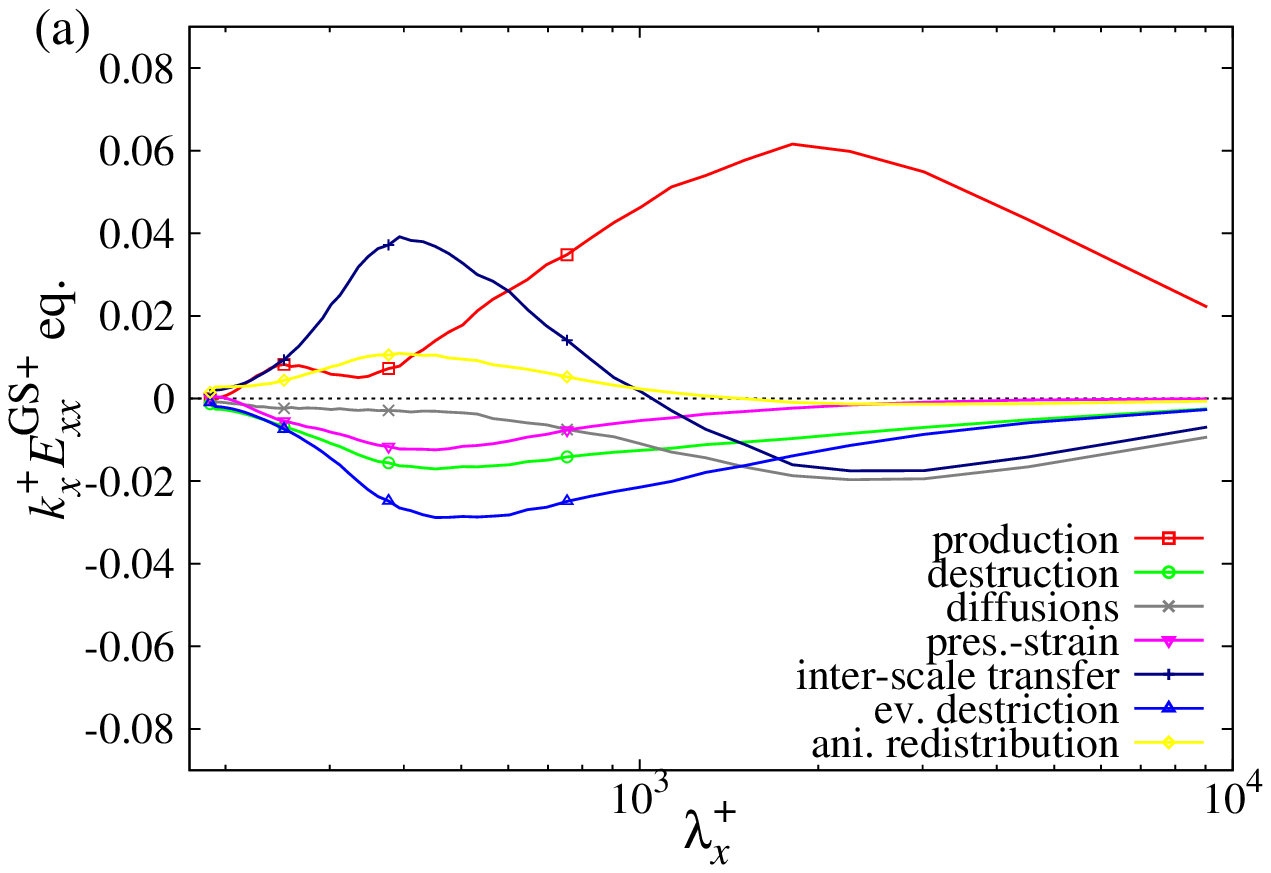}
 \includegraphics[width=0.5\textwidth]{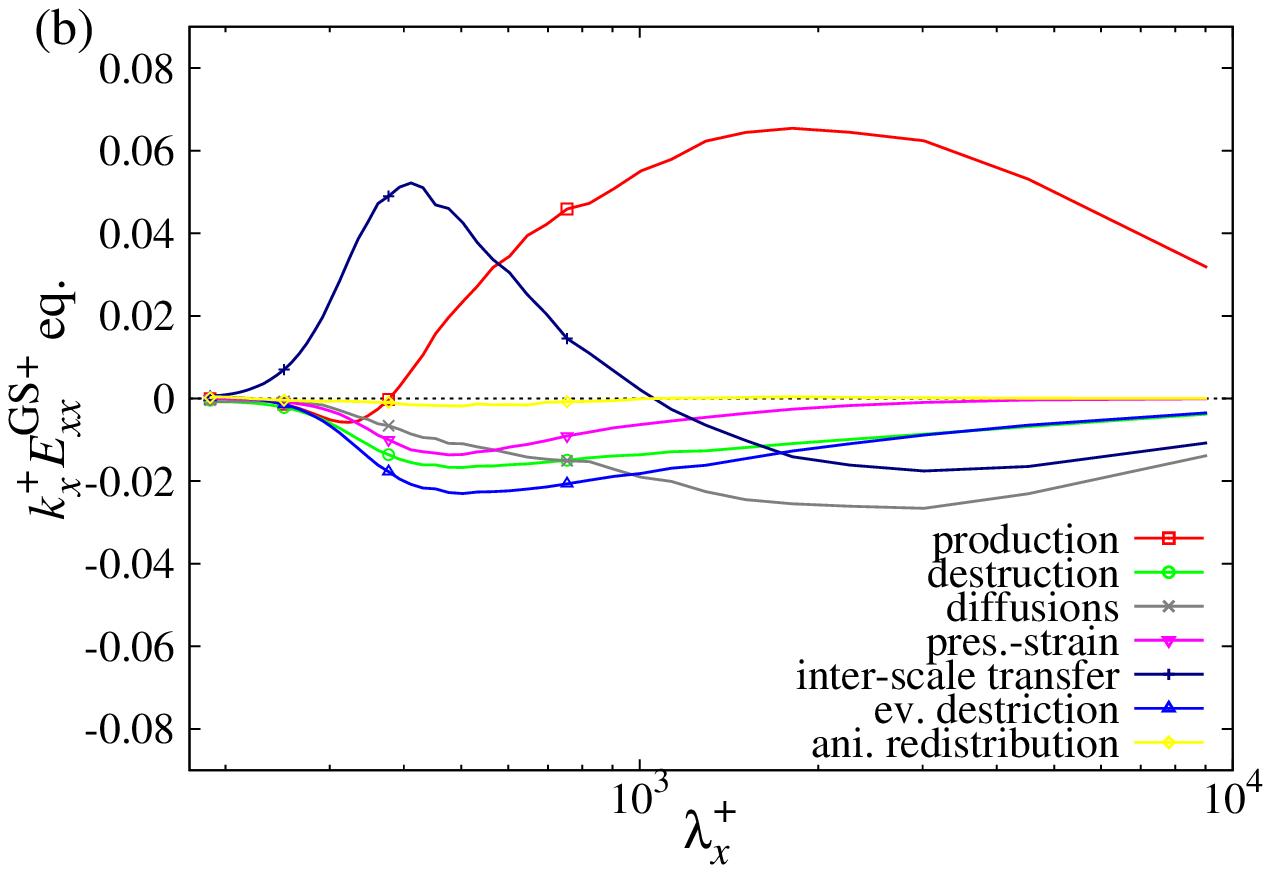}
 \includegraphics[width=0.5\textwidth]{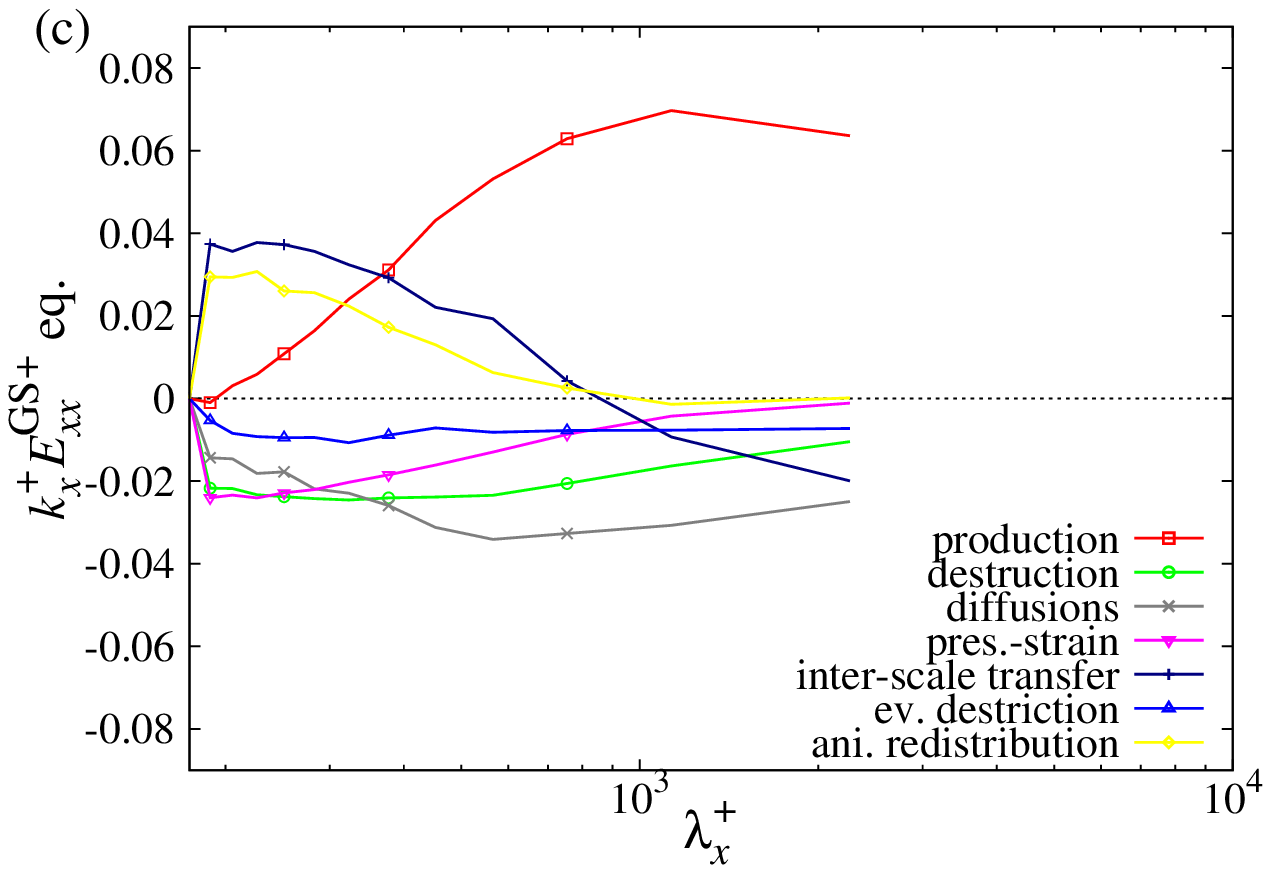}
\caption{\label{fig:12} Budget for the streamwise component of the GS Reynolds spectrum $E^\mathrm{GS}_{xx}$ normalized by viscous unit for (a) IA180LD, (b) IA-CL180LD, and (c) f-DNS at $y^+=20$ for $\mathrm{Re}_\tau = 180$}
\end{figure}

Figure~\ref{fig:12} shows the budget for the streamwise component of the GS Reynolds stress spectrum $E^\mathrm{GS}_{xx}$ for representative cases at $y^+=20$ for $\mathrm{Re}_\tau = 180$. The production term $\check{P}^\mathrm{GS}_{xx}$ for IA in Fig.~\ref{fig:12}(a) provides a positive value close to the cut-off wavelength scale owing to the reproduction of $E^\mathrm{GS}_{xy}$, because $\check{P}^\mathrm{GS}_{xx}$ reads
\begin{align}
\check{P}^\mathrm{GS}_{xx} = -2 E^\mathrm{GS}_{xy} \frac{\partial U_x}{\partial y}.
\label{eq:4.8}
\end{align}
Moreover, the anisotropic redistribution term $\check{\xi}^\mathrm{EAT}_{xx}$ is likewise positive in the entire wavelength range for IA. Although the profiles of each term are different between IA and f-DNS, IA succeeds in enhancing the GS streamwise velocity fluctuation in the low-wavelength region close to the cut-off scale. In contrast, in IA-CL, the anisotropic redistribution term $\check{\xi}^\mathrm{EAT}_{xx}$ contributes little to the budget in the entire wavelength range, and the production becomes negative close the cut-off wavelength scale. This leads to the rapid decay of the spectrum in the low-wavelength region, as shown in Fig.~\ref{fig:8}(a). Furthermore, the increase in the velocity fluctuation in the low-wavelength region leads to the enhancement of the dissipation. In fact, the eddy-viscosity destruction for IA in Fig.~\ref{fig:12}(a) is larger than that for IA-CL in Fig.~\ref{fig:12}(b). This accounts for the difference between the profiles of the GS streamwise velocity fluctuation for IA and IA-CL in Fig.~\ref{fig:5}(a). Hence, the overestimation of the GS streamwise velocity fluctuation can be restored by the increase of the turbulence close the cut-off wavelength scale.

\subsubsection{\label{sec:level4.1.4}Summary of analysis through budget for the GS Reynolds stress spectrum}

We summarize the physical role of the EAT in the SMM in turbulent channel flows. The EAT in IA model enhances the wall-normal component of the velocity fluctuation through the anisotropic redistribution term. This feeds the streamwise and shear components of the GS Reynolds stress in the low-wavelength region close to the cut-off scale, through the production term. Consequently, IA succeeds in activating turbulence in the entire wavelength range. To achieve this activation, the scale-similarity model for the SGS-Reynolds term is more appropriate than the Clark term, as the former is based on a higher-order spatial derivative than the latter. Hence, the scale-similarity model for the SGS-Reynolds term is efficient in enhancing the turbulence close to the cut-off scale. 

Anderson and Domaradzki\cite{ad2012} discussed the physics of inter-scale energy transfer among the largest, small resolved, and sub-grid scales, through scale-similarity models. They showed that the modified Leonard-like term yields an excessive dissipation directly from the largest resolved scales, which is unphysical in the concept of localness in scale of energy transfer. Such a non-local energy transfer does not occur in the SMM because the energy transfer between the GS and SGS due to the EAT is prohibited, although the EAT can redistribute energy between mean and the GS turbulent kinetic energies, as shown in Appendix~\ref{sec:b}. A failure of IA-CL may be partly attribute to the similar mechanism suggested by Anderson and Domaradzki\cite{ad2012}, because the Clark term retains only the leading term of the scale-similarity model for the SGS stress, as shown in Eq.~(\ref{eq:2.6}) and (\ref{eq:2.8}). Namely, the Clark term may be dominant in the largest resolved scale, although the statistical contribution of $\check{\xi}^\mathrm{EAT}_{xx}$ is negligible, as shown in Figs.~\ref{fig:12}(b). In contrast, the scale-similarity model for the SGS-Reynolds term (\ref{eq:2.17}) is composed only by the small resolved scale velocity, which more correlates to the small resolved scale than the largest scale. This is observed from $\check{\xi}^\mathrm{EAT}_{xx}$ in Figs.~\ref{fig:12}(a). Thereby, IA does not suffer from the non-local property of the scale-similarity model suggested by Anderson and Domaradzki\cite{ad2012}.

We also investigate the model where the EAT enters with a negative coefficient, $\tau^\mathrm{sgs}_{ij}|_\mathrm
{tl} = - 2 \nu^\mathrm{sgs} \overline{s}_{ij} - \tau^\mathrm{eat}_{ij}$, through the scale-similarity model for the SGS-Reynolds term, because this term yields a negative correlation on the energy transfer between the GS and SGS fields around an elliptic Burgers vortex\cite{kobayashi2018}. However, this model excessively overestimates the mean velocity (not shown). Moreover, it violates the realizability conditions\cite{schumann1977,vremanetal1994realizability} for the SGS stress, as $\langle \tau^\mathrm{sgs}_{xx} \rangle < 0$. This is because it usually yields $\tau^\mathrm{a}_{xx}/\tau^\mathrm{a}_{\ell \ell} > 2/3$ with Eq.~(\ref{eq:2.17}) in turbulent channel flows, resulting in $\langle \tau^\mathrm{sgs}_{xx} \rangle \simeq \langle 2 k^\mathrm{sgs} [1/3 - (\tau^\mathrm{a}_{xx}/\tau^\mathrm{a}_{\ell \ell}-1/3)] \rangle < 0$. Furthermore, the form of the EAT provided in Eq.~(\ref{eq:2.16}) is likewise a key element. The present simulation of IA shows that $\langle 2k^\mathrm{sgs}/\tau^\mathrm{a}_{\ell \ell} \rangle \gg 1$. Hence, the contribution of the scale-similarity model is enhanced as compared with the case when it is simply added. These results suggest that it is not sufficient to simply add the EAT to improve the SGS model. Instead, we should properly express the EAT, which efficiently excites turbulence close to the cut-off scale.

\subsection{\label{sec:level4.2}Streak structures in stabilized mixed models}

\begin{figure}
 \centering
 \includegraphics[width=0.5\textwidth]{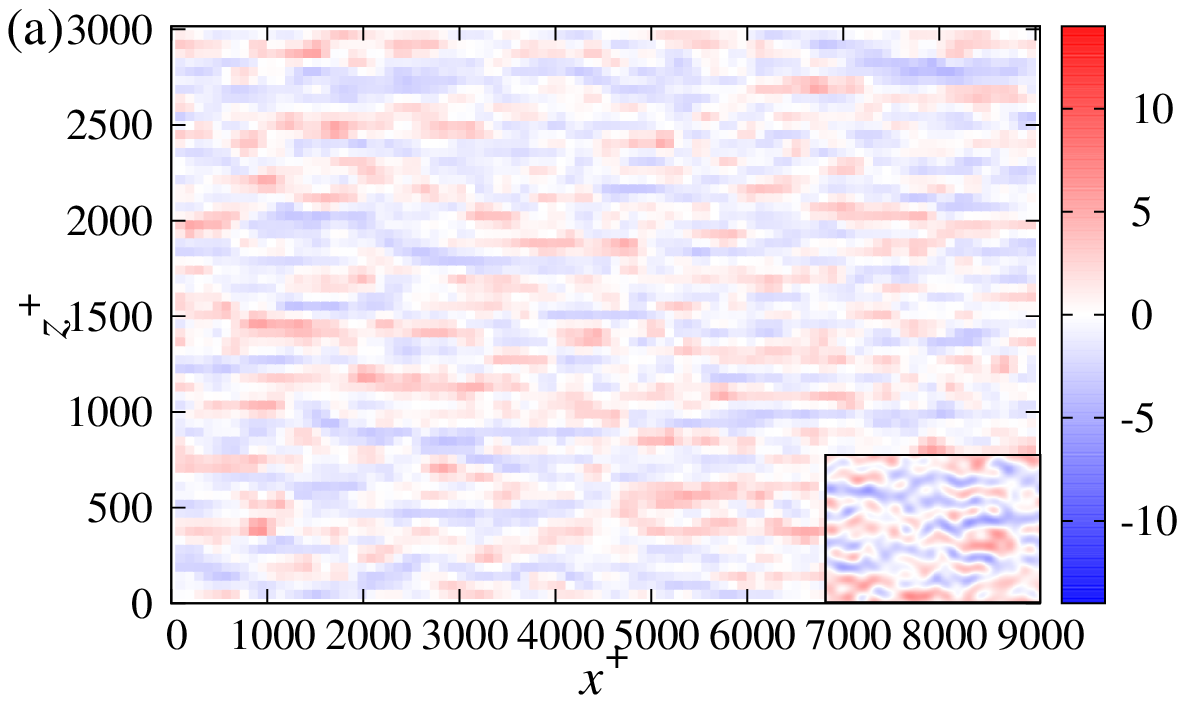}
 \includegraphics[width=0.5\textwidth]{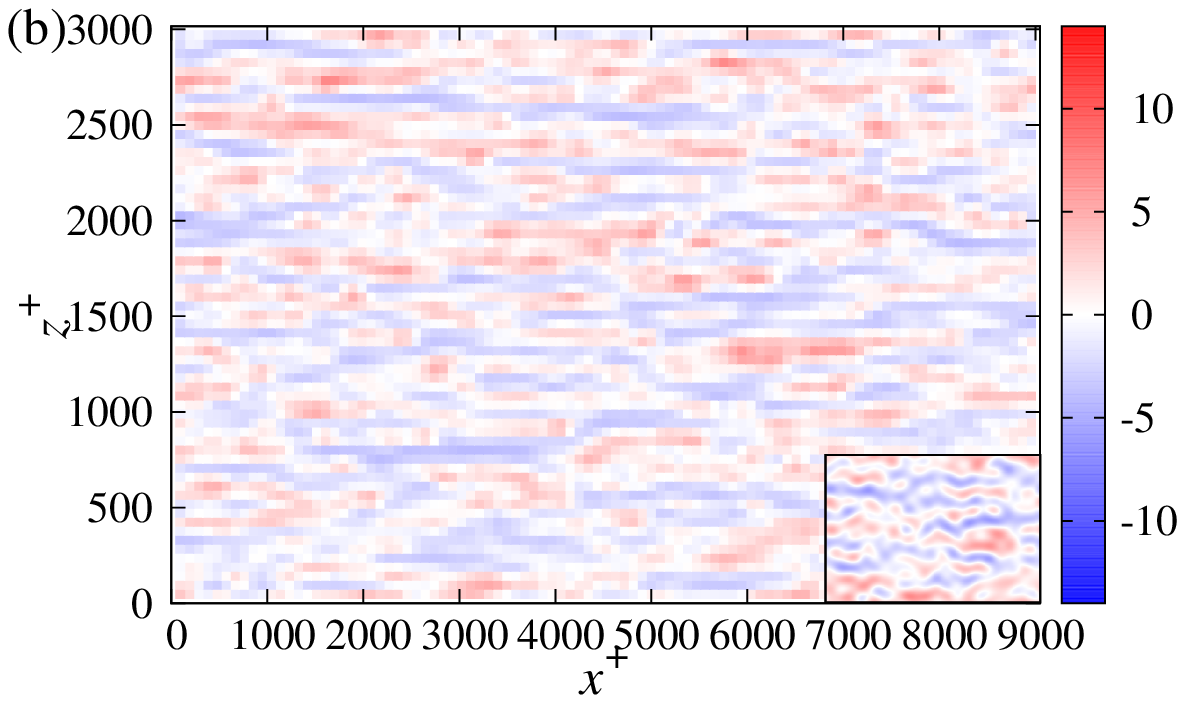}
 \includegraphics[width=0.5\textwidth]{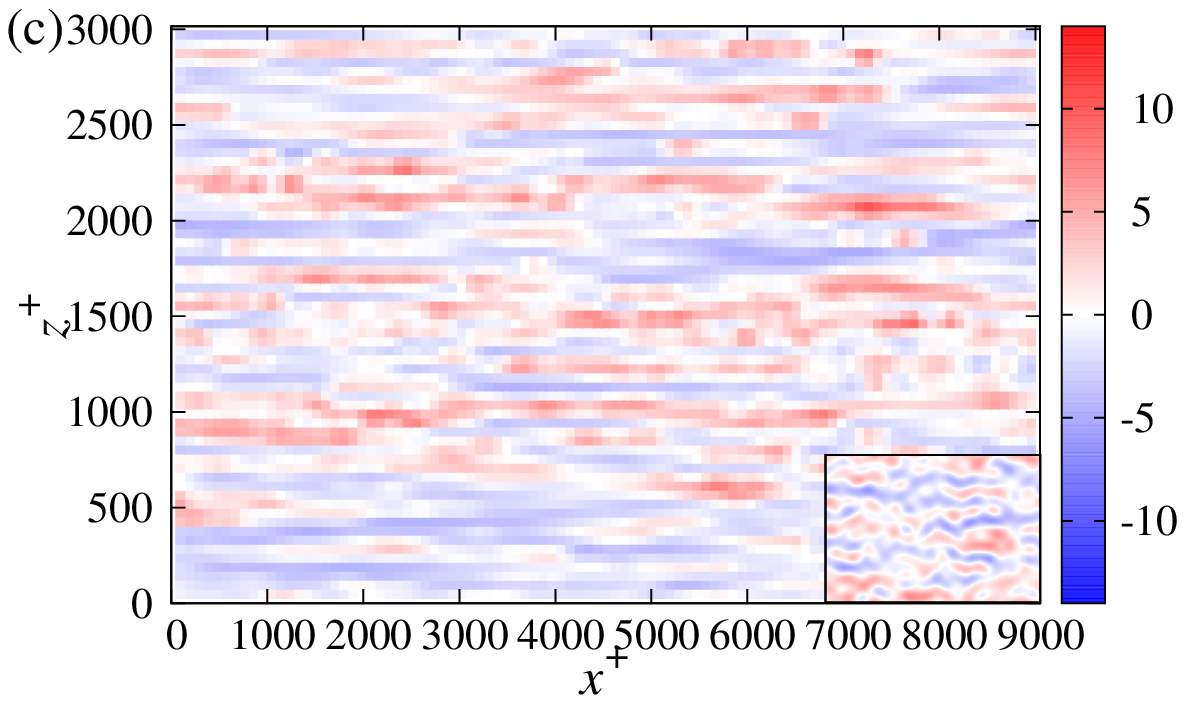}
\caption{\label{fig:13} Contour map of instantaneous streamwise velocity fluctuation $\overline{u}_x-\langle \overline{u}_x \rangle_{x\text{--}z\text{-plane}}$ for (a) IA180LD, (b) IA-CL180LD, (c) IA-LNcs42-180LD at $y^+=20$ for $\mathrm{Re}_\tau=180$. The bottom right inset in each figure shows the result for f-DNS with the exact domain size.}
\end{figure}

As shown in Fig.~\ref{fig:8}(a), all SGS models except for IA result in the rapid decay of the spectrum of the streamwise turbulent fluctuation in the low-wavelength region. The spectrum accumulated in the high-wavelength region corresponds to the flow structure excessively elongated in the streamwise direction. Figure~\ref{fig:13} shows the contour map of the instantaneous streamwise velocity fluctuation $\overline{u}_x-\langle \overline{u}_x \rangle_{x\text{--}z\text{-plane}}$ at $y^+=20$ for $\mathrm{Re}_\tau=180$. Figure~\ref{fig:13}(c) indicates that IA-LNcs42 predicts excessively elongated streamwise streak structures. In Figs.~\ref{fig:13}(a) and (b), IA and IA-CL predict reasonable streak structures in comparison with f-DNS, although IA-CL exhibits the rapid decay of the spectrum in the low-wavelength region. To quantitatively evaluate the length of the streaks, we investigate the streamwise velocity correlation function $C^\mathrm{GS}_{xx}$ defined by
\begin{align}
C^\mathrm{GS}_{xx} (r_x,y) = \frac{\langle \overline{u}_x' (\bm{x}+r_x \mathbf{e}_x) \overline{u}_x' (\bm{x}) \rangle}{\langle \overline{u}_x'{}^2 \rangle},
\label{eq:4.9}
\end{align}
where $\mathbf{e}_x$ denotes the unit vector in the $x$ direction. Figure~\ref{fig:14} shows the streamwise velocity correlation at $y^+ = 20$ for $\mathrm{Re}_\tau=180$. IA reproduces a rapid decay of the correlation, while IA-LNcs42 predicts a more moderate decay. In LR, the decay of IA-LNcs42 is considerably more moderate, and the correlation does not decrease sufficiently within this domain size. This situation is similar for DSM. IA and IA-CL predict the rapid decay even in LR. IA-CL likewise predicts the rapid decay, albeit it is slightly more moderate compared with IA or f-DNS. Nevertheless, IA-CL overestimates the streamwie component of the GS velocity fluctuation, as shown in Fig.~\ref{fig:5}(a). In this sense, the SMM using the scale-similarity model for the SGS-Reynolds term is the most appropriate among the present cases in reproducing the turbulent structures including its intensity in the wall-bounded turbulent shear flow.

\begin{figure}
 \centering
 \includegraphics[width=0.5\textwidth]{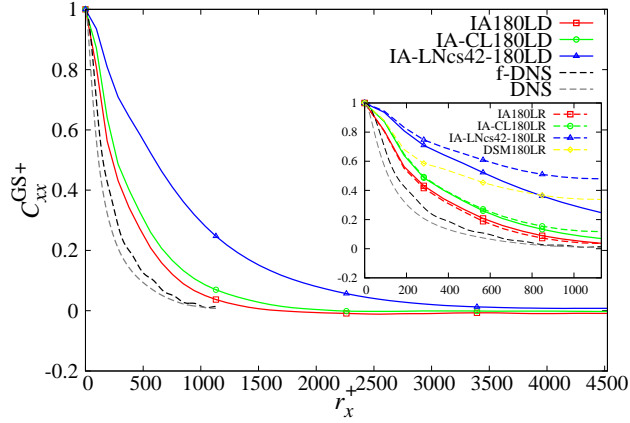}
\caption{\label{fig:14} Profile of streamwise velocity correlation $C^\mathrm{GS}_{xx}$ at $y^+=20$ for $\mathrm{Re}_\tau=180$. The right screen depicts LR cases. The result for the non-filtered DNS is plotted for reference.}
\end{figure}

\section{\label{sec:level5}Conclusions}

We investigated the physical role of various scale-similarity models in the SMM\cite{abe2013,ia2017}. The SMM proposed by Abe\cite{abe2013} adopted the scale-similarity model for the SGS-Reynolds term although, other scale-similarity models yield a better correlation with the exact SGS stress. In the present study, we applied various scale-similarity models to the SMM and evaluated their performance in turbulent channel flows. As previously shown in previous studies\cite{abe2013,oa2013,abe2014,abe2019,kobayashi2018}, the SMM using the scale-similarity model for the SGS-Reynolds term is less sensitive to the grid resolution than the conventional eddy-viscosity models in the prediction of the mean velocity. In particular, it can predict the near-wall mean velocity profile even in coarse grid resolutions at both low- and high-Reynolds numbers. In various SMMs, the original model using the SGS-Reynolds term provides the best prediction of the Reynolds stress, whereas other models overestimate the GS streamwise velocity fluctuation. We also investigated Lumley's invariant map\cite{hanjaliclaunderbook} to quantitatively evaluate the anisotropy of the GS and SGS turbulent stress. The result indicates that the original model predicts a similar near-wall behavior as the filtered DNS. The GS velocity fluctuations for the eddy-viscosity models result in a nearly one-component turbulence in the vicinity of the solid wall, instead of the conventional two-component state. Moreover, the eddy-viscosity model cannot predict the anisotropy of the SGS stress, which reflects the isotropic property of the eddy-viscosity model. A critical difference between various scale-similarity models is found in the spectra of the GS Reynolds stress close to the cut-off scale. The original SMM using the scale-similarity model for the SGS-Reynolds term succeeds in predicting the large intensities of the spectra close to the cut-off scale in accordance with the filtered DNS, whereas other models predict a rapid decay of the spectra in the low-wavelength region. The success of the scale-similarity model for the SGS-Reynolds term relies on the property that it is expressed by the higher-order spatial derivative, unlike other scale-similarity models.

To investigate the behavior of the models close to the cut-off scale, we analyzed the budget equation for the GS Reynolds stress spectrum\cite{mizuno2016,ka2018,lm2019}. As a result, it was shown that the scale-similarity model for the SGS-Reynolds term plays a role in enhancing the wall-normal component of the GS velocity fluctuation close to the cut-off scale. This leads to the enhancement of the streamwise and shear components of the GS Reynolds stress in that scale through the production term. Hence, the activation of turbulence close to the cut-off scale is achieved. Owing to these properties, the streak structures observed in wall-bounded turbulent flows are successfully reproduced. Although the SMM employing the scale-similarity model for the SGS-Reynolds term does not predict the overall profiles of the budget of the GS Reynolds stress spectrum obtained from the filtered DNS, it predicts both the statistics and structures in the wall-bounded turbulent flow at the coarse grid resolution. For further development of SGS models, one should consider how to reproduce the turbulence structures including the low-wavelength region close to the cut-off scale.

\begin{acknowledgments}
The authors would like to acknowledge Prof. K. Abe for valuable discussions. K. I. is grateful to Prof. F. Hamba for fruitful comments and discussions. The work of H. K. is supported in part by Keio Gijuku Academic Development Funds. We would also like to thank the referees for valuable comments for improvement of this paper.
\end{acknowledgments}

\appendix

\makeatletter
\renewcommand{\theequation}{\thesection\arabic{equation}}
\@addtoreset{equation}{section}
\makeatother

\section{\label{sec:a}Contribution of the eddy-viscosity term in SGS normal stress}

Figure~\ref{fig:a1} shows the profile of the eddy-viscosity term and the EAT in the normal component of the SGS stress for IA180LR. The eddy-viscosity term is negligible compared with the EAT. IA-CL and f-DNS exhibit a similar result (not shown). The eddy-viscosity term in IA-LNcs42 is also negligible compared with the SGS kinetic energy $k^\mathrm{sgs}$, which can be found in the Lumley invariant map in Fig.~\ref{fig:7}(b). Hence, the eddy-viscosity model cannot predict the anisotropy in turbulent channel flows.

\begin{figure}
 \centering
 \includegraphics[width=0.5\textwidth]{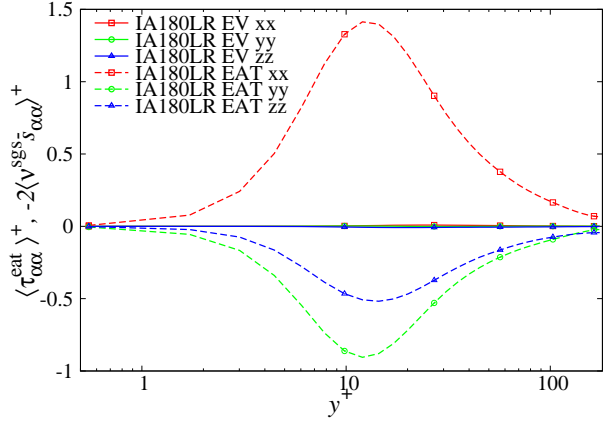}
\caption{\label{fig:a1} Profile of the eddy-viscosity term and EAT in the normal components of SGS stress for IA180LR. Solid lines denote the eddy-viscosity term, while dashed lines denote the EAT.}
\end{figure}

\section{\label{sec:b}Budget for the GS Reynolds stress}

The budget equation for the GS Reynolds stress yields
\begin{align}
& \frac{\partial R^\mathrm{GS}_{ij}}{\partial t} + \frac{\partial}{\partial x_\ell} (U_\ell R^\mathrm{GS}_{ij})
\nonumber \\
& = P^\mathrm{GS}_{ij} - \varepsilon^\mathrm{GS}_{ij}
+ D^\mathrm{t,GS}_{ij} + \Phi^\mathrm{GS}_{ij} + D^\mathrm{p,GS}_{ij}
+ D^\mathrm{v,GS}_{ij}
\nonumber \\
& \hspace{1em}
- \varepsilon^\mathrm{EV}_{ij} + \xi^\mathrm{EAT}_{ij} + D^\mathrm{SGS}_{ij}.
\label{eq:b1}
\end{align}
Terms on the right-hand side are similar to those in Eq.~(\ref{eq:4.1}). Note that the inter-scale transfer term vanishes when it is summed over the wavenumbers:
\begin{align}
\sum_{k_x=0}^{k_x^\mathrm{max}} \check{T}^\mathrm{GS}_{ij} \Delta k_x = 0.
\label{eq:b2}
\end{align}
The trace of the pressure--strain correlation $\Phi^\mathrm{GS}_{ij}$ should disappear due to incompressibility:
\begin{align}
\Phi^\mathrm{GS}_{ii} = 0.
\label{eq:b3}
\end{align}
Therefore, it is sometimes referred to as the redistribution term, which plays a role of the redistribution of intensities among normal stress components. In contrast, the trace of the anisotropic redistribution term does not vanish, $\xi^\mathrm{EAT}_{ii} \neq 0$, even though $\tau^\mathrm{eat}_{ij}$ does not exchange energy between the GS and SGS fields, as shown in Eq.~(\ref{eq:2.14}). This is because
\begin{align}
\xi^\mathrm{eat}_{ii} = 2 \left< \tau^\mathrm{eat}_{ij} \overline{s}_{ij} \right> - 2\left< \tau^\mathrm{eat}_{ij} \right> S_{ij} = - 2 \left< \tau^\mathrm{eat}_{ij} \right> S_{ij} \neq 0.
\label{eq:b4}
\end{align}
The mean kinetic energy equation reads
\begin{align}
\frac{\partial}{\partial t} \left( \frac{1}{2} U_i U_i \right) = 
- \left< \tau^\mathrm{eat}_{ij} \right> S_{ij} + \cdots.
\label{eq:b5}
\end{align}
Therefore, $\langle \tau^\mathrm{eat}_{ij} \rangle S_{ij}$ is interpreted as the energy transfer between the mean and SGS kinetic energies, while $\xi^\mathrm{EAT}_{ii}/2$ is that between the GS turbulent and SGS kinetic energies. Equation~(\ref{eq:b4}) indicates that the amount energy transfer between the mean and SGS kinetic energies must be equal to that between GS and SGS turbulent kinetic energies. In other words, a change of the mean kinetic energy must be compensated as the GS turbulent kinetic energy. Thereby, $\xi^\mathrm{EAT}_{ii}/2$ plays a role of the redistribution between the mean and GS turbulent kinetic energies. Hence, we named $\xi^\mathrm{EAT}_{ij}$ the anisotropic redistribution term.

Figure~\ref{fig:b1} shows the profile of the trace part of the eddy-viscosity destruction $-\varepsilon^\mathrm{EV}_{ii}/2$ and the anisotropic redistribution term $\xi^\mathrm{EAT}_{ii}/2$ for various SMMs in LR at $\mathrm{Re}_\tau = 180$. For all cases, the contribution of $\xi^\mathrm{EAT}_{ii}$ is relatively small compared with $\varepsilon^\mathrm{EV}_{ii}$. However, IA provides a positive $\xi^\mathrm{EAT}_{ii}$ at $y^+=20$ in the same manner as f-DNS, supporting the increase in the GS velocity fluctuations. The success of IA may partly lie in the property of EAT, which enhances the GS velocity fluctuations in the buffer layer $10 < y^+ < 50$.

\begin{figure}
 \centering
 \includegraphics[width=0.5\textwidth]{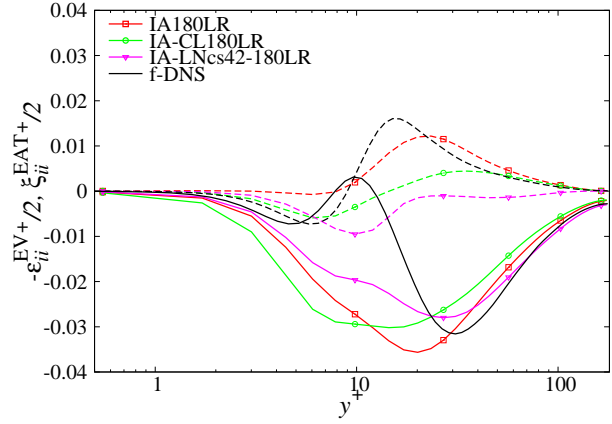}
\caption{\label{fig:b1}Profile of trace part of the eddy-viscosity destruction $-\varepsilon^\mathrm{EV}_{ii}/2$ and the anisotropic redistribution term $\xi^\mathrm{EAT}_{ii}/2$ for various SMMs in LR at $\mathrm{Re}_\tau = 180$.}
\end{figure}

\section*{DATA AVAILABILITY}

The data that support the findings of this study are available from the corresponding author upon reasonable request.

\bibliography{ref}

\end{document}